\documentclass[12pt]{article}
\pdfoutput=1

\usepackage{gensymb}
\usepackage{array} 
\usepackage{amssymb}
\usepackage{graphics,graphpap}
\usepackage{graphicx}
\usepackage{color}
\usepackage{graphicx}
\usepackage{dcolumn}
\usepackage{epsfig}
\usepackage{epstopdf}
\usepackage{bbm}
\usepackage{amsmath}
\usepackage{amsfonts}
\usepackage{textcomp}
\usepackage{subfigure}
\usepackage{setspace}
\usepackage{slashed}
\usepackage{tikz}
\usetikzlibrary{arrows,decorations.pathreplacing}

\usepackage{todonotes}

\usepackage{cite}

\usepackage{url}

\usepackage{booktabs}

\usepackage{float}

\newcommand{\eff}[3]{\overset{(#1)}{\kappa}_{\hspace{-.5ex} #3}^{\raisebox{-.8ex}{\scriptsize$(#2#2)$}}}
\renewcommand{\Re}{\mathrm{Re}}
\renewcommand{\Im}{\mathrm{Im}}

\newcommand{\beq}{\begin{eqnarray}}
\newcommand{\eeq}{\end{eqnarray}}

\newcommand{\bmp}{\noindent\begin{minipage}{16cm}}
\newcommand{\emp}{\end{minipage}\vskip 7mm} 

\def\drawbox#1#2{\hrule height#2pt
        \hbox{\vrule width#2pt height#1pt \kern#1pt
              \vrule width#2pt}
              \hrule height#2pt}

\def\Asym#1#2{\vcenter{\vbox{\drawbox{#1}{#2}
              \kern-#2pt 
              \drawbox{#1}{#2}}}}

\def\simge{\mathrel{%
   \rlap{\raise 0.511ex \hbox{$>$}}{\lower 0.511ex \hbox{$\sim$}}}}

\def\simle{\mathrel{
   \rlap{\raise 0.511ex \hbox{$<$}}{\lower 0.511ex \hbox{$\sim$}}}}

\def\s#1{\setbox0=\hbox{$#1$}%
\rlap{\ifdim\wd0>.7em\kern.22\wd0\else\kern.1\wd0\fi /}#1}

\begin{document}

\begin{titlepage}
\title{\vspace*{-2.0cm}
\hfill {\small MPP-2015-162}\\[20mm]
\bf\Large
Running of Radiative Neutrino Masses:\\ The Scotogenic Model -- REVISITED
\\[5mm]\ }

\author{
Alexander Merle\thanks{email: \tt amerle@mpp.mpg.de}~~~and~~Moritz Platscher\thanks{email: \tt mplat@mpp.mpg.de}
\\ \\
{\normalsize \it Max-Planck-Institut f\"ur Physik (Werner-Heisenberg-Institut),}\\
{\normalsize \it F\"ohringer Ring 6, 80805 M\"unchen, Germany}\\
}
\date{\today}
\maketitle
\thispagestyle{empty}

\begin{abstract}
\noindent
A few years ago, it had been shown that effects stemming from renormalisation group running can be quite large in the scotogenic model, where neutrinos obtain their mass only via a 1-loop diagram (or, more generally, in many models in which the light neutrino mass is generated via quantum corrections at loop-level). We present a new computation of the renormalisation group equations (RGEs) for the scotogenic model, thereby updating previous results. We discuss the matching in detail, in particular in what regards the different mass spectra possible for the new particles involved. We furthermore develop approximate analytical solutions to the RGEs for an extensive list of illustrative cases, covering all general tendencies that can appear in the model. Comparing them with fully numerical solutions, we give a comprehensive discussion of the running in the scotogenic model. Our approach is mainly top-down, but we also discuss an attempt to get information on the values of the fundamental parameters when 
inputting the low-energy measured quantities in a bottom-up manner. This work serves the basis for a full parameter scan of the model, thereby relating its low- and high-energy phenomenology, to fully exploit the available information.
\end{abstract}

\end{titlepage}

\section{\label{sec:intro}Introduction}

Neutrinos are probably our best handles to detect physics beyond the Standard Model (SM) of elementary particle physics, most certainly in the absence of new physics signals at the Large Hadron Collider (LHC) and without a clear detection of a Dark Matter (DM) particle with well-defined properties. In the last decades, we have experienced tremendous successes of experimental neutrino physics. Starting with the first observations of neutrino oscillations~\cite{Fukuda:1998mi,Eguchi:2002dm}, which proved that at least some light neutrinos must have a non-zero mass, we have by now obtained a fairly complete picture of leptonic mixing~\cite{Ahn:2006zza,Abe:2012gx,An:2012eh,Ahn:2012nd,Abe:2012tg}: all three leptonic mixing angles $\theta_{ij}$, which describe the mismatch between the neutrino mass and flavour bases, have been measured and we have determined two mass square differences which constrain the neutrino mass spectrum. Yet, we do not know the neutrino mass ordering (i.e., which mass eigenstate is the 
lightest), we do not know whether neutrinos are equal to their own antiparticles (related to an observation of neutrinoless double beta decay), and we do not know their absolute mass. The only information we have on the latter is that is can at most amount to about 1~eV~\cite{Lobashev:2000vb,Kraus:2004zw}.

One possibility to explain the smallness of neutrino masses is by tree-level suppression mechanisms, such as the famous type-I
seesaw~\cite{Minkowski:1977sc,Yanagida:1979as,GellMann:1980vs,Glashow:1979nm,Mohapatra:1979ia,Schechter:1980gr}. However, a generic problem at least with the simplest setting is that it is hardly
testable, since the new fields involved are generically very heavy. An alternative route to go is to generate light neutrino masses at loop-level instead~\cite{Farzan:2012ev}. In such models, the
neutrino mass vanishes exactly at tree-level, but it is generated by loop-diagrams as a small correction to the tree-level Lagrangian. The easiest such models generate a neutrino mass at
1-loop~\cite{Zee:1980ai,Ma:2006}, see~\cite{Bonnet:2012kz} for a general classification, although
2-loop~\cite{Zee:1985id,Babu:1988ki,Kajiyama:2013zla,Aoki:2013gzs,Baek:2014awa,King:2014uha,Farzan:2014aca}\footnote{Also for this case a general classification is available, see
Ref.~\cite{Sierra:2014rxa}.} or 3-loop~\cite{Gustafsson:2012vj,Gustafsson:2014vpa,Ahriche:2014cda,Chen:2014ska,Hatanaka:2014tba} mass generation is possible, too.\footnote{In principle one could go to
even more loops, see Ref.~\cite{Duerr:2011zd} for a 4-loop example, however, at that point the light neutrino masses would become too small to agree with the experimental bounds.}

The most attractive point of radiative neutrino mass models is that they intrinsically connect the phenomenology of light neutrinos to other sectors. The reason is that the particles contained in the loop-diagram generating the neutrino mass are typically not ``invisible'', but rather they have quantum numbers that make them observable. For example, the Zee-Babu model~\cite{Zee:1985id,Babu:1988ki} contains new scalar fields that are electrically charged and could thus be detected at colliders~\cite{Nebot:2007bc,Herrero-Garcia:2014hfa}. Another instance is the Ma-model~\cite{Ma:2006} studied in this paper, in which the loop contains particles that could act as DM. In general, radiative models prove interesting from a phenomenological point of view because such connections to other sectors comprise a handle to distinguish them when combining data from different sectors. A generic example is to use input from both low- and high-energy experiments and to exploit their complementarity to strongly constrain 
settings with a radiative neutrino mass~\cite{King:2014uha}. However, radiative models are also intrinsically interesting because of their structure forcing light neutrino masses to depend on many qualitatively different model parameters at the same time. This is the key for renormalisation group running influencing radiative models in a very non-trivial way.

Renormalisation group running of such radiative models is still a comparatively young field, first mentioned in Ref.~\cite{Kniehl:1996bd}. In fact, at least for the so-called scotogenic model~\cite{Ma:2006},\footnote{The curious name of this model derives from Ancient Greek and can be roughly translated into English as ``generated by darkness'', which refers to the DM candidates in this model being part of the loop-diagram that generates a non-zero active neutrino mass.} which is arguably the most simple setting with a 1-loop neutrino mass and full agreement with experiments up to now, the pioneering work on the renormalisation of radiative neutrino masses has only been presented in 2012~\cite{Bouchand:2012}.\footnote{See also the later Ref.~\cite{Babu:2014kca} for the Zee-Babu model.} In this reference, the general tendencies of the renormalisation of radiative neutrino masses have been worked out:
\begin{itemize}

\item {\bf The running can be (very) strong:}\\
This is easily understood, since loop diagrams depend on products of couplings such that, if the two factors $a_{1,2}$ in a product $a_1 \times a_2$ both receive corrections of the form $a_i \to a_i + \Delta a_i$, their product receives a correction $\Delta (a_1 \times a_2) = a_1 \times \Delta a_2 + a_2 \times \Delta a_1$, which can result in enhancements if the terms add up. This fact is trivially reflected in the light neutrino mass matrix for the model discussed here, cf.\ Eq.~\eqref{eq:MaMass}.

\item {\bf There is no inconsistency:}\\
One may ask whether it at all makes sense to work out a 1-loop correction to a 1-loop diagram leading to a light neutrino mass. However, one has to keep in mind that, in fact, what is corrected by the RGEs is not the mass itself but rather the Lagrangian of the model. If any diagram of arbitrary order is then computed using the RG-improved Lagrangian parameters, one consequently obtains an improved result.

\item {\bf The origin of lepton number violation becomes clearer:}\\
A certain set of the model parameters (in the scotogenic model discussed here these are the neutrino Yukawa coupling matrix $h_{ij}$, the right-handed (RH) neutrino masses $M_k$, and the 4-scalar coupling $\lambda_5$, to be explained in a moment in Sec.~\ref{sec:Overview}) drives the lepton number violation (LNV) but, if one of them is set to zero, one could in fact define a conserved version of the lepton number. Thus, one would expected these LNV parameters to be naturally small in the 't Hooft sense~\cite{'tHooft:1979bh}, since the symmetry of the Lagrangian would be increased if any of these parameters was set to zero. This is reflected in the corresponding RGEs, as they only allow for \emph{multiplicative} corrections to the decisive couplings; thus, if one of the LNV parameters is zero at \emph{any} scale, it can \emph{never} be generated radiatively. This tendency is quite generic for radiative neutrino mass models: while running effects can considerably change certain observables, they are typically 
not powerful enough to break lepton number in the first place, at least unless the model contains a scalar whose radiatively generated vacuum expectation value would do the job. Instead, most radiative neutrino mass models contain one or more sectors which intrinsically break lepton number, and this breaking is then only \emph{translated} into the light neutrino sector. More technically, this is reflected in the light neutrino mass being proportional to all LNV couplings, cf.\ Eq.~\eqref{eq:MaMass} for the case at hand.

\end{itemize}
Furthermore, Ref.~\cite{Bouchand:2012} has discussed several technical aspects such as how to correctly match different effective field theories (EFT) for loop-realisations of the Weinberg operator~\cite{Weinberg:1979sa} or how to correctly integrate out the heavy RH neutrinos present in the scotogenic model.

In this work, we will revisit the renormalisation group running in the scotogenic model in a more illustrative way and thereby extend, confirm, and update previous results. This manuscript presents a follow-up to~\cite{Bouchand:2012} which adds several new and important aspects to what has been known before. First, while the discussion of the matching in Ref.~\cite{Bouchand:2012} was rather technical, we add a more intuitive picture by dropping some of the more formal expressions in favour of graphically matching the different EFTs and directly relating the diagrams to the corresponding formula. Second, while~\cite{Bouchand:2012} relied on a purely numerical analysis of the RGEs, we present approximate analytical solutions to the equations for several illustrative cases, wherever possible. We relate the approximate analytical results to the numerical computations presented, which enables the reader to get an intuitive understanding of the behaviour of the quantities involved. And third, by presenting results 
for several limiting cases not discussed in~\cite{Bouchand:2012} (all of which are consistent with the low-energy neutrino data), we develop an intuitive feeling for the different regimes that can appear, depending on the region in parameter space that is considered.

Note that in this paper we focus on presenting \emph{illustrative} examples for pedagogical reasons, which serve to clearly exhibit the  various effects running can have in the scotogenic model. This means in particular that we compute the renormalisation group evolution from a very high to a comparatively low energy scale, as the effects are visible more clearly in this case. In reality, however, one may be interested only in a comparison between collider and low-energy data sets, in which case the running can be confined to this range. Of course, the tools presented also serve for such a case, so that it should not be any problem for the reader to restrict our considerations to such a case. Furthermore, different readers may regard some of the examples presented as more ``generic'' or ``natural'' than others. While of course such a viewpoint may be very well motivated in a concrete setting, we would like to stress that none of the scenarios presented is experimentally excluded (possibly up to the neutrino 
oscillation parameters, which is just what we want to investigate). This is done purposely, given that we deliberately aim at presenting what running in the scotogenic model can do, and what it cannot. A detailed phenomenological study where all possible bounds are taken into account in greatest detail is left for future work, in favour of first presenting an illustrative study which allows the reader to get a global understanding of the running of radiative neutrino masses.\\

This paper is structured as follows: In Sec.~\ref{sec:Overview} we give a brief overview of the scotogenic model's general features and in Sec.~\ref{sec:analytic} we derive analytical equations for the running neutrino mass and mixing parameters. In Sec.~\ref{sec:EFT} we discuss the resulting EFTs and describe the matching procedure in great detail. We discuss the results we have obtained from our analysis in Sec.~\ref{sec:Analysis} before concluding in Sec.~\ref{sec:Summary}. The 1-loop RGEs can be found in App.~\ref{app:Appendix_RGE}, while Apps.~\ref{app:AnalyticDerivation} and~\ref{app:MixingAngles} give some details needed for our computations.

\section{\label{sec:Overview}Model Overview}

The scotogenic model is an extension of the SM using several RH Majorana neutrinos $N_{1,2,\,\ldots}$, which are SM gauge singlets and can therefore have non-zero masses $M_{1,2,\,\ldots}$, even before electroweak symmetry breaking (EWSB). At least two of them are needed to obtain the two non-zero active neutrino masses which are required to explain the observed oscillations of neutrino flavours~\cite{Xing:2007uq}. We will consider the ``next-to-minimal'' case with three RH neutrinos if not stated otherwise. In addition, there is a second scalar doublet $\eta$, with SM quantum numbers identical to those of the Higgs doublet. Both types of new fields (scalar and fermionic) are odd under a discrete $\mathbb{Z}_2$ symmetry which is imposed in order to prevent tree-level neutrino masses and the decay of the lightest $\mathbb{Z}_2$-odd particle, which constitutes a DM candidate if electrically neutral.

The Lagrangian is given by~\cite{Ma:2006}:
\begin{equation}\label{eq:lagrangian}
  \mathcal{L} = \mathcal{L}_{\rm SM} + \frac{i}{2} \overline{N_i} \slashed{\partial} N_i - \frac{1}{2}\left( \overline{N_i} M_{ij} N^\mathcal{C}_j + h.c.\right)
		  + \left(D_\mu \eta\right)^\dag \left( D^\mu \eta \right) - \left( h_{ij} \overline{N}_i\, \tilde{\eta}^\dag\, \ell_{L\, j} + h.c. \right) - V(\phi,\eta),
\end{equation} 
where $\tilde{\eta}=i\sigma_2 \eta^*$ and a sum over repeated indices is implied. The scalar potential is:
\begin{equation}\label{eq:potential}
\begin{aligned}
  V(\phi,\eta) =& m_H^2 \phi^\dag \phi + m_\eta^2 \eta^\dag \eta + \frac{\lambda_1}{2} \left(\phi^\dag \phi \right)^2 + \frac{\lambda_2}{2} \left(\eta^\dag \eta\right)^2 		
		  +\\& + \lambda_3 \left(\phi^\dag \phi\right) \left(\eta^\dag \eta\right) + \lambda_4 \left(\phi^\dag \eta\right)^2 + \frac{\lambda_5}{2}\left[ \left(\phi^\dag \eta \right)\left(\phi^\dag \eta\right) + h.c. \right],
\end{aligned}
\end{equation}
where we take $\lambda_5$ to be real since any phase can be absorbed into $\eta$. Eqs.~\eqref{eq:lagrangian} and \eqref{eq:potential} contain all terms allowed by the symmetries of the model, which in particular do not yield a Dirac mass term for the active neutrinos, since $\eta$ cannot develop a vacuum expectation value (VEV) due to the $\mathbb{Z}_2$ symmetry. The RH neutrino masses emerge from the matrix $M$ upon diagonalising it via field redefinitions.

Upon EWSB, the Higgs field $\phi$ develops a VEV $\left\langle \phi \right\rangle = \left(0,v\right)^T$, and one identifies the physical scalar fields in $\phi = \left( 0 , v + \frac{h}{\sqrt{2}}\right)^T$ and $\eta = \left(\eta^+,\left(\eta_R+ i \eta_I\right)/\sqrt{2}\right)^T$ with the following masses:
\begin{subequations}\label{eq:scalarMasses}
  \begin{align}
    m_h^2 &= 2 \lambda_1 v^2 = - 2 m_H^2,\\
    m_\pm^2 &= m_\eta^2 + v^2 \lambda_3,\\
    m_R^2 &= m_\eta^2 + v^2 \left(\lambda_3 + \lambda_4 + \lambda_5\right),\\
    m_I^2 &= m_\eta^2 + v^2 \left(\lambda_3 + \lambda_4 - \lambda_5\right).
  \end{align}
\end{subequations}

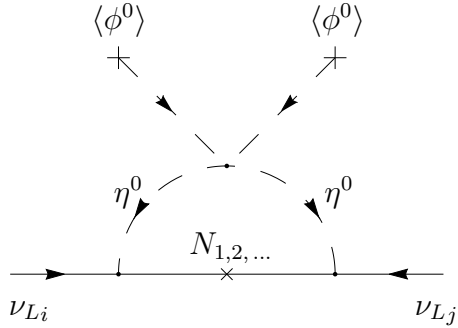
\begin{figure}[t]
  \centering
  \begin{picture}(60,60)
    \put(0,0){\includegraphics[width=60mm]{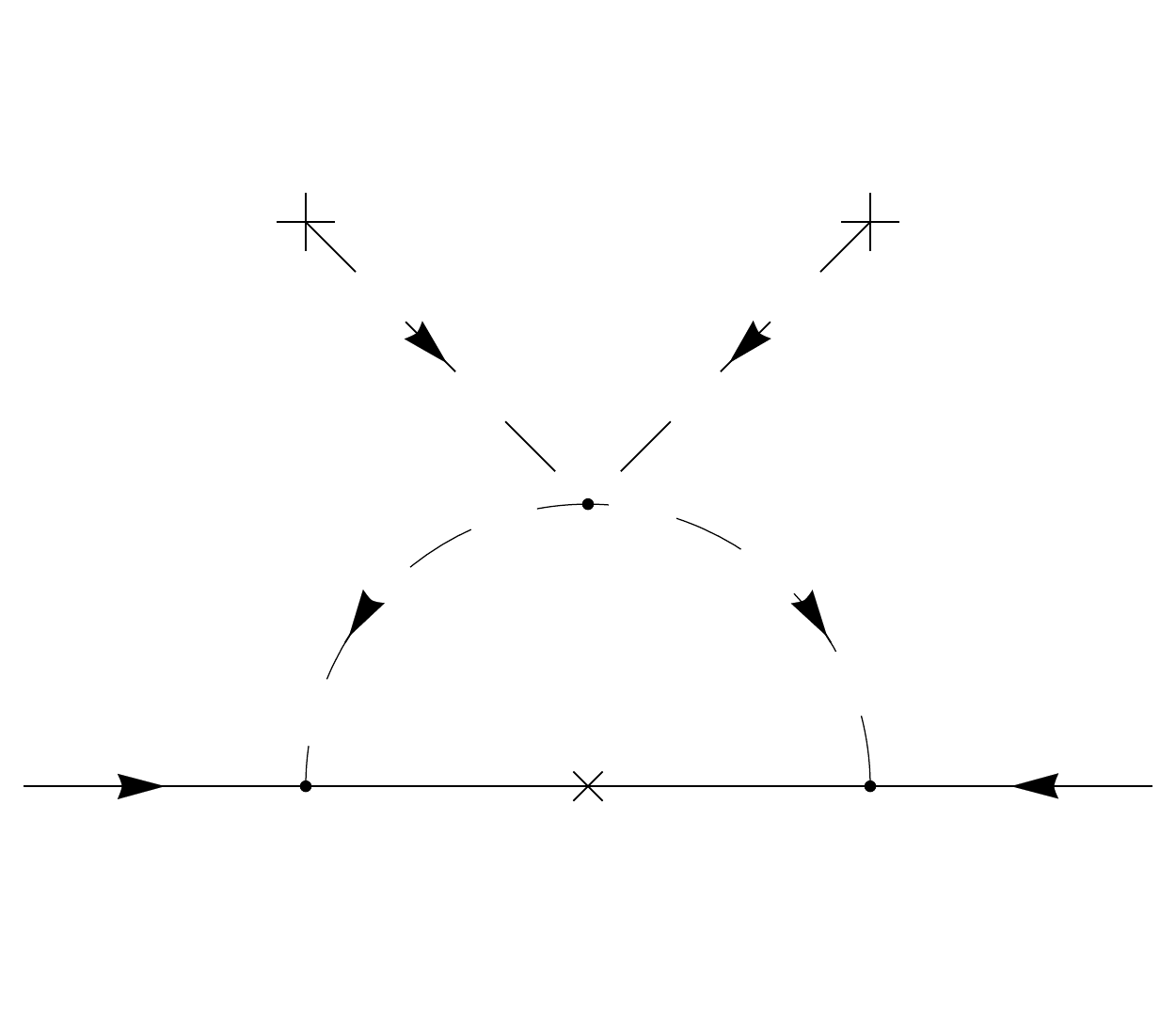}}
    \put(1,6){${\nu_L}_i$}
    \put(55,6){${\nu_L}_j$}
    \put(25,14){$N_{1,2,\,\ldots}$}
    \put(12,44){$\left\langle\phi^0\right\rangle$}
    \put(41,44){$\left\langle\phi^0\right\rangle$}
    \put(15,21){$\eta^0$}
    \put(43,21){$\eta^0$}
  \end{picture}
  \caption{\label{fig:MaMass}Light neutrino mass generation in the scotogenic model via the exchange of $\eta^0$ and $N_{1,2,\,\ldots}$.}
\end{figure}

\noindent With these degrees of freedom one obtains for the active neutrino mass matrix from the diagram shown in Fig.~\ref{fig:MaMass}:
\begin{align} \label{eq:MaMass}
 \left(\mathcal{M}_\nu\right)_{ij} =& \sum_{k=1}^3\frac{M_k h_{ki} h_{kj}}{32 \pi^2} \left\lbrace \frac{m_R^2}{m_R^2-M_k^2} \log\left(\frac{m_R^2}{M_k^2}\right) - \frac{m_I^2}{m_I^2-M_k^2} \log\left(\frac{m_I^2}{M_k^2}\right) \right\rbrace \nonumber\\
 \equiv&\sum_{k=1}^3\frac{M_k h_{ki} h_{kj}}{32 \pi^2} g(M_k,m_R,m_I),
\end{align}
which reduces to the expression:
\begin{equation}
 \left(\mathcal{M}_\nu\right)_{ij} = - \sum_{k=1}^3 v^2 \frac{\lambda_5}{(4\pi)^2} h_{ki}M_k^{-1} f(M_k, m_0) h_{kj},
\end{equation}
if the two neutral scalars are nearly degenerate in mass, i.e., if $\lambda_5 \ll 1$. This choice is motivated by the fact that, if $\lambda_5 = 0$, we can define a global $U(1)$ lepton number symmetry such that $\lambda_5$ is naturally small in the 't Hooft sense~\cite{'tHooft:1979bh}. Here we have defined $m_0^2\equiv(m_R^2+m_I^2)/2$, and the function $f$ is given by:
\begin{equation} \label{eq:Deff}
  f(M_k,m_0) \equiv \frac{M_k^4}{(M_k^2-m_0^2)^2}\log{\left(\frac{m_0^2}{M_k^2}\right)} + \frac{M_k^2}{M_k^2-m_0^2}.
\end{equation}
Note that our expression for $\mathcal{M}_\nu$ differs from the original one reported in~\cite{Ma:2006} by a factor of 2. This factor was missed in the original reference, but it is required due to rescaling the real scalar fields by a factor of $1/\sqrt{2}$ in order for them to be canonically normalised. Also, we wish to remark that it is \emph{not} straightforward to define the functions $f$ or $g$ in a basis where the RH neutrino mass matrix $M$ is not diagonal. In order to obtain the active neutrino masses and mixing angles for any given energy scale, we need to either use numerical tools, or make further assumptions that lead to simplifications of these functions.

\section{Analytical Formulae} \label{sec:analytic}

In the following derivation of analytical RGEs, we will closely follow previous analyses performed e.g.\ for the seesaw type-I~\cite{Antusch:2005gp, Mei:2005qp}, type-II~\cite{Chao:2006ye,Schmidt:2007nq}, type-III~\cite{Chakrabortty:2008zh}, the inverse seesaw~\cite{Bergstrom:2010qb}, or for the Weinberg operator~\cite{Casas:1999tg, Chankowski:1999xc, Antusch:2003kp}. The general problem for the case of a radiative model is that, in calculating the active neutrino mass matrix, we are relying on a  perturbative method which in turn requires the RH Majorana mass matrix $M$ to be diagonal (and real). However, owing to the renormalisation group flow, if this is true at one scale it need not be true at another. Even if we achieve to define the loop-function for matrix valued arguments, the problem becomes manifest when we try to compute the derivative of the mass matrix, the latter being schematically written as $h^T g(M,m_R,m_I) h$: a chain rule for a matrix-valued function $g$ is needed, however it 
cannot be found in general unless $\left[M,M^\prime\right]=0$, where $M^\prime\equiv\frac{\textrm{d}M}{\textrm{d}t}$ and $t\equiv\log\left(\frac{\mu}{\mu_0}\right)$.\footnote{Here, $\mu$ is the renormalisation scale and $\mu_0$ is some reference scale that is needed to make the logarithm dimensionless.} We can therefore only give analytic expressions in cases where either the loop-function $g$ takes a simple form whose derivative can be handled or where its running may be ignored completely.\footnote{As we will see in Sec.~\ref{sec:Analysis}, however, ignoring the running can be rather problematic.}

Let us make one comment on the values of the RH neutrinos before we discuss the different limiting cases. In the absence of a concrete mass generation mechanism for the RH neutrino masses, their values can in principle be \emph{arbitrary}. While historically, having $SO(10)$ Grand Unified Theories (GUTs) in mind~\cite{Ross:1985ai}, RH neutrino masses have been assumed to be much larger than the electroweak scale, there is in fact no physics reason for this. For one thing, experiments have not found any sign of a GUT, making the argument for huge RH neutrino masses considerably weaker.\footnote{In some sense it was not too strong from the very beginning, though, given that some GUTs, such as those based on $SU(5)$~\cite{Ross:1985ai} do not even suggest heavy RH neutrinos~\cite{Merle:2011yv}.} On the contrary, the argument based on 't Hooft naturalness, see Sec.~\ref{sec:intro}, in fact suggest very light RH neutrinos, since taking their masses to zero would increase the symmetry of the Lagrangian. In the 
scotogenic model, RH neutrino masses are typically taken to be somewhere at the GeV to TeV scale for phenomenological reasons (see, e.g., Refs.~\cite{Kubo:2006yx,Sierra:2008wj,Suematsu:2009ww}), and indeed there are arguments for either choice. We thus take on an independent point of view and simply discuss several possibilities for the RH neutrino mass. All the examples we present are experimentally allowed, but different readers may consider one or the other scenario to be better motivated. However, we would like to stress that this paper comprises no phenomenological analysis but rather an \emph{illustrative} and \emph{general} study of the running in the scotogenic model, i.e., our focus lies most on showing examples which demonstrate the different effects of the running, so that we do not assume any a priori relation between the inert scalar and RH neutrino masses. A stricter phenomenological parameter scan is left for future work.

Aiming to find a suitable classification, some useful limiting cases beyond the assumption $\lambda_5 \ll 1$ are:

\vbox{
\renewcommand{\labelenumi}{\roman{enumi})}
\begin{enumerate}
  \item scalar mass dominates: $m_0^2 \gg M_{1,2,\,\ldots}^2 \Rightarrow \mathcal{M}_\nu \simeq \frac{\lambda_5 v^2}{16 \pi^2}\, h^T \frac{M}{m_0^2} \, h$,
  \item similar masses: $m_0^2 \simeq M_{1,2,\,\ldots}^2 \Rightarrow \mathcal{M}_\nu \simeq \frac{\lambda_5 v^2}{32 \pi^2}\, h^T M^{-1} h \simeq \frac{\lambda_5 v^2}{32 \pi^2}\, h^T m_0^{-1} h $,
  \item fermion masses dominate: $m_0^2 \ll M_{1,2,\,\ldots}^2 \Rightarrow\mathcal{M}_\nu \simeq -\frac{\lambda_5 v^2}{16 \pi^2} h^T M^{-1} \left( 1 - \log\left(\frac{M^2}{m_0^2}\right) \right) h$.
\end{enumerate}
}

\begin{table}[t]
  \renewcommand{\arraystretch}{1.5}
  \begin{center}
    \begin{tabular}{l|cc}
      Limiting case & $C$ & $P$\\ \toprule
      i) $m_0^2 \gg M_{1,2,\,\ldots}^2$ & $ C_{\lambda_5} + C_h - C_{m_\eta^2}$ & $\frac{5}{2}\left( h^\dag h \right) + \frac{1}{2} \left(Y_e^\dag Y_e \right)$\\
      \midrule
      ii) $m_0^2 \simeq \boldsymbol{M_{1,2,\,\ldots}^2}$ & $C_{\lambda_5} + C_h$ & $\frac{1}{2}\left( h^\dag h \right) + \frac{1}{2} \left(Y_e^\dag Y_e \right)$\\
      ii) $\boldsymbol{m_0^2} \simeq M_{1,2,\,\ldots}^2$ & $C_{\lambda_5} + C_h -\frac{1}{2} C_{m_\eta^2}$ & $\frac{3}{2}\left( h^\dag h \right) + \frac{1}{2} \left(Y_e^\dag Y_e \right)$\\
      \midrule
      iii) $m_0^2 \ll M_{1,2,\,\ldots}^2$ & $C_{\lambda_5} + C_h$ & $\frac{3}{2}\left( h^\dag h \right) + \frac{1}{2} \left(Y_e^\dag Y_e \right)$\\
      \bottomrule
    \end{tabular}
  \end{center}
  \caption{\label{tab:analyticalParameters} Quantities appearing in the analytical RGEs of neutrino masses and  mixing angles for the three different mass hierarchies. In case~ii), $m_0^2 \simeq M_{1,2,\,\ldots}^2$, we can express the mass matrix in terms of $m_0$ or $M$. Which one is used is indicated by bold letters. In all expressions we take $m_0^2 \simeq m_\eta^2$.}
\end{table}

In the first case we get a suppression due to the large quantity $m_0^2 \equiv \frac{m_R^2+m_I^2}{2}\simeq m_\eta^2$, while in the second and third case this happens due to large $M$. In cases~i) and~ii), we have eliminated the loop-function in favour of a much simpler approximation. Case~iii), on the other hand, still suffers from the aforementioned problem due to the RH mass matrix appearing inside a logarithm. However, since the running is logarithmically suppressed, one might be led to the conclusion that it could potentially be negligible.

We are now in a position to compute the RGE for the active neutrino mass matrix, which is generally of the form (see App.~\ref{app:AnalyticDerivation}):
\begin{equation}
  (4\pi)^2 \mathcal{M}_\nu^\prime = C \mathcal{M}_\nu + P^T \mathcal{M}_\nu + \mathcal{M}_\nu P\,,
\end{equation}
where $C$ is a flavour-blind function of $t$ and $P$ a matrix with potentially non-trivial flavour structure. While $C$ may exclusively influence the running of the mass eigenvalues, only $P$ induces running of the mixing angles. Our findings for $C$ and $P$ for the three different cases defined above, assuming $m_0^2 \simeq m_\eta^2$, are summarised in Tab.~\ref{tab:analyticalParameters}. We use the abbreviations $C_\Gamma \equiv (4\pi)^2 \Gamma^\prime / \Gamma$ for $\Gamma \in \left\lbrace \lambda_5, m_\eta^2 \right\rbrace$ (see App.~\ref{app:Appendix_RGE} for the corresponding RGEs) and: 
\begin{equation}
  C_h \equiv 2 T_\nu - \frac{3}{2}\left( g_1^2+3g_2^2\right),
\end{equation}
which is simply twice the flavour diagonal part of $ (4\pi)^2 h^{-1}h^\prime$ [see Eq.~\eqref{eq:neutrinoRG}]. 

Following the methods described in the aforementioned references, which are reviewed in App.~\ref{app:AnalyticDerivation}, we have derived analytical expressions for the neutrino mixing parameters. To this end we assume dominant neutrino Yukawa couplings, i.e.\ $P=\alpha_h h^\dag h + \alpha_e Y_e^\dag Y_e \simeq \alpha_h h^\dag h$ and $h^\dag h = \textrm{diag}(h_1^2,h_2^2,h_3^2)$, where $Y_e$ is the charged lepton Yukawa matrix.\footnote{In case~ii), $m_0^2 \simeq M_{1,2,\,\ldots}^2$, we only find such a simple form for $P$ if we take $h$ to be real or if we replace $M^{-1}$ in favour of $m_0^{-1}$. We find that the latter approximation is in better agreement with our numerical treatment.}

The expressions we obtain are:
\begin{subequations}\label{eq:MassesAnalytic}\allowdisplaybreaks
  \begin{align}
    \begin{split}
      (4\pi)^2 m_1^\prime &= m_1 \left[ C + 2\alpha_h\left[ c_{12}^2 c_{13}^2 h_1^2 + s_{23}^2(c_{12}^2s_{13}^2h_2^2+s_{12}^2 h_3^2) +\right. \right.\\ 
      &\qquad + \left.\left. c_{23}^2(s_{12}^2h_2^2+c_{12}^2s_{13}^2h_3^2)\right]+ \alpha_h s_{13}^2 \cos\delta \sin(2\theta_{12})\sin(2\theta_{23})(h_2^2-h_3^2) \right],
    \end{split}\\
    \begin{split}
      (4\pi)^2 m_2^\prime &= m_2 \left[ C + 2\alpha_h\left[ s_{12}^2 c_{13}^2 h_1^2 + s_{23}^2(s_{12}^2s_{13}^2h_2^2+c_{12}^2 h_3^2) +\right. \right.\\ 
      &\qquad + \left.\left. c_{23}^2(c_{12}^2h_2^2 + s_{12}^2s_{13}^2h_3^2)\right] + \alpha_h s_{13}^2 \cos\delta \sin(2\theta_{12})\sin(2\theta_{23})(h_3^2-h_2^2) \right],
    \end{split}\\
    \textrm{and \hspace{6mm}} & \nonumber \\
    (4\pi)^2 m_3^\prime &= m_3 \left[ C + 2\alpha_h \left[ s_{13}^2 h_1^2 + c_{13}^2(s_{23}^2h_2^2+c_{23}^2h_3^2)\right]\right]
  \end{align}
\end{subequations}
for the masses, where we have made use of the standard abbreviations $s_{ij}\equiv \sin \theta_{ij}$ and $c_{ij}\equiv \cos \theta_{ij}$ with mixing angles $\theta_{ij}$. 

For the mixing angles we find:
\begin{subequations}\label{eq:MixingAnglesAnalytic}\allowdisplaybreaks
  \begin{align}
    (4\pi)^2 \theta_{12}^\prime &= \frac{\alpha_h}{2} \sin(2\theta_{12})\frac{\left|m_1e^{i\phi_1}+m_2e^{i\phi_2}\right|^2}{\Delta m_{21}^2}\left( h_1^2 - c_{23}^2h_2^2 - s_{23}^2h_3^2 \right) + \mathcal{O}(\theta_{13}), \label{eq:theta12Analytic}\\
    (4\pi)^2 \theta_{23}^\prime &= \frac{\alpha_h}{2} \frac{\sin(2\theta_{23})(h_2^2-h_3^2)}{\Delta
m_{32}^2} \left[ c_{12}^2 \left|m_2 e^{i\phi_2} + m_3\right|^2 + s_{12}^2 \frac{\left|m_1
e^{i\phi_1} + m_3\right|^2}{1+\zeta} \right] + \mathcal{O}(\theta_{13}), \label{eq:theta23Analytic}
\\
    \textrm{and \hspace{6mm}} & \nonumber \\
    \begin{split}
      (4\pi)^2 \theta_{13}^\prime &= \frac{\alpha_h}{2} (h_3^2-h_2^2) \sin(2\theta_{12}) \sin(2\theta_{23}) \frac{m_3}{(1+\zeta)\Delta m_{32}^2} \times \\
      &\qquad \times \left[ m_1 \cos(\delta - \phi_1) - (1+\zeta) m_2 \cos(\delta - \phi_2) - \zeta
m_3 \cos\delta \right] + \mathcal{O}(\theta_{13})\label{eq:theta13Analytic},
    \end{split}
  \end{align}
\end{subequations}
where we have used the abbreviation $\zeta \equiv \frac{\Delta m_{21}^2}{\Delta m_{32}^2}$. Note that, since there is no mechanism that explains leptonic mixing in this model, we have to impose certain values of the mixing angles at the input scale by hand. As we will see in Sec.~\ref{sec:Analysis}, a non-diagonal RH neutrino mass matrix $M$ is one possibility to do so, however, Eqs.~\eqref{eq:MixingAnglesAnalytic} tell us that a purely radiative generation of all mixing angles is \emph{not} possible. One may also wonder why there is running among the mixing angles even though $h^\dag h$ is diagonal. The reason is simply that the diagonal entries in this quantity have to be different such that $\left[ \mathcal{M}_\nu ,P \right] \neq 0$, which then induces the running of the mixing angles.
 
For the phases we obtain:
\begin{subequations}\allowdisplaybreaks
  \begin{align}
    (4\pi)^2 \delta^\prime &= \frac{\alpha_h(h_2^2-h_3^2)}{2\theta_{13}}\delta^{(-1)} + 2 \alpha_h
\delta^{(0)} + \mathcal{O}(\theta_{13}), \label{eq:deltaEQ}\\
    \begin{split}
      4\pi^2 \phi_1^\prime &= \alpha_h \left[c_{12}^2 \frac{m_1m_2}{\Delta m_{21}^2} \sin(\phi_2-\phi_1)\left(h_1^2-c_{23}^2 h_2^2 - s_{23}^2 h_3^2\right) + \right.\\
	&\qquad \left. + \cos{\left(2\theta_{23}\right)}\frac{h_3^2-h_2^2}{\Delta m_{32}^2}\left( s_{12}^2\frac{m_1m_3}{1+\zeta}\sin\phi_1 + c_{12}^2m_2m_3 \sin\phi_2\right)\right] + \mathcal{O}(\theta_{13}),
    \end{split}\\
    \textrm{and \hspace{6mm}} & \nonumber \\
    \begin{split}
      4\pi^2 \phi_2^\prime &= \alpha_h \left[s_{12}^2 \frac{m_1m_2}{\Delta m_{21}^2} \sin(\phi_2-\phi_1)\left(h_1^2-c_{23}^2h_2^2-s_{23}^2h_3^2\right) + \right.\\ & \qquad \left. + \cos(2\theta_{23})\frac{h_3^2-h_2^2}{\Delta m_{32}^2}  \left( s_{12}^2\frac{m_1 m_3}{1+\zeta} \sin\phi_1 + c_{12}^2 m_2 m_3 \sin\phi_2\right)\right] + \mathcal{O}(\theta_{13}).
    \end{split}
  \end{align}
\end{subequations}
In the expression for $\delta^\prime$, we have abbreviated:
\begin{subequations}\allowdisplaybreaks
\begin{align}
  \begin{split}
    \delta^{(-1)} &\equiv \sin(2\theta_{12})\sin(2\theta_{23})\frac{m_3}{(1+\zeta) \Delta m_{32}^2} \times \\
      &\qquad \times \left[ m_1 \sin(\delta - \phi_1) - (1+\zeta) m_2 \sin(\delta - \phi_2) - \zeta m_3 \sin\delta \right]
  \end{split} \\
    \textrm{and \hspace{6mm}} & \nonumber \\
  \begin{split}
    \delta^{(0)} &\equiv c_{12}^2 \frac{m_1m_3}{(1+\zeta)\Delta m_{32}^2}\sin(\phi_1-2\delta)\left(h_1^2-s_{23}^2 h_2^2 - c_{23}^2 h_3^2\right) + \\
    &\qquad + s_{12}^2\frac{m_2m_3}{\Delta m_{32}^2}\sin(\phi_2-2\delta)\left(h_1^2-s_{23}^2 h_2^2 - c_{23}^2 h_3^2\right) + \\
    &\qquad + \frac{m_1m_2}{\Delta m_{21}^2}\sin(\phi_2-\phi_1) \left(h_1^2-c_{23}^2 h_2^2 - s_{23}^2 h_3^2\right) + \\
    &\qquad + \cos(2\theta_{23})\frac{m_3(h_3^2-h_2^2)}{\Delta m_{32}^2}\left[s_{12}^2\frac{m_1\sin\phi_1}{(1+\zeta)}+c_{12}^2m_2\sin\phi_2\right].
  \end{split} 
\end{align}
\end{subequations}
Note that the equations for the mixing angles and (Majorana) phases have been approximated to first order in $\theta_{13}$ to get more compact expressions, while the $\theta_{13}$-dependence of the masses is exact. It is evident from Eq.~\eqref{eq:deltaEQ} that for $\theta_{13} \to 0$ we run into trouble, because $\delta$ is ill-defined at this point. However, as it was pointed out in~\cite{Antusch:2003kp}, $\delta$ can be continued to the point $\theta_{13}=0$ by demanding $\delta^\prime$ to remain finite. In practice we can avoid this subtlety by exploiting this fact and choosing a very small ($\theta_{13} \sim 10^{-3}$) instead of vanishing $\theta_{13}$ whenever this is needed. 

As we can see from Eqs.~\eqref{eq:MixingAnglesAnalytic}, the running of the mixing angles $\theta_{23}$ and $\theta_{13}$ becomes large if the difference $(h_2^2-h_3^2)$ is large. For $\theta_{12}$ to run significantly, we must instead ensure that the quantity $(h_1^2 - c_{23}^2h_2^2 - s_{23}^2h_3^2)$ is sizable. Clearly, the enhancement or suppression of the running due to the mixing parameters themselves is similar to previous analyses done for different models. Especially if the neutrino masses are hierarchical, i.e.\ $m_1 \ll m_2 \ll m_3$ for normal or $m_3 \ll m_1 \ll m_2$ for inverted ordering, the running is strongly suppressed, while for the case of nearly degenerate masses, $m_1 \simeq m_2 \simeq m_3$, we get large enhancements due to the inverse powers of mass square differences.

\section{Effective Operators} \label{sec:EFT}

In order to use the convenient MS renormalisation scheme and at the same time extract low energy observables, we must take care of the decoupling of heavy degrees of freedom by hand, as otherwise perturbation theory breaks down~\cite{Appelquist:1974tg, Collins1984renormalization}. We have done so taking into account effective operators of mass dimension up to $d=5$ at 1-loop level. As is well-known, in the SM the only relevant $d=5$ operator respecting all SM gauge symmetries is the \emph{Weinberg operator}~\cite{Weinberg:1979sa}, which generalises in a model with $N$ Higgs doublets to:
\begin{equation} \label{eq:Weinberg}
  \mathcal{L}_{\rm eff}=\frac{1}{4} \kappa_{ij}^{(kl)} \overline{\ell_L^\mathcal{C}}_i^a \epsilon_{ab} \phi^{(k)\,b} {\ell_L}_j^c \epsilon_{cd} \phi^{(l)\,d} + h.c.
\end{equation}
Here, $\phi^{(k)}\ ( k=1,\ldots, N)$ is one of the $N$ scalar doublets (our convention is $\phi^{(1)}=\phi$ and $\phi^{(2)}=\eta$ for the $N=2$ scotogentic model) and a sum over repeated indices is implied. Note that, in the case of the scotogentic model, we have an additional \emph{exact} $\mathbb{Z}_2$ symmetry imposed on our Lagrangian and therefore the operators $\kappa_{ij}^{(12)}$ and $\kappa_{ij}^{(21)}$ vanish exactly.\footnote{It is understood that, by calling $\kappa$ an operator, we are referring to the operator multiplying it.}

In the present situation, we are interested in the values of the neutrino-related observables at the electroweak scale, i.e.\ at the $Z$-boson mass $M_Z\simeq 91\,\textrm{GeV}$.\footnote{Below this scale we expect only little running of the mixing parameters, since the Weinberg operator is the only source of light neutrino masses and therefore the values at $M_Z$ carry over to lower energies to a good approximation.} Thus, we may encounter several mass thresholds between the input scale, which is chosen to be equal to the GUT scale $M_{\rm GUT}=10^{16}\,\textrm{GeV}$, and $M_Z$. Note that we only use the GUT scale as one possible example for a high scale, while we do not assume any GUT-inspired structure, and in particular no unification of the gauge couplings. Part of the reason for this choice is that we simply want to demonstrate what running effects could possibly do in the scotogenic model, and this is illustrated much more easily when running over several orders of magnitude in energy. On the other 
hand, a high scale which is at least somewhat close to the GUT scale is physically motivated when taking into account that the structure of the neutrino flavour sector may be explained by discrete symmetries, which are typically imposed at some high scale close to $M_{\rm GUT}$~\cite{Altarelli:2010gt,King:2013eh,King:2014nza}. As we will show in the explicit examples presented in Sec.~\ref{sec:Analysis}, very simple leptonic mixing patterns at the high scale could translate into phenomenologically valid regions when running to a lower energy scale, which may be of interest for a significant part of the readership. It should be clear, though, that in general running effects are suppressed if the running only extends over a few orders of magnitude in energy scale. Nevertheless, as already argued in Sec.~\ref{sec:intro}, the running in models with a radiative neutrino mass can be very strong. Thus, the values of certain observables could be modified even if, e.g., comparing their values at collider physics 
energies to low-energy measurements. We will in fact present several examples with strong running. Ultimately, we do in this paper provide the general tools to analyse the running, which should enable any inclined reader to make use of them when e.g.\ comparing the constraints on the scotogenic model arising from different data sets taken at different energies.

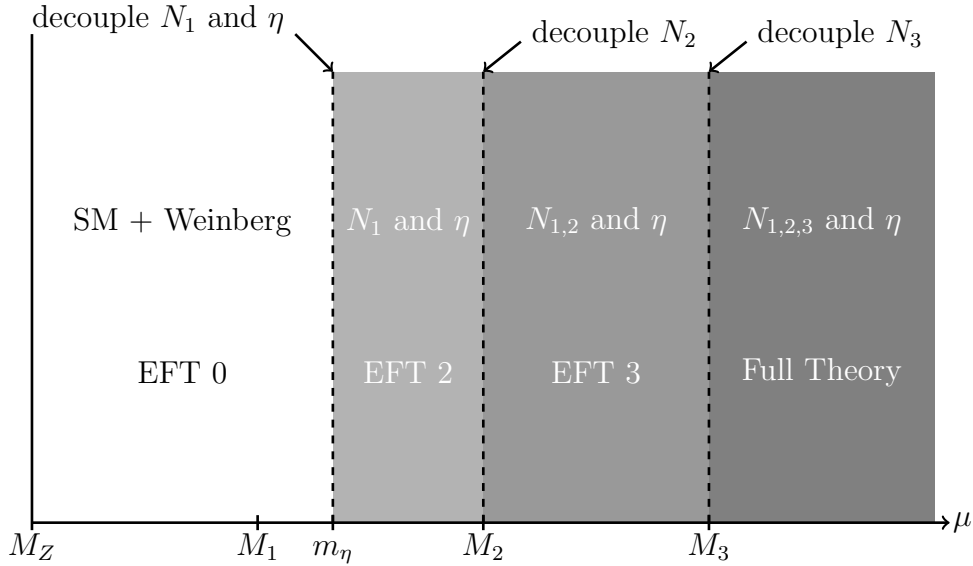
\begin{figure}[t]
  \centering
  \begin{tikzpicture}
    \fill[gray!60] (4,0) rectangle (6,6);
    \fill[gray!80] (6,0) rectangle (9,6);
    \fill[gray] (9,0) rectangle (12,6);
    \draw[line width = 1.] (0,-.1) -- (0,6.5);
    \draw[line width = 1., ->] (0,0) -- (12.2,0);
    \draw[line width = 1] (3,.1) -- (3,-.1);
    \draw[line width = 1] (4,.1) -- (4,-.1);
    \draw[line width = 1] (6,.1) -- (6,-.1);
    \draw[line width = 1] (9,.1) -- (9,-.1);
    \draw[line width = 1., dashed] (4,0) -- (4,6);
    \draw[line width = 1., dashed] (6,0) -- (6,6);
    \draw[line width = 1., dashed] (9,0) -- (9,6);
    \node[anchor=north] at (0,0) {$M_Z$}; 
    \node[anchor=north] at (3,0) {$M_1$}; 
    \node[anchor=north] at (6,0) {$M_2$}; 
    \node[anchor=north] at (9,0) {$M_3$}; 
    \node[anchor=north] at (4,-.1) {$m_\eta$}; 
    \node[anchor=west] at (12.1,0) {$\mu$};
    \node[color=white] at (10.5,4) {$N_{1,2,3}$ and $\eta$};
    \node[color=white] at (7.5,4) {$N_{1,2}$ and $\eta$};
    \node[color=white] at (5,4) {$N_{1}$ and $\eta$};
    \node at (2,4) {SM + Weinberg};
    \node[color=white] at (10.5,2) {Full Theory};
    \node[color=white] at (7.5,2) {EFT 3};
    \node[color=white] at (5,2) {EFT 2};
    \node at (2,2) {EFT 0};
    \draw[line width = 1, ->] (3.5,6.5) --  (4,6);
    \draw[line width = 1, ->] (6.5,6.4) --  (6,6);
    \draw[line width = 1, ->] (9.5,6.4) --  (9,6);
    \node[anchor=east] at (3.5,6.7) {decouple $N_1$ and $\eta$};
    \node[anchor=west] at (6.5,6.5) {decouple $N_2$};
    \node[anchor=west] at (9.5,6.5) {decouple $N_3$};
  \end{tikzpicture}
  \caption{\label{fig:EFT} Schematics of mass thresholds and the resulting EFTs for the case of three RH neutrinos and a mass hierarchy $M_1 < m_\eta < M_2 < M_3$ among the new particles. Below the scalar threshold the remaining RH neutrinos no longer couple to the SM fields, such that they are effectively decoupled even above their mass thresholds.}
\end{figure}

The thresholds we consider are the RH neutrinos $N_{1,2,\,\ldots}$ and the inert scalars that emerge from the doublet $\eta$. The scheme of resulting EFTs is depicted in Fig.~\ref{fig:EFT} for an example mass hierarchy. At each mass threshold $M_i$ or $m_\eta$, we must remove the corresponding field and its couplings from the theory, and match them to the effective operators.

Note that, in the scotogentic model, the active neutrino mass matrix is generated at 1-loop level, which conversely means that there is no tree-level expression for it. This is different e.g.\ in the various realisations of the seesaw mechanism where the neutrino mass is generated via tree-level diagrams or, equivalently, by diagonalising the (active plus sterile) neutrino mass matrix~\cite{Minkowski:1977sc, Yanagida:1979as, Glashow:1979nm, GellMann:1980vs, Mohapatra:1979ia, Schechter:1980gr, Wetterich:1981bx, Lazarides:1980nt, Mohapatra:1980yp, Cheng:1980qt, Foot:1988aq}. When such a theory is renormalised, one can exploit the fact that counter terms of the broken electroweak phase can be obtained from the symmetric phase by simple algebraic relations~\cite{Collins1984renormalization}. The same holds for the $\beta$ functions and all calculations can be performed in the symmetric phase of the theory, which is much simpler. 

However, this is in general not true if we are considering a loop-level neutrino mass. This becomes obvious if we translate the above reasoning with Lagrangian parameters and counter terms into a diagrammatic language: the broken phase degrees of freedom are linear combinations of those before symmetry breaking. A tree-level diagram in the broken phase can therefore be obtained by a combination of diagrams in the unbroken phase. If loops are involved, however, the degrees of freedom propagating in the loops cannot be ``combined'' in such a simple manner -- we have to take into account the \emph{physical} degrees of freedom from the very beginning.\footnote{Since the masses of the inert scalars are in turn just a linear combination of Lagrangian parameters, we can calculate their running masses in the symmetric phase.} In short, this means that the matching of couplings has to be carried out in the broken phase of the theory.

\subsection{Integrating out the RH neutrinos}

Suppose that we have $k$ RH neutrinos in the full theory. As we run down to lower energy scales, we will encounter the mass threshold $\mu = M_k$ of the heaviest RH neutrino $N_k$, and we integrate it out as discussed in detail below. Thereby, we obtain what we will call EFT~$k$. Integrating out the next-to-heaviest field yields the EFT labelled $(k-1)$, and so on. In the general EFT $n$, where $(n-1)$ RH neutrinos are left, the active neutrino mass matrix takes the following diagrammatic form, where we have contributions from the Weinberg operator $\eff{n}{1}{}$, the Weinberg-like operator $\eff{n}{2}{}$, and the remaining RH neutrinos:\footnote{We assume ordered Majorana masses, i.e.\ $M_1 < M_2 <\cdots < M_n$.}

\begin{equation} \label{eq:effectiveMassDiagram}
 -i \stackrel{(n)}{\mathcal{M}}_{\nu\, ij} = 
\begin{minipage}[c]{4cm}
\begin{picture}(40,30)
  \put(0,0){\includegraphics[width=4cm]{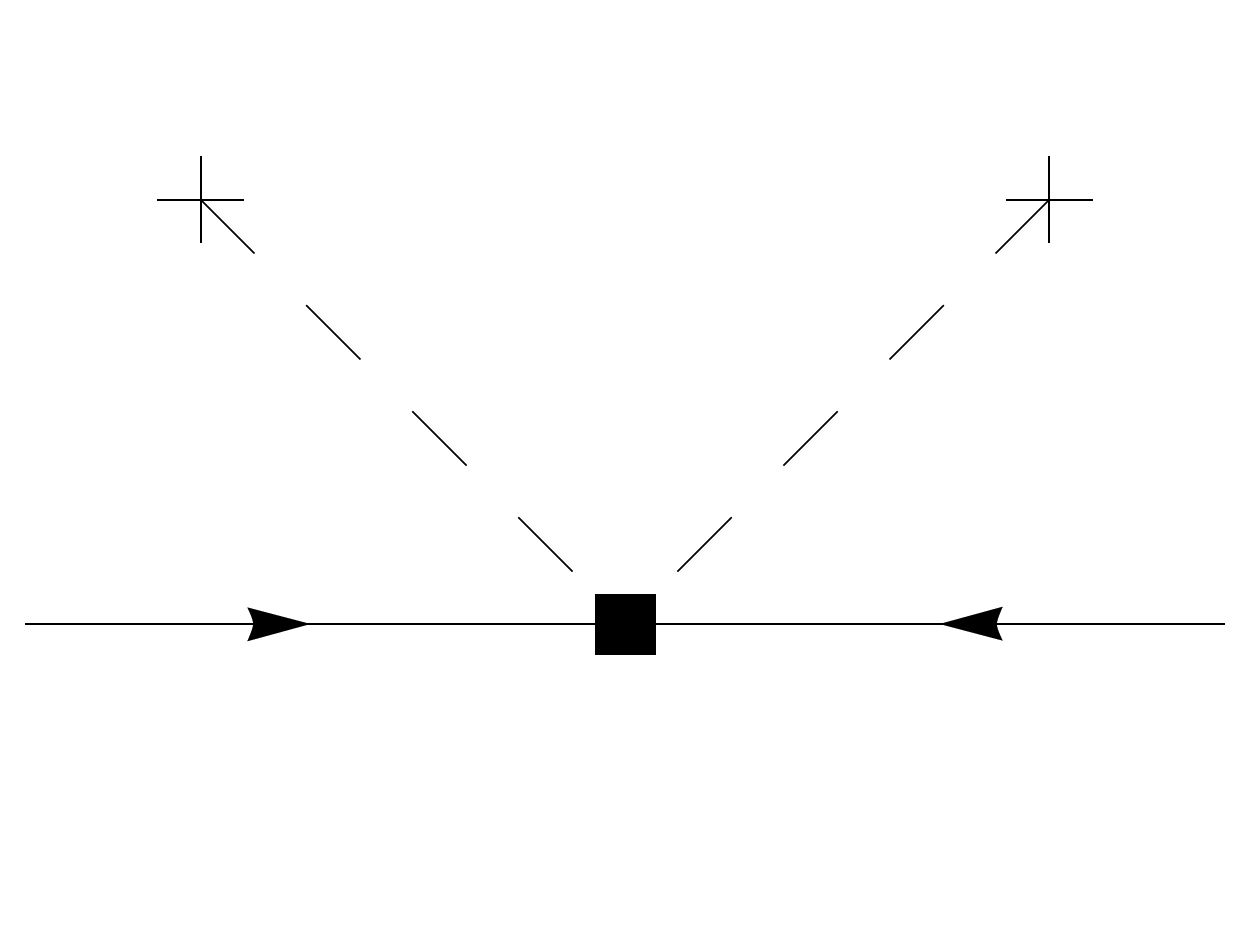}}
  \put(0,6){\text{\footnotesize ${\nu_L}_j$}}
  \put(36,6){\text{\footnotesize ${\nu_L}_i$}}
  \put(3.5,28){\text{\footnotesize $\left\langle\phi^0\right\rangle$}}
  \put(30.5,28){\text{\footnotesize $\left\langle\phi^0\right\rangle$}}
  \put(16,3.5){\text{\footnotesize $\overset{(n)}{\kappa}^{\raisebox{-.9ex}{$\scriptstyle(11)$}}$}}
\end{picture}
\end{minipage}
 +
\begin{minipage}[c]{4cm}
\begin{picture}(40,30)
 \put(0,0){\includegraphics[width=4cm]{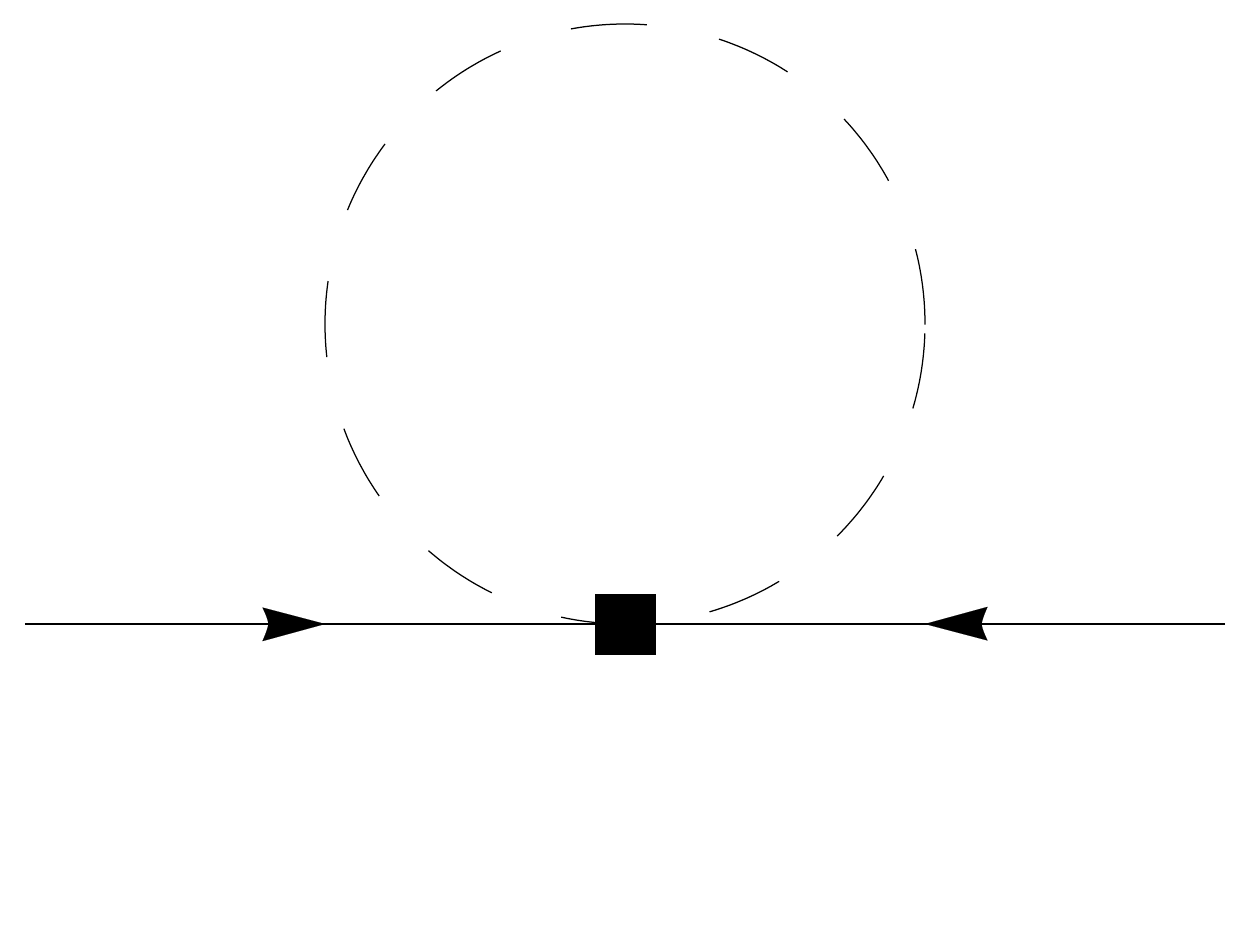}}
  \put(0,6){\text{\footnotesize ${\nu_L}_j$}}
  \put(36,6){\text{\footnotesize ${\nu_L}_i$}}
  \put(18,33){\text{\footnotesize $\eta_{R/I}$}}
  \put(16,3.5){\text{\footnotesize $\overset{(n)}{\kappa}^{\raisebox{-.9ex}{$\scriptstyle(22)$}}$}}
\end{picture}
\end{minipage}
 +
\begin{minipage}[c]{4cm}
\begin{picture}(40,30)
 \put(0,0){\includegraphics[width=4cm]{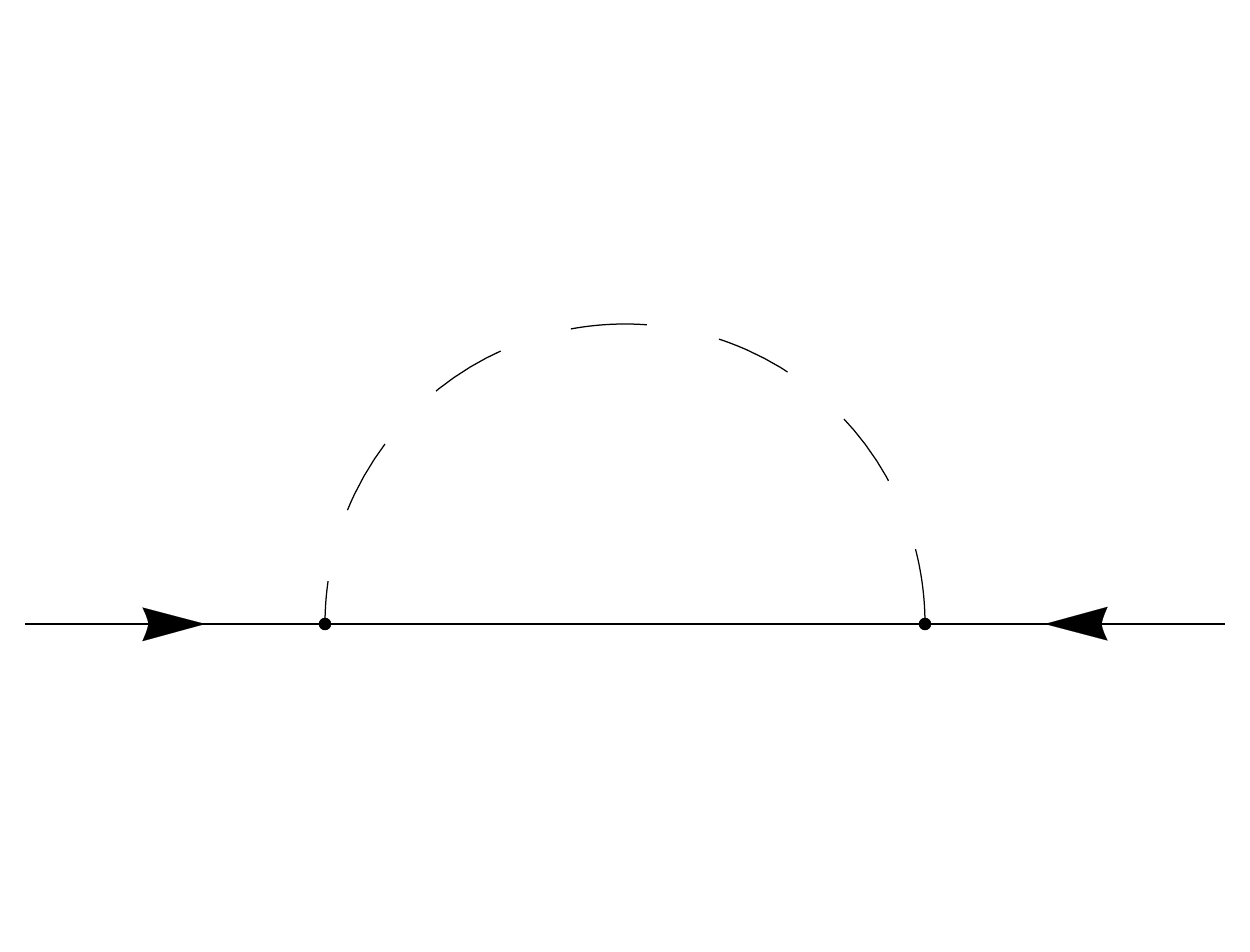}}
  \put(0,6){\text{\footnotesize ${\nu_L}_j$}}
  \put(36,6){\text{\footnotesize ${\nu_L}_i$}}
  \put(13,6){\text{\footnotesize $N_{1,\,\ldots,\,n-1}$}}
  \put(18,23){\text{\footnotesize $\eta_{R/I}$}}
\end{picture}
\end{minipage},
\end{equation}
where the operators $\eff{n}{l}{}$ have to be matched at all of the previous $(k-n+1)$ thresholds. It might seem as though the first diagram gives a tree-level neutrino mass, however, the matching of the Weinberg operator will reveal that only loop-suppressed diagrams enter the corresponding matching condition, such that no loop-diagrams with insertions of $\eff{n}{1}{}$ have to be considered in what follows. In writing $\eta_{R/I}$, it is understood that the diagrams with $\eta_R$ and $\eta_I$ in the loop have to be calculated separately and then summed.

The corresponding analytic expression is:
\begin{equation} \label{eq:effectiveMass}
  \begin{split}
    \stackrel{(n)}{\mathcal{M}}_{\nu\, ij} = -\frac{v^2}{2} \eff{n}{1}{ij} & - \frac{1}{32 \pi^2} \left[ \frac{\eff{n}{2}{ij}}{2} \left(m_R^2\log\frac{m_R^2}{\mu^2} - m_I^2\log\frac{m_I^2}{\mu^2}\right)\right. - \\ 
      &\hspace{25mm} -\left. \sum_{\ell<n} \stackrel{(n)}{h}_{\hspace{-.3ex}\ell i} \overset{(n)}{M_\ell}\ g(\overset{(n)}{M_\ell},m_R,m_I) \overset{(n)}{h}_{\hspace{-.3ex}\ell j} \right],
  \end{split}
\end{equation}
where $g$ is defined in Eq.~\eqref{eq:MaMass}. Note that the mass matrix in the full theory is finite \emph{without} renormalisation. This has to be true because there is no counter term available to cancel a potential divergence of the diagram in Fig.~\ref{fig:MaMass} and only renormalisable vertices had been used. More technically, in the broken phase this is realised by cancellations of the divergent parts in the diagrams for $\eta_R$ and $\eta_I$ in the loop, stemming from the factor $i$ in the expansion $\sqrt{2}\ \eta^0 = (\eta_R + i \eta_I)$. At the same time the explicit dependence of the mass matrix in the EFTs on $\mu$ is signaling that the expression is not independent of the renormalisation scheme which we employ, since we have to renormalise the divergent (centre diagram) contribution to the mass matrix  (a discussion of this can be found in Ref.~\cite{Pilaftsis:2002nc}). Of course, once we compare the values of a physical observable between two different scales, the prediction will be the same, 
no matter which scheme is used.

In practice, we input Yukawa couplings $h_{ij}$ and a RH mass matrix $M$ at the GUT scale and run them down to the first mass threshold $\mu_*=M_k(\mu_*)$. Since the running may introduce off-diagonal elements in the Majorana mass matrix, we must then diagonalise $M$ and cancel the row and column corresponding to the largest eigenvalue in $M$ to obtain the \emph{effective} mass matrix $\overset{(k)}{M}$. Note that, since $M$ must be symmetric, it can be diagonalised by a unitary matrix $V$ according to $V^T M V = \textrm{diag}\left(M_1,\ldots, M_k\right)$. By appropriate field redefinitions, we find that the Yukawa matrix in this basis reads $\tilde{h} = V^T h$. Removing the row that couples the heaviest (in the Majorana mass eigenbasis) RH neutrino $N_k$ to the lepton doublet yields the new \emph{effective} Yukawa matrix $\overset{(k)}{h}$. This procedure is continued down to the final threshold and it is identical to that typically applied to the type-I seesaw mechanism (see e.g.\ Ref.~\cite{Antusch:2002rr}
 for further details). 

Note that in the effective theories the mass matrix may consist of up to three independent parts [cf.\ the three diagrams in Eq.~\eqref{eq:effectiveMassDiagram}], whose running will in general be different. Thereby, the running of the mixing angles may be amplified, even in the case where each contribution only has flavour-diagonal running. Such \emph{threshold effects} have their origin in divergences that occur in the effective theory, which are however absent in the full theory -- an effect that is well-known, see e.g.\ Ref.~\cite{Antusch:2005gp}. Again, Eq.~\eqref{eq:effectiveMassDiagram} serves as an example: while the right-most diagram is finite, the one left to it -- which we obtain by removing the heaviest RH neutrino -- is divergent.

\begin{figure}[t]
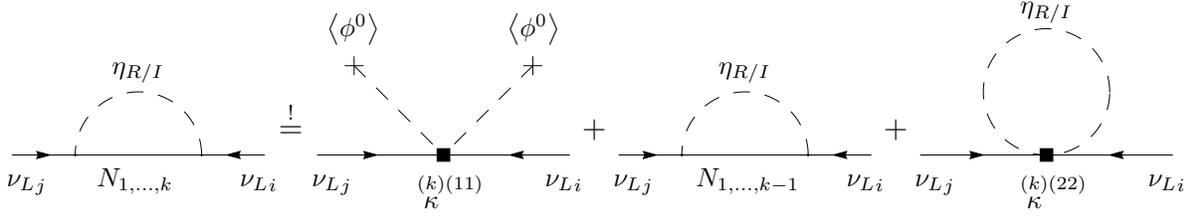

\begin{equation*}
\begin{gathered}
\begin{minipage}[c]{3.5cm}
\begin{picture}(35,25)
 \put(0,0){\includegraphics[width=3.5cm]{loop1phi}}
  \put(0,5){\text{\footnotesize ${\nu_L}_j$}}
  \put(31,5){\text{\footnotesize ${\nu_L}_i$}}
  \put(12,5){\text{\footnotesize $N_{1, \ldots, k}$}}
  \put(14,20){\text{\footnotesize $\eta_{R/I}$}}
\end{picture}
\end{minipage}
\overset{!}{=} 
\begin{minipage}[c]{3.5cm}
\begin{picture}(35,25)
  \put(0,0){\includegraphics[width=3.5cm]{weinbergphi}}
  \put(2,25){\text{\footnotesize $\left\langle\phi^0\right\rangle$}}
  \put(26,25){\text{\footnotesize $\left\langle\phi^0\right\rangle$}}
  \put(0,5){\text{\footnotesize ${\nu_L}_j$}}
  \put(31,5){\text{\footnotesize ${\nu_L}_i$}}
  \put(14,2){\text{\footnotesize $\stackrel{(k)}{\kappa}^{\raisebox{-.9ex}{$\scriptstyle(11)$}}$}}
\end{picture}
\end{minipage}
+
\begin{minipage}[c]{3.5cm}
\begin{picture}(35,25)
 \put(0,0){\includegraphics[width=3.5cm]{loop1phi}}
  \put(0,5){\text{\footnotesize ${\nu_L}_j$}}
  \put(31,5){\text{\footnotesize ${\nu_L}_i$}}
  \put(11,5){\text{\footnotesize $N_{1, \ldots, k-1}$}}
  \put(14,20){\text{\footnotesize $\eta_{R/I}$}}
\end{picture}
\end{minipage}
+
\begin{minipage}[c]{3.5cm}
\begin{picture}(35,25)
 \put(0,0){\includegraphics[width=3.5cm]{loop2phi}}
  \put(0,5){\text{\footnotesize ${\nu_L}_j$}}
  \put(31,5){\text{\footnotesize ${\nu_L}_i$}}
  \put(14,28){\text{\footnotesize $\eta_{R/I}$}}
  \put(14,2){\text{\footnotesize $\stackrel{(k)}{\kappa}^{\raisebox{-.9ex}{$\scriptstyle(22)$}}$}}
\end{picture}
\end{minipage}
\end{gathered}
\end{equation*}
\caption{Diagrammatic matching of the operator $\kappa^{(11)}$ at the mass threshold $\mu_*=M_k(\mu_*)$. Note that no insertions of $\kappa^{(11)}$ have to be considered in loops, since this would yield effective 2-loop contributions.}\label{fig:matchingk11} 
\end{figure}

\begin{figure}[t]
\begin{equation*}
\begin{gathered}
\begin{minipage}[c]{3.5cm}
\begin{picture}(35,35)
 \put(0,0){\includegraphics[width=3.5cm]{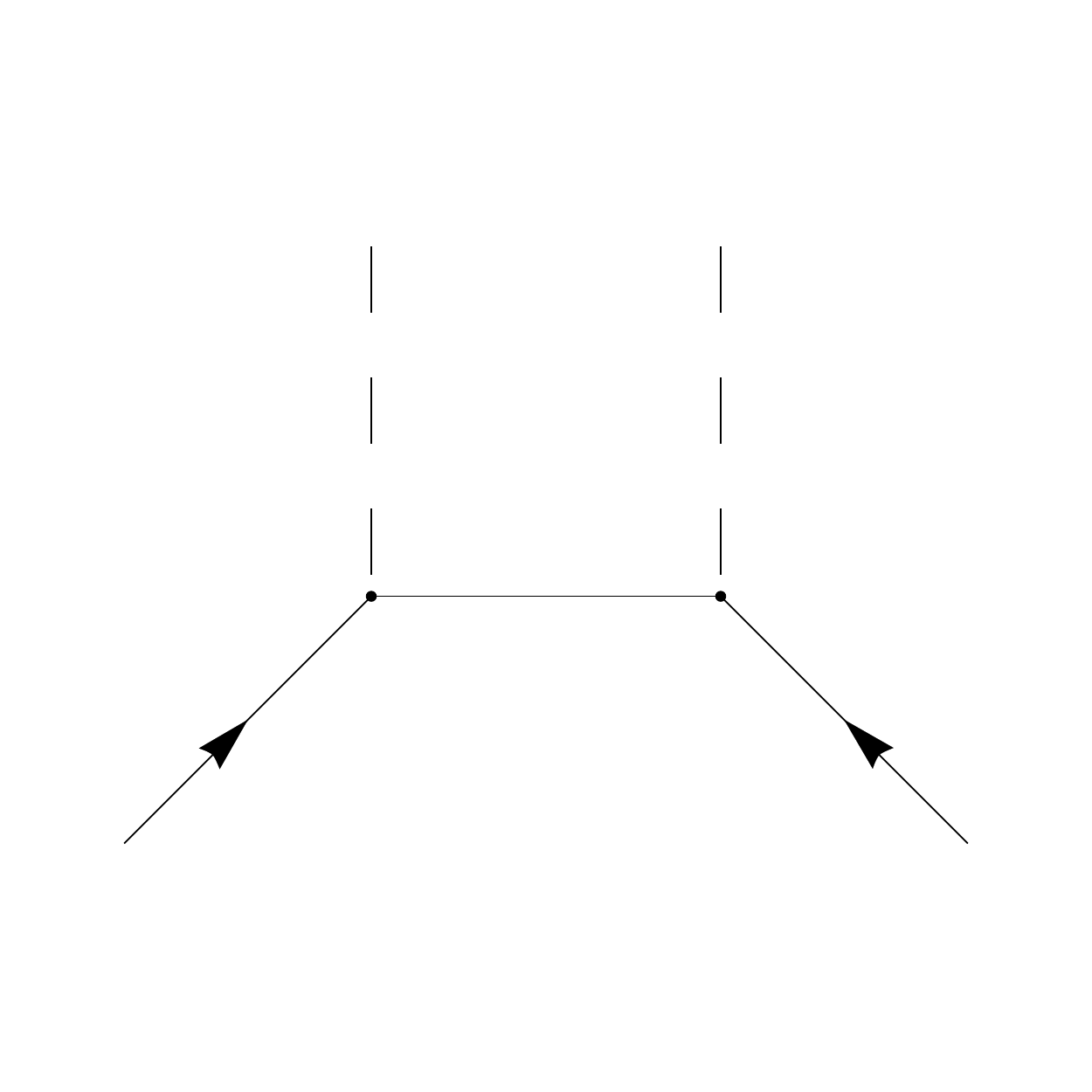}}
  \put(0,6){\text{\footnotesize ${\nu_L}_j$}}
  \put(32,6){\text{\footnotesize ${\nu_L}_i$}}
  \put(15,12){\text{\footnotesize $N_k$}}
  \put(11,29){\text{\footnotesize $\eta_R$}}
  \put(22,29){\text{\footnotesize $\eta_R$}}
\end{picture}
\end{minipage}
 +
\begin{minipage}[c]{3.5cm}
\begin{picture}(35,35)
 \put(0,0){\includegraphics[width=3.5cm]{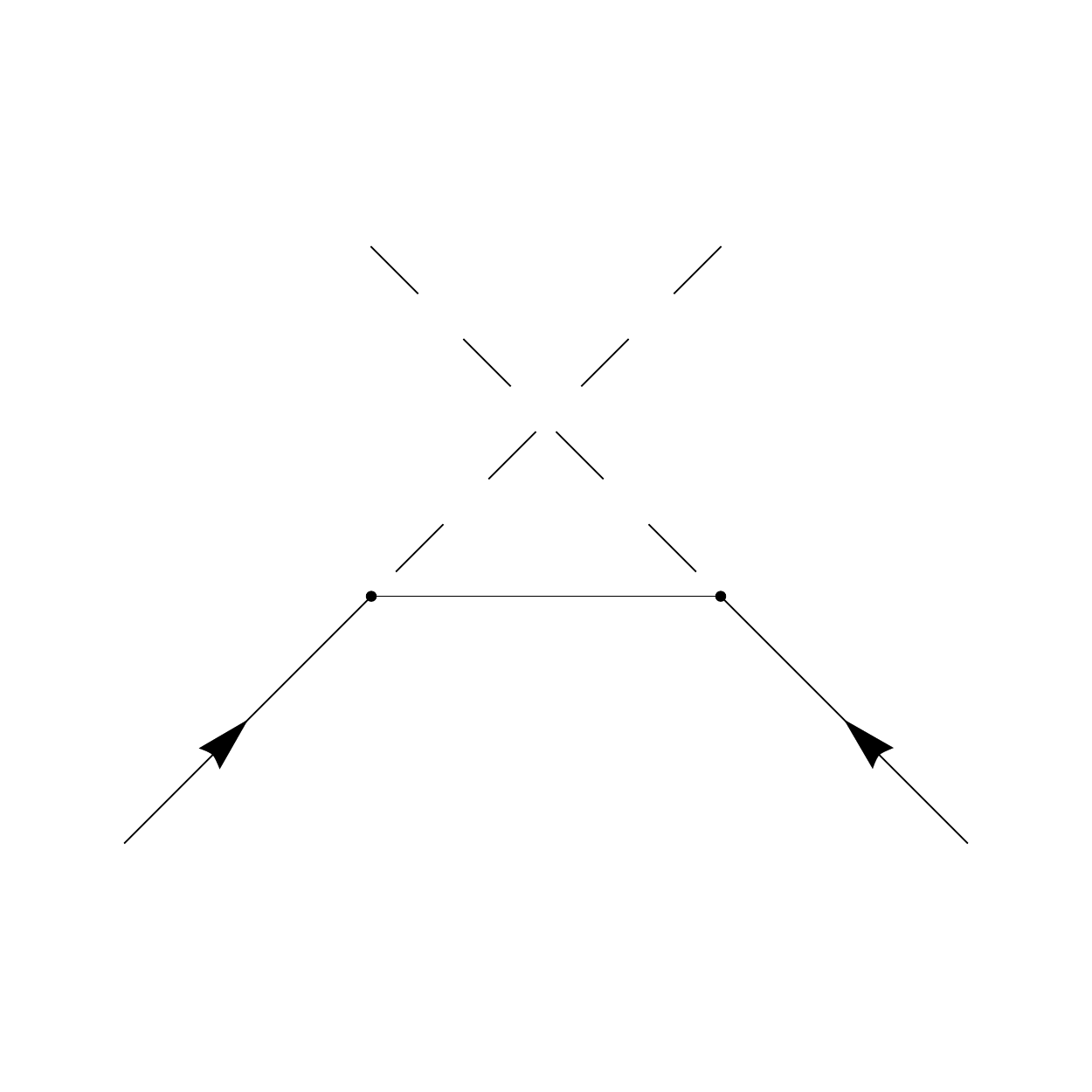}}
  \put(0,6){\text{\footnotesize ${\nu_L}_j$}}
  \put(32,6){\text{\footnotesize ${\nu_L}_i$}}
  \put(15,12){\text{\footnotesize $N_k$}}
  \put(11,29){\text{\footnotesize $\eta_R$}}
  \put(22,29){\text{\footnotesize $\eta_R$}}
\end{picture}
\end{minipage}
 +
\begin{minipage}[c]{3.5cm}
\begin{picture}(35,35)
 \put(0,0){\includegraphics[width=3.5cm]{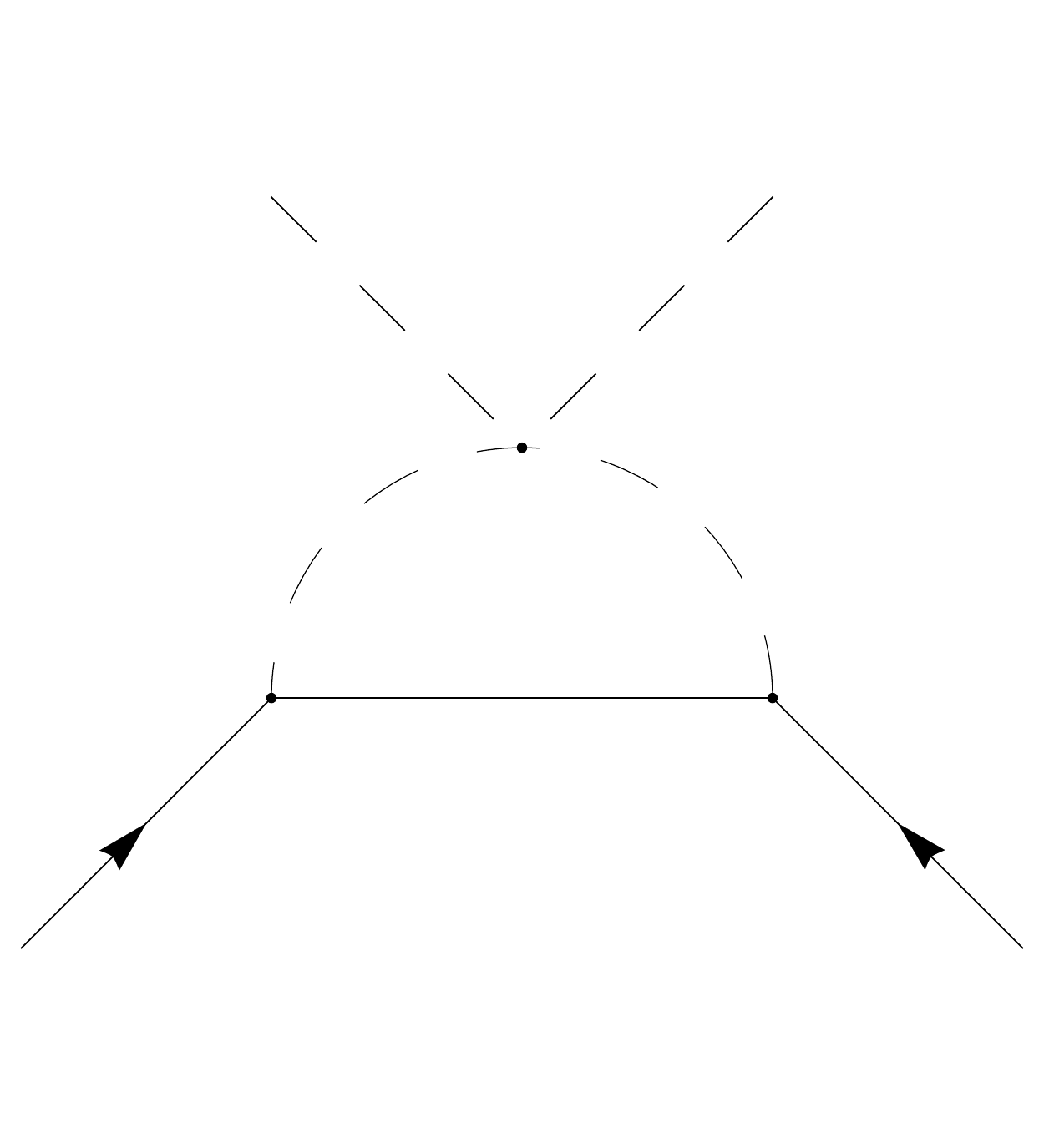}}
  \put(-3,5){\text{\footnotesize ${\nu_L}_j$}}
  \put(35,5){\text{\footnotesize ${\nu_L}_i$}}
  \put(15,11){\text{\footnotesize $N_k$}}
  \put(7,33){\text{\footnotesize $\eta_R$}}
  \put(25,33){\text{\footnotesize $\eta_R$}}
  \put(3,20){\text{\footnotesize $\eta_{R/I}$}}
  \put(25,20){\text{\footnotesize $\eta_{R/I}$}}
\end{picture}
\end{minipage}\\
 \overset{!}{=}\\
\begin{minipage}[c]{3.5cm}
\begin{picture}(35,35)
  \put(0,0){\includegraphics[width=3.5cm]{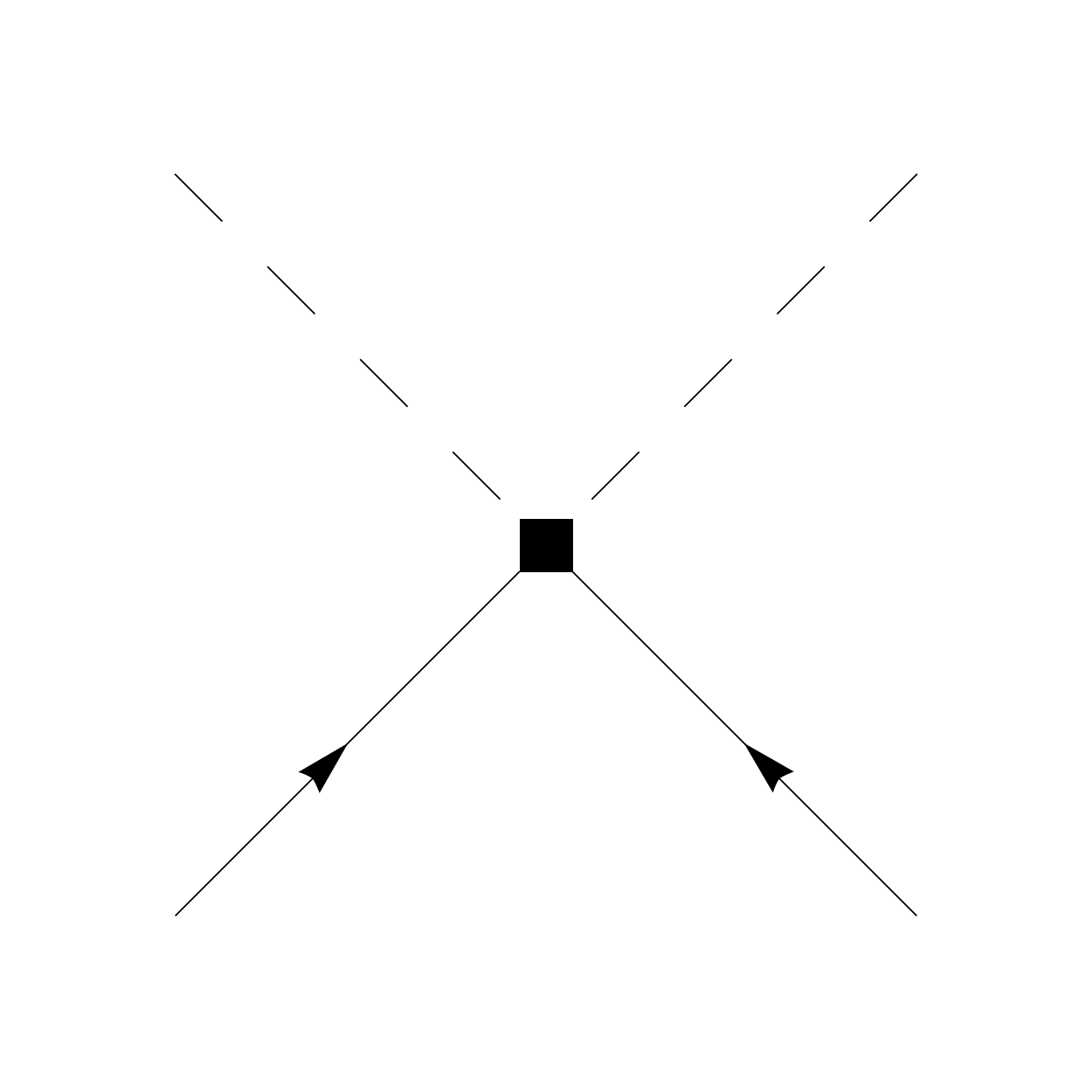}}
  \put(2,4){\text{\footnotesize ${\nu_L}_j$}}
  \put(30,4){\text{\footnotesize ${\nu_L}_i$}}
  \put(2,30){\text{\footnotesize $\eta_R$}}
  \put(30,30){\text{\footnotesize $\eta_R$}}
  \put(22,16){\text{\footnotesize $\stackrel{(k)}{\kappa}^{\raisebox{-.9ex}{$\scriptstyle(22)$}}$}}
\end{picture}
\end{minipage}
 +
\begin{minipage}[c]{3.5cm}
\begin{picture}(35,50)
 \put(0,0){\includegraphics[width=3.5cm]{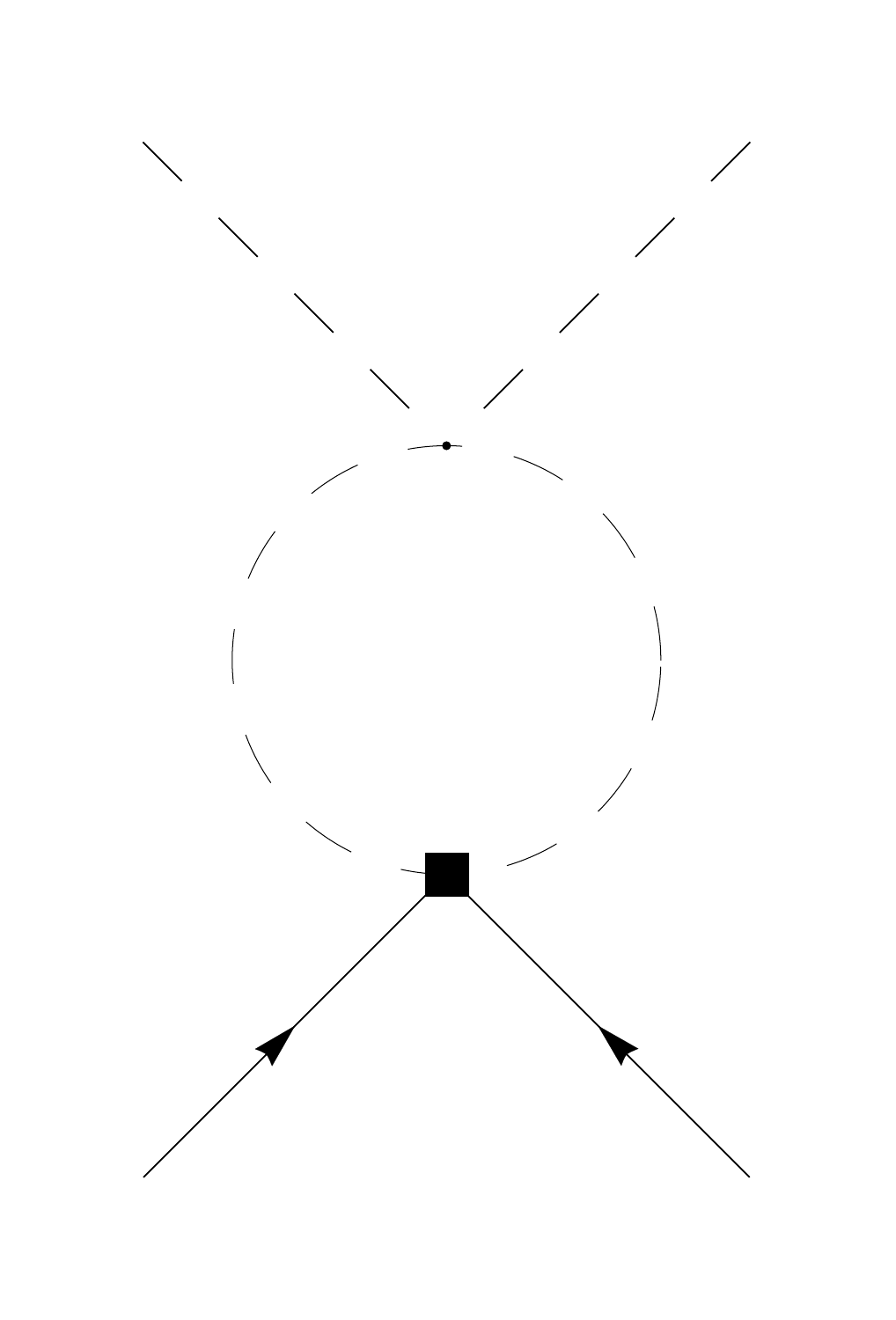}}
  \put(2,4){\text{\footnotesize ${\nu_L}_j$}}
  \put(30,4){\text{\footnotesize ${\nu_L}_i$}}
  \put(30,47){\text{\footnotesize $\eta_R$}}
  \put(2,47){\text{\footnotesize $\eta_R$}}
  \put(1.5,25){\text{\footnotesize $\eta_{R/I}$}}
  \put(27,25){\text{\footnotesize $\eta_{R/I}$}}
  \put(14,20){\text{\footnotesize $\stackrel{(k)}{\kappa}^{\raisebox{-.9ex}{$\scriptstyle(22)$}}$}}
\end{picture}
\end{minipage}
\end{gathered}
\end{equation*} 
\caption{Diagrammatic matching of the operator $\kappa^{(22)}$ at the mass threshold $\mu_*=M_k(\mu_*)$. Again no loop diagrams with an insertion of $\kappa^{(11)}$ have to be considered, since they result in effective 2-loop contributions.}\label{fig:matchingk22} 
\end{figure}

We now turn to the matching conditions for the effective operators at the RH neutrino thresholds. For simplicity, we only consider the matching at the heaviest mass threshold $M_k$ -- the generalisation to others is straightforward. The matching condition for the operator $\kappa^{(11)}$ at the first threshold is shown diagrammatically in Fig.~\ref{fig:matchingk11}. Observe that the contributions from the fields $N_{1,\,\ldots,\,k-1}$ cancel exactly, since they do not ``feel'' that a heavier particle has been integrated out. Note also that we have to match the operator $\kappa^{(22)}$ simultaneously since it appears in the last diagram of Fig.~\ref{fig:matchingk11}. The corresponding matching condition is shown in Fig.~\ref{fig:matchingk22}, where canceling diagrams are disregarded.

Since we are working in perturbation theory, we also need to treat the $\kappa^{(ll)}$ as perturbative quantities which receive corrections from all loop-levels. This suggests that we could write:
\begin{equation}
  \eff{k}{l}{ij} = \eff{k}{l}{ij,\, 0} + \epsilon\, \eff{k}{l}{ij,1} + \mathcal{O}\left(\epsilon^2\right),
\end{equation}
where $\epsilon$ is some small (loop-suppression) factor, e.g. $(4\pi)^{-2}$, and $l=1,2$.

By this approach, the leading $\mathcal{O}\left(\epsilon^0\right)$ matching conditions can be calculated from Figs.~\ref{fig:matchingk11} and \ref{fig:matchingk22} for vanishing external momenta (no sum over $k$):
\begin{subequations}
\begin{align}
  0 &\overset{!}{=} i \frac{v^2}{2} \eff{k}{1}{ij,\, 0} \quad \textrm{and}\\
  2 \times i \frac{h_{ki}h_{kj}}{2 M_k} &\overset{!}{=} i \frac{\eff{k}{2}{ij,\, 0}}{2}\,,
\end{align}
\end{subequations}
and thus:
\begin{subequations}
\begin{align}
  \eff{k}{1}{ij,\, 0} & = 0 \quad \textrm{and} \label{eq:treek11} \\
  \eff{k}{2}{ij,\, 0} & = 2\, \frac{h_{ki}h_{kj}}{M_k}\,.\label{eq:treek22} 
\end{align}
\end{subequations}
At $\mathcal{O}\left(\epsilon^1\right)$ we obtain, after applying some simplifications and using the tree-level matching conditions \eqref{eq:treek22} and \eqref{eq:treek11} for the 1-loop diagrams:
\begin{subequations}
\begin{align}
  \eff{k}{1}{ij,1} & = -\frac{h_{ki}h_{kj}}{v^2 M_k}\left[ \frac{m_R^4}{m_R^2-M_k^2}\log\frac{m_R^2}{M_k^2} - \frac{m_I^4}{m_I^2-M_k^2}\log\frac{m_I^2}{M_k^2} \right], \\
  \eff{k}{2}{ij,1} & = \lambda_2 \frac{h_{ki}h_{kj}}{M_k} \left[ 3\left(f(M_k,m_R)+\log\frac{M_k^2}{m_R^2}\right) - \left(f(M_k,m_I)+\log\frac{M_k^2}{m_I^2}\right) \right],
\end{align}
\end{subequations}
with the auxiliary function $f$ being given explicitly in Eq.~\eqref{eq:Deff}.

Putting all the pieces together and generalising to the $n$\textsuperscript{th} threshold, we find the matching conditions:
\begin{subequations}
\begin{align}
  \left.\eff{n}{1}{ij}\right|_{\mu=M_n}  =& \left.\eff{n+1}{1}{\hspace{-1ex}ij}\right|_{\mu=M_n} 
	- \frac{2}{v^2}\frac{1}{32\pi^2}\frac{h_{ni}h_{nj}}{M_n}\left[ \frac{m_R^4}{m_R^2-M_n^2}\log\frac{m_R^2}{M_n^2} - \frac{m_I^4}{m_I^2-M_n^2}\log\frac{m_I^2}{M_n^2} \right]_{\mu=M_n}, \label{eq:matchingk11}\\
  \begin{split}
    \left.\eff{n}{2}{ij}\right|_{\mu=M_n}  =& \left.\eff{n+1}{2}{\hspace{-1ex}ij}\right|_{\mu=M_n} + 
	  \frac{h_{ni}h_{nj}}{M_n}\left[ 2 + \frac{\lambda_2}{(4\pi)^2}\left\lbrace 3\left(f(M_n,m_R)+\log\frac{M_n^2}{m_R^2}\right) - \right.\right.\\ 
	  &\hspace{55mm}- \left.\left. \left(f(M_n,m_I)+\log\frac{M_n^2}{m_I^2}\right) \right\rbrace\right]_{\mu=M_n}. \label{eq:matchingk22}
  \end{split}
\end{align}
\end{subequations}
In these expressions it is understood that all quantities (especially $M$ and $h$) are to be evaluated in the $n$\textsuperscript{th} effective theory. Also, since the operator $\kappa^{(22)}$ only appears in a 1-loop graph for the mass matrix, see Eq.~\eqref{eq:effectiveMassDiagram}, it is sufficient to consider Eq.~\eqref{eq:matchingk22} to leading order only.

\subsection{Integrating out the inert scalars}

\begin{figure}[t]
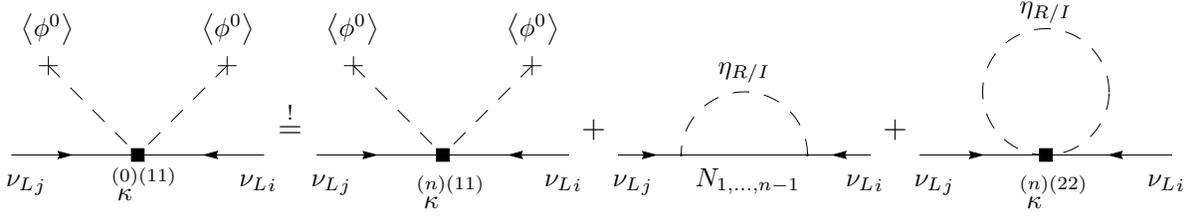

\begin{equation*}
\begin{minipage}[c]{3.5cm}
\begin{picture}(35,25)
  \put(0,0){\includegraphics[width=3.5cm]{weinbergphi}}
  \put(2,25){\text{\footnotesize $\left\langle\phi^0\right\rangle$}}
  \put(26,25){\text{\footnotesize $\left\langle\phi^0\right\rangle$}}
  \put(0,5){\text{\footnotesize ${\nu_L}_j$}}
  \put(31,5){\text{\footnotesize ${\nu_L}_i$}}
  \put(14,3){\text{\footnotesize $\stackrel{(0)}{\kappa}^{\raisebox{-.9ex}{$\scriptstyle(11)$}}$}}
\end{picture}
\end{minipage}
\overset{!}{=}
\begin{minipage}[c]{3.5cm}
\begin{picture}(35,25)
  \put(0,0){\includegraphics[width=3.5cm]{weinbergphi}}
  \put(2,25){\text{\footnotesize $\left\langle\phi^0\right\rangle$}}
  \put(26,25){\text{\footnotesize $\left\langle\phi^0\right\rangle$}}
  \put(0,5){\text{\footnotesize ${\nu_L}_j$}}
  \put(31,5){\text{\footnotesize ${\nu_L}_i$}}
  \put(14,2){\text{\footnotesize $\stackrel{(n)}{\kappa}^{\raisebox{-.9ex}{$\scriptstyle(11)$}}$}}
\end{picture}
\end{minipage}
 +
\begin{minipage}[c]{3.5cm}
\begin{picture}(35,25)
 \put(0,0){\includegraphics[width=3.5cm]{loop1phi}}
  \put(0,5){\text{\footnotesize ${\nu_L}_j$}}
  \put(31,5){\text{\footnotesize ${\nu_L}_i$}}
  \put(11,5){\text{\footnotesize $N_{1,\ldots,n-1}$}}
  \put(14,20){\text{\footnotesize $\eta_{R/I}$}}
\end{picture}
\end{minipage}
 +
\begin{minipage}[c]{3.5cm}
\begin{picture}(35,25)
 \put(0,0){\includegraphics[width=3.5cm]{loop2phi}}
  \put(0,5){\text{\footnotesize ${\nu_L}_j$}}
  \put(31,5){\text{\footnotesize ${\nu_L}_i$}}
  \put(14,28){\text{\footnotesize $\eta_{R/I}$}}
  \put(14,2){\text{\footnotesize $\stackrel{(n)}{\kappa}^{\raisebox{-.9ex}{$\scriptstyle(22)$}}$}}
\end{picture}
\end{minipage}
\end{equation*} 
\caption{Diagrammatic matching of the operator $\kappa^{(11)}$ at the mass threshold $\mu_*=m_\eta(\mu_*)$.}\label{fig:matchingscalar} 
\end{figure}

Finally, we must consider the possibility that we have to integrate out the inert scalars, as it is shown in Fig.~\ref{fig:matchingscalar}. Since by Eq.~\eqref{eq:scalarMasses} their mass scale is set by the value of either $m_\eta^2$ or $v^2$, we integrate them out at a common scale $\mu_*=m_\eta(\mu_*)$ if this threshold is encountered between the high input scale and $M_Z$. The resulting EFT carries the label ``0'' in Fig.~\ref{fig:EFT}. 

In this case the matching will have several effects: one is that the scalar sector is reduced to the simple SM form with only one Higgs doublet. This also modifies the gauge coupling RGEs and thereby all other RGEs, which has to be taken into account in the numerical analysis. Also, we remove the operator $\kappa^{(22)}$ and all neutrino Yukawa couplings $h_{ij}$. Therefore, once the inert scalars are integrated out, the RH neutrinos ``decouple'' from the remaining fields and the active neutrino masses are generated exclusively through the Weinberg operator $\kappa^{(11)}$ (cf.~Fig.~\ref{fig:EFT}). 

Assuming that $(n-1)$ RH neutrinos are still present, the matching condition shown in Fig.~\ref{fig:matchingscalar} gives: 
\begin{equation}\begin{split} \label{eq:matchingscalar}
  \left.\eff{0}{1}{ij}\right|_{\mu=m_\eta} &= \left.\eff{n}{1}{ij}\right|_{\mu=m_\eta} + 
    \frac{1}{32\pi^2} \left[ \frac{1}{v^2}\eff{n}{2}{ij} \left(m_R^2 \log\frac{m_R^2}{m_\eta^2} - m_I^2 \log\frac{m_I^2}{m_\eta^2} \right) - \right.\\ 
    & \hspace{4.15cm }\left. - \frac{2}{v^2} \sum_{\ell<n} \stackrel{(n)}{h}_{\hspace{-.4ex}\ell i} \overset{(n)}{M_\ell}\ g(\overset{(n)}{M_\ell},m_R,m_I) \overset{(n)}{h}_{\hspace{-.4ex}\ell j}\right]_{\mu=M_n}.
\end{split}\end{equation}
Once this threshold is encountered, no further thresholds need to be taken into account, since the RH neutrinos are decoupled. It is interesting to observe that the contribution of the last diagram in Fig.~\ref{fig:matchingscalar} to Eq.~\eqref{eq:matchingscalar} is proportional to the logarithm $\log\frac{m_{R/I}^2}{m_\eta^2}$ -- which is small. In fact, if we matched at a scale $m_R \simeq m_I$ instead, this contribution would vanish and the matching condition would simplify. However, this would require a modification of the numerical determination of the threshold, so we refrain from this alternative approach.

\section{\label{sec:Analysis}Numerical Analysis \& Analytical Results}

For our numerical study we have used the full set of RGEs given in App.~\ref{app:Appendix_RGE}, neglecting all SM Yukawa couplings but that of the top and considering the case of three RH
neutrinos. As we have verified, this is a good approximation as long as the running of the mixing angles is driven by the neutrino Yukawa couplings rather than the charged lepton Yukawa couplings.
This can also be seen from Eqs.~\eqref{eq:MixingAnglesAnalytic} using the replacement $\alpha_h h_{1,2,3}^2 \mapsto \alpha_h h_{1,2,3}^2 +\alpha_e y_{e,\mu,\tau}^2$.

The purpose of this section is twofold: on the one hand, we illustrate how the running of parameters translates into running of physical observables such as neutrino masses and mixing angles. We will see that running effects can be large without fine-tuning parameters. On the other hand, we wish to compare the results predicted by our analytical formulae to those of a more detailed numerical study. Since the analytic equations are valid assuming $\theta_{13}$ to be small, we will impose a bimaximal mixing pattern~\cite{Barger:1998ta} at the high input scale (which we take to be equal to the GUT scale for definiteness) such that $\theta_{13}=0^\circ$ and $\theta_{12}=\theta_{23}=45^\circ$. Due to the potentially large running of the mixing angles this can yield the experimentally measured values at the electroweak scale. 

In doing so we need to ensure that we meet all requirements for the analytic formulae to be applicable. This means especially that $h^\dag h$ must be diagonal. We may achieve this by starting off with a diagonal neutrino Yukawa matrix $h = \mathrm{diag}\left(h_1,h_2,h_3\right)$ at the high scale, but leaving the RH mass matrix $M$ arbitrary. We can then transform into a basis where $M$ is diagonal. The procedure is the following:

Let us start with the active neutrino mass matrix, which can be diagonalised as:
\begin{equation}
  U^T \mathcal{M}_\nu U = D_\nu \equiv \mathrm{diag}\left(m_1,m_2,m_3\right),
\end{equation}
where $U$ is unitary. In a basis where the charged lepton Yukawa matrix is diagonal, it is just the PMNS matrix.

With our expression for the mass matrix in the scotogenic model, Eq.~\eqref{eq:MaMass}, we get in the limit $\lambda_5 \ll 1$:
\begin{equation} 
  D_\nu \simeq U^T h^T \left(-v^2 \frac{\lambda_5}{(4\pi)^2}M^{-1} f(M,m_\eta)\right) h\, U \equiv \left( U^T h^T \right) \Lambda\, \left(h\, U\right),
\end{equation}
which we can solve for $\Lambda$. In general, this matrix will not be diagonal but, since it is symmetric, we can diagonalise it with the help of another unitary matrix $V$:
\begin{align}
  \textrm{diag}\left(\Lambda_1,\Lambda_2,\Lambda_3\right) \equiv& D_\Lambda = V^T \Lambda V = V^T \left(U^T h^T \right)^{-1} D_\nu\, (h\, U)^{-1} V \nonumber\\
  =& {\left(V^\dag h\right)^T}^{-1} U^* D_\nu\, U^\dag \left(V^\dag h\right)^{-1}. \label{eq:inputYukawaDiag}
\end{align}
Thus, if we fix $m_\eta$, the three active masses, mixing angles, and phases at the GUT scale, we may solve Eq.~\eqref{eq:inputYukawaDiag} for the $\Lambda_{1,2,3}$ (and thereby $M_{1,2,3}$). In this basis the Yukawa matrix is of the form $\tilde{h} = V^\dag \textrm{diag}\left(h_1,h_2,h_3\right)$ such that ${\tilde{h}}^\dag \tilde{h}$ is indeed diagonal (as assumed in our analytical estimates).

This procedure is similar but not identical to the Casas-Ibarra parametrisation~\cite{Casas:2001sr}, and it allows us to impose that $h^\dag h$ is diagonal, which is needed for the analytic equations to be applicable. If this is not required, we may instead fix the RH mass matrix and use the Casas-Ibarra result:
\begin{equation} \label{eq:CasasIbarra}
  h = \sqrt{\Lambda^{-1}} R \sqrt{D_\nu} U^\dag,
\end{equation}
where now we assume $\Lambda$ (or equivalently $M$) to be diagonal, while $R$ is a (complex) orthogonal but otherwise arbitrary matrix and $U$ is defined as above.

For the scalar couplings we find a suitable choice to be:
\begin{equation}
  \left.\left( \lambda_1, \lambda_2,\ \lambda_3,\ \lambda_4,\ \lambda_5\right)\right|_{\mu=M_\textrm{GUT}} = \left(-0.01,\ 0.1,\ 10^{-9},\ 10^{-9},\ 10^{-9} \right),
\end{equation}
where at large energy scales we need to violate vacuum stability ($\lambda_1<0$) in order to reproduce the measured Higgs mass and self-coupling, as it is the case in the SM. It is a known shortcoming of inert doublet models that it is not simple to ensure stability of the scalar potential up to the Planck scale, see Ref.~\cite{Das:2015mwa}. Since we focus on illustrating the consequences of the running in the scotogenic model for the neutrino sector, we have chosen the above values in order to avoid the additional difficulty of having to deal with an unsuitably chosen potential as well. This could distract the reader from the actual messages we would like to bring across. However, we would like to point out that of course the running of the scalar potential may introduce all kinds of additional difficulties when trying to find a realistic choice of parameters (the parity problem~\cite{Merle:2015gea} of the scotogenic model being just one example).

\subsection{Dominant scalar mass}

\begin{figure}[p]
  \subfigure[mixing angles]{\includegraphics[width=.5\textwidth]{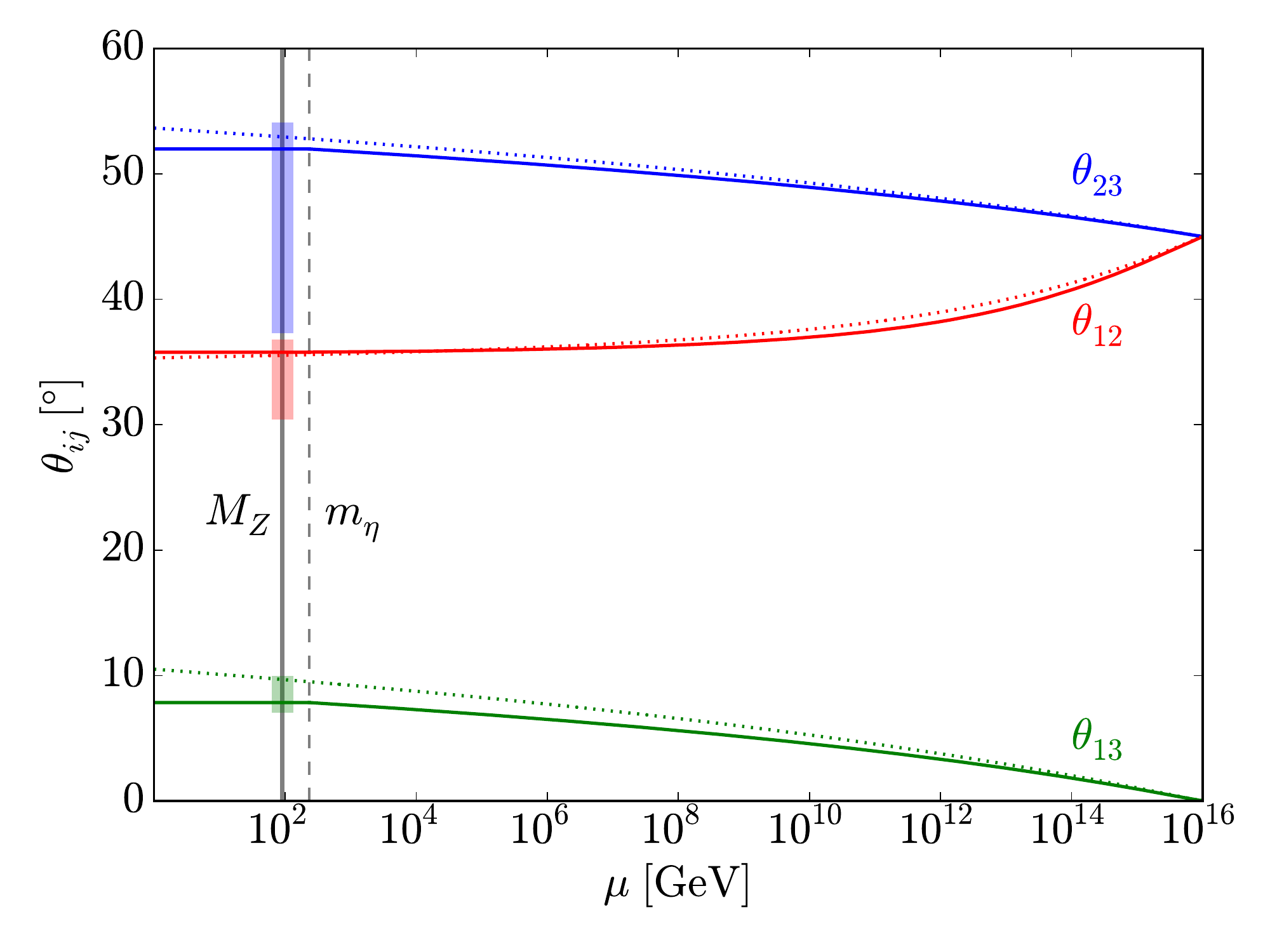}\label{fig:angles_scalarDom1}}
  \subfigure[masses]{\includegraphics[width=.5\textwidth]{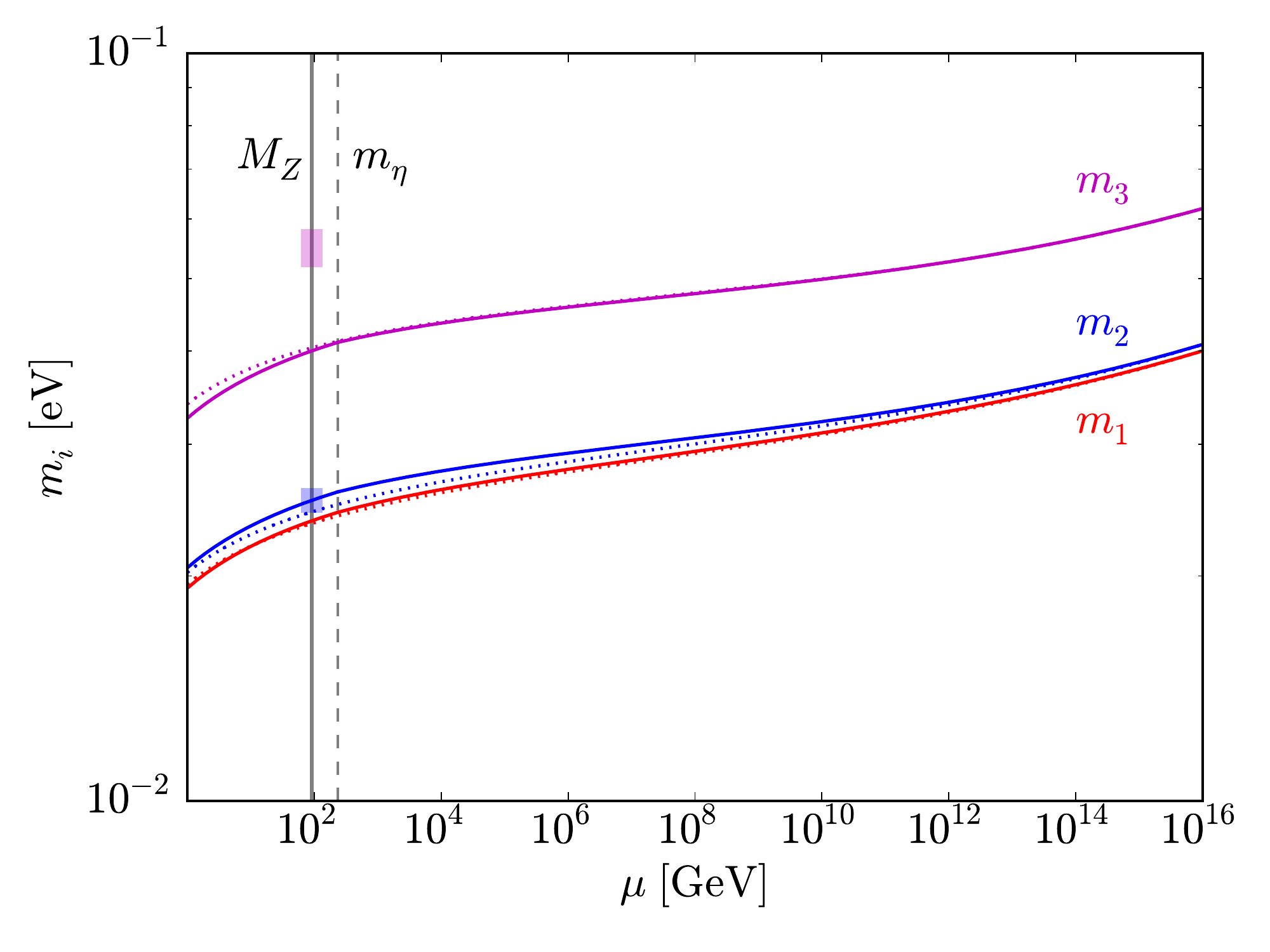}}
  \subfigure[mass square differences (absolute)]{\includegraphics[width=.5\textwidth]{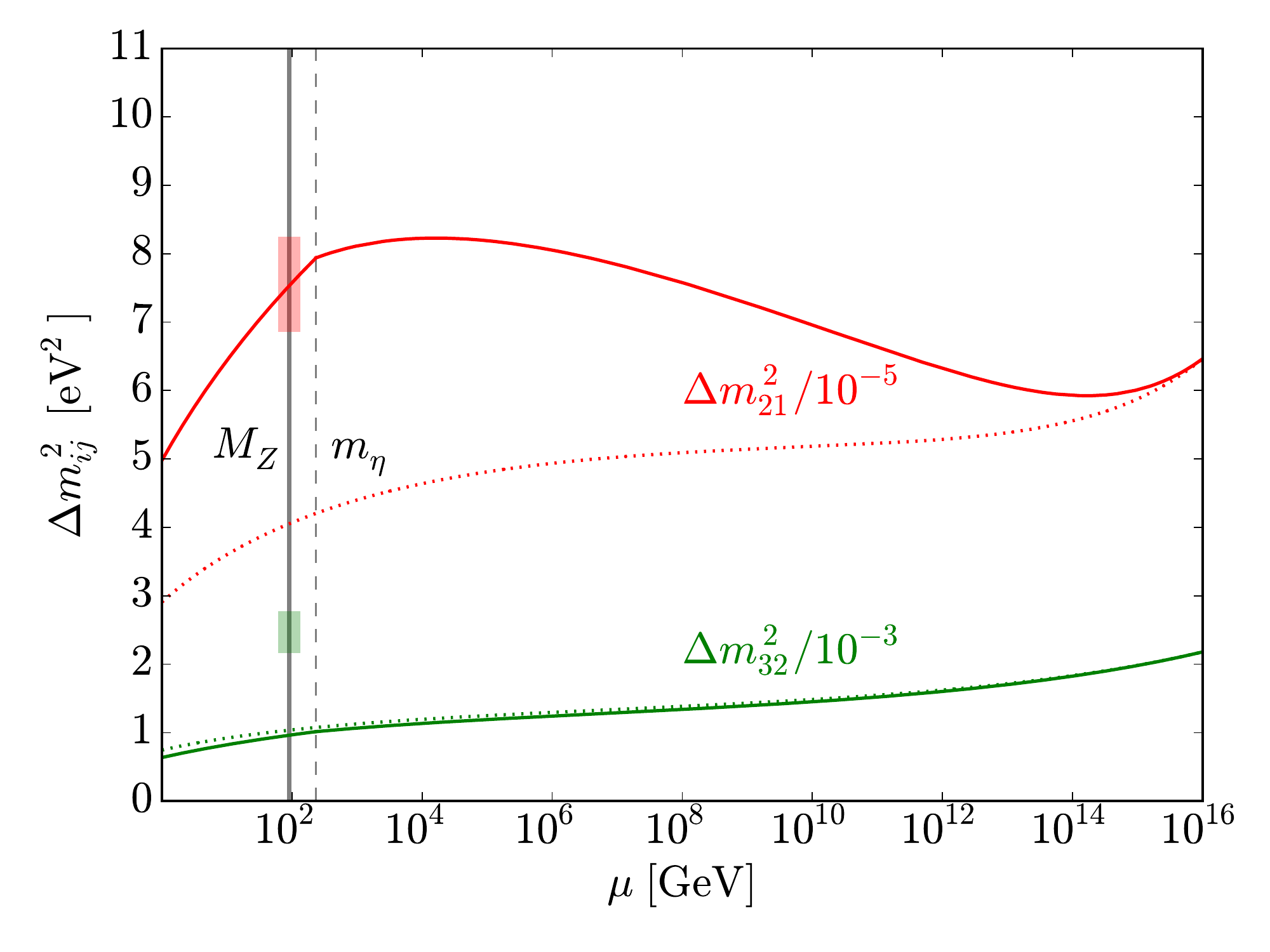}\label{fig:runningDiff_scalarDom1}}
  \subfigure[mass square differences (relative)]{\includegraphics[width=.5\textwidth]{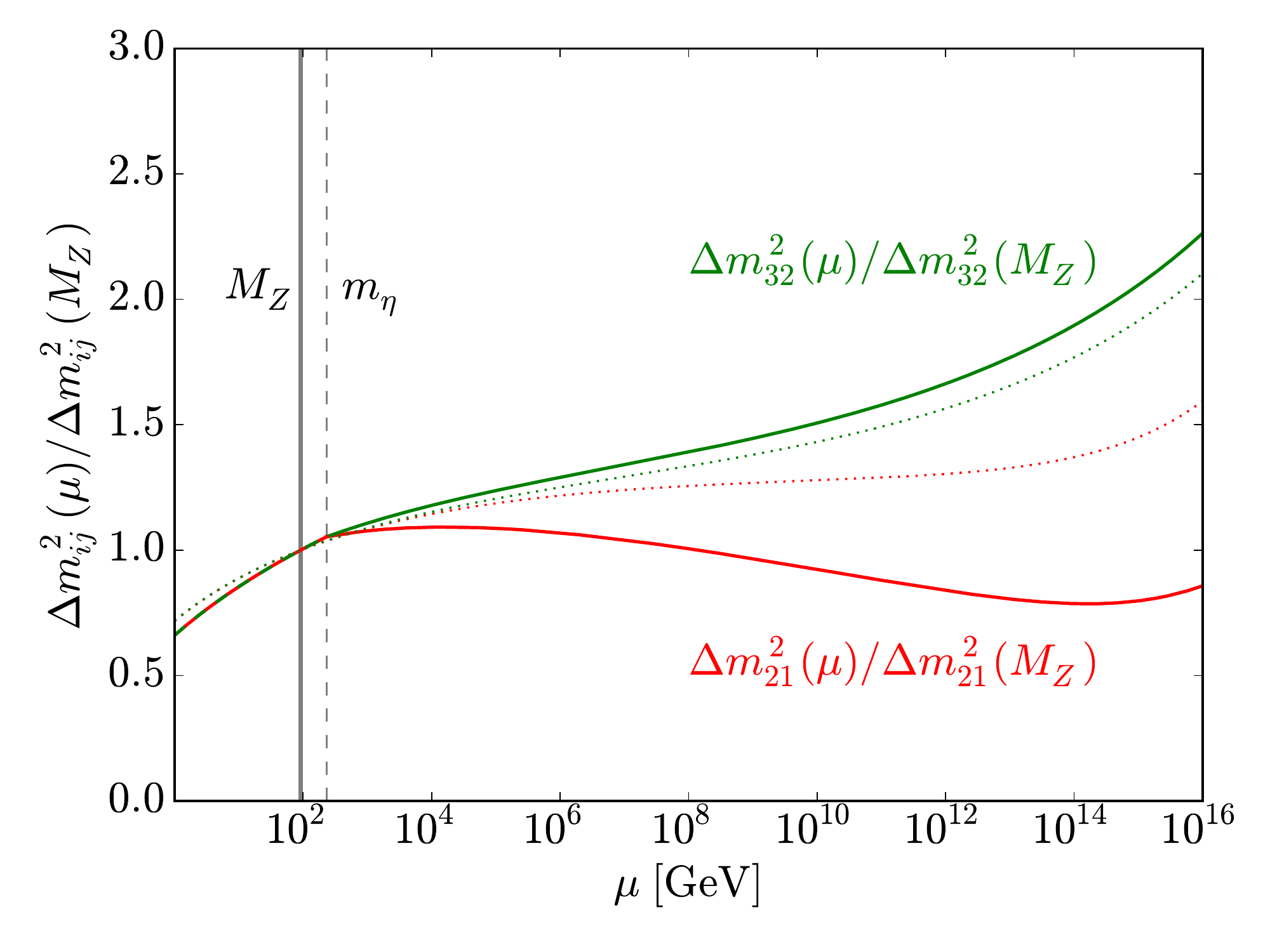}\label{fig:runningDiff_scalarDom2}}
  \caption{\label{fig:running_ScalarDom}Running mixing angles and active neutrino masses for heavy scalars [case~i) in section \ref{sec:analytic}]. Solid lines are numerical results and the dotted lines originate from our analytical equations. The coloured areas indicate the experimental $3\sigma$ ranges for the mixing angles and mass square differences in normal mass ordering.}
\end{figure}

In our first example, we show the results of both analytical and numerical computations for the case~i), i.e.\ $m_\eta^2 \gg M_k^2$. The Yukawa couplings are of $\mathcal{O}(1)$ and we choose a scalar mass parameter $m_\eta=350 \textrm{ GeV}$ at $M_\textrm{GUT}$. The RH neutrino masses are fixed according to the above procedure and take values $\lesssim 100 \textrm{ GeV}$. Besides, we have chosen Majorana phases $\phi_1 = \frac{3 \pi}{2}$ and $\phi_2 = \frac{3\pi}{4}$, while the Dirac phase is $\delta = 0$ at the high input scale.

Note that such a scenario can be under tight constraints coming from collider physics~\cite{Agashe:2014kda} and lepton flavour violating (LFV) processes~\cite{Sierra:2008wj,Toma:2013zsa}, since many such processes ultimately yield lower bounds on the RH neutrino and/or inert scalar masses, which can in particular be dangerous for RH neutrino masses around the $Z$-pole. However, given the nature of LFV diagrams at both low and high energies (where the outer states are comprised of SM leptons while the virtual particles involved are the new scalars and/or fermions of the scotogenic model), what is actually constrained is only a \emph{combination} of certain masses and couplings. While in a detailed phenomenological study all the collider and/or LFV bounds have to be included, the approach we take here for illustrative purposes is to assume the neutrino Yukawa couplings small enough that the bounds are no problem.

Fig.~\ref{fig:running_ScalarDom} shows the results for this example. First, let us emphasise that indeed strong running effects can be achieved without fine-tuning parameters to artificially small or large values. This is true for both mixing angles and masses. Note also that below the scalar threshold, i.e.\ in the SM effective theory, the mixing angles no longer run.\footnote{This would still be approximately the case had we not neglected all SM particles but the top quark: while Eqs.~\eqref{eq:MixingAnglesAnalytic} tell us that the running will then be induced by lepton Yukawa couplings, they are however very small and can only cause very little running (cf.~\cite{Antusch:2003kp}).}

\begin{figure}[p]
  \subfigure[mixing angles]{\includegraphics[width=.5\textwidth]{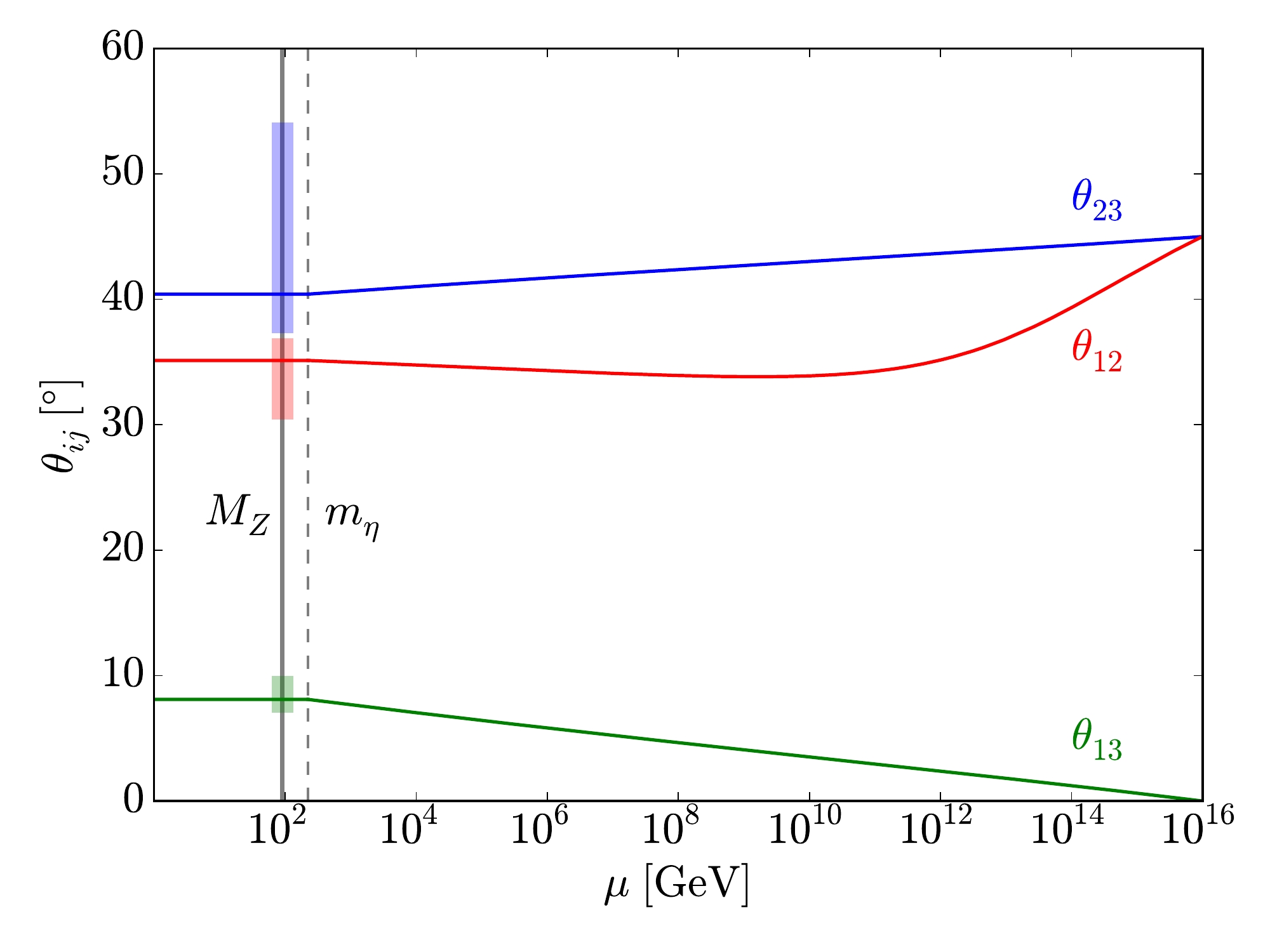} \label{fig:angles_scalarDom2}}
  \subfigure[masses]{\includegraphics[width=.5\textwidth]{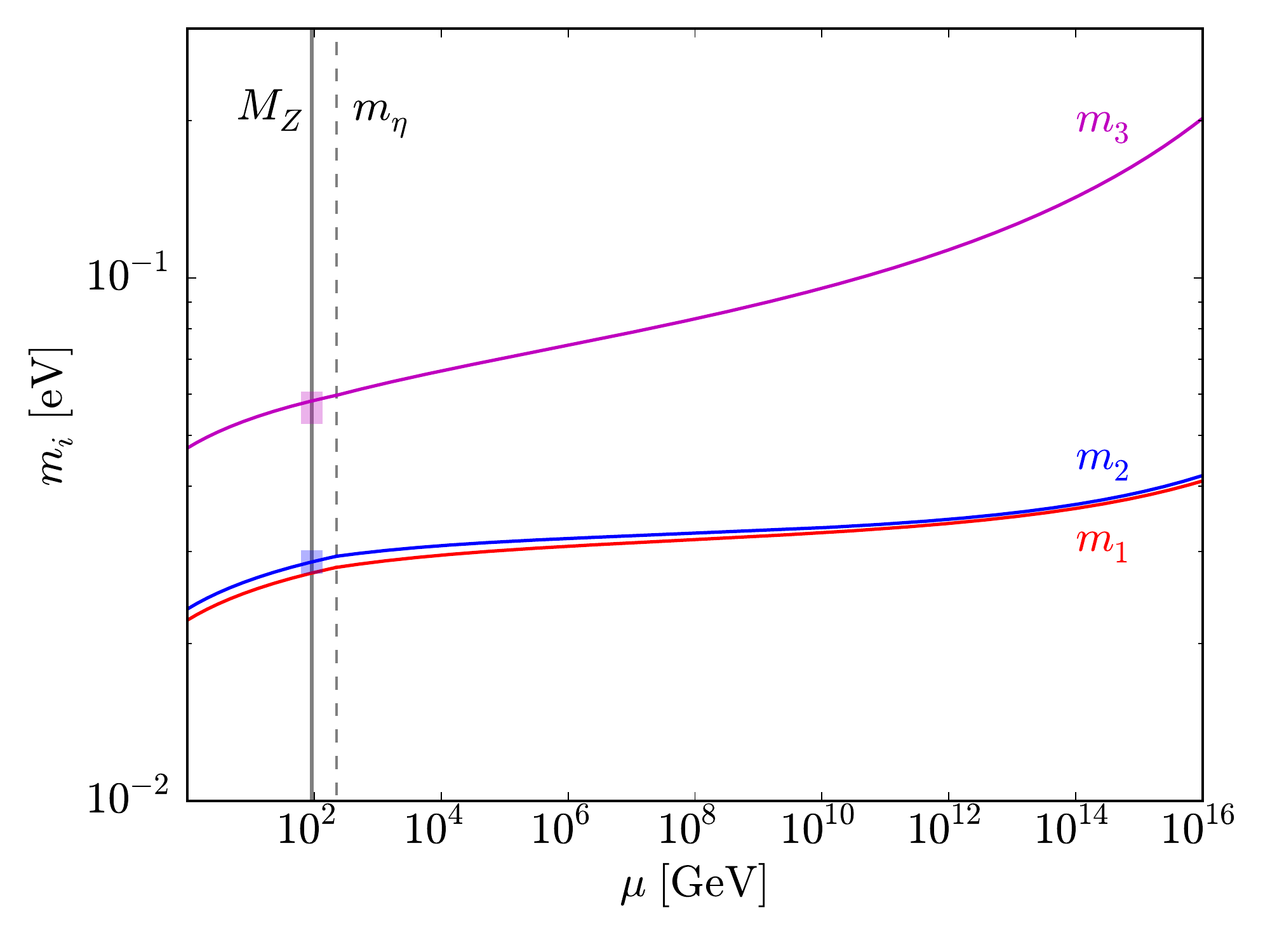} \label{fig:masses_scalarDom2}}
  \subfigure[mass square differences (absolute)]{\includegraphics[width=.5\textwidth]{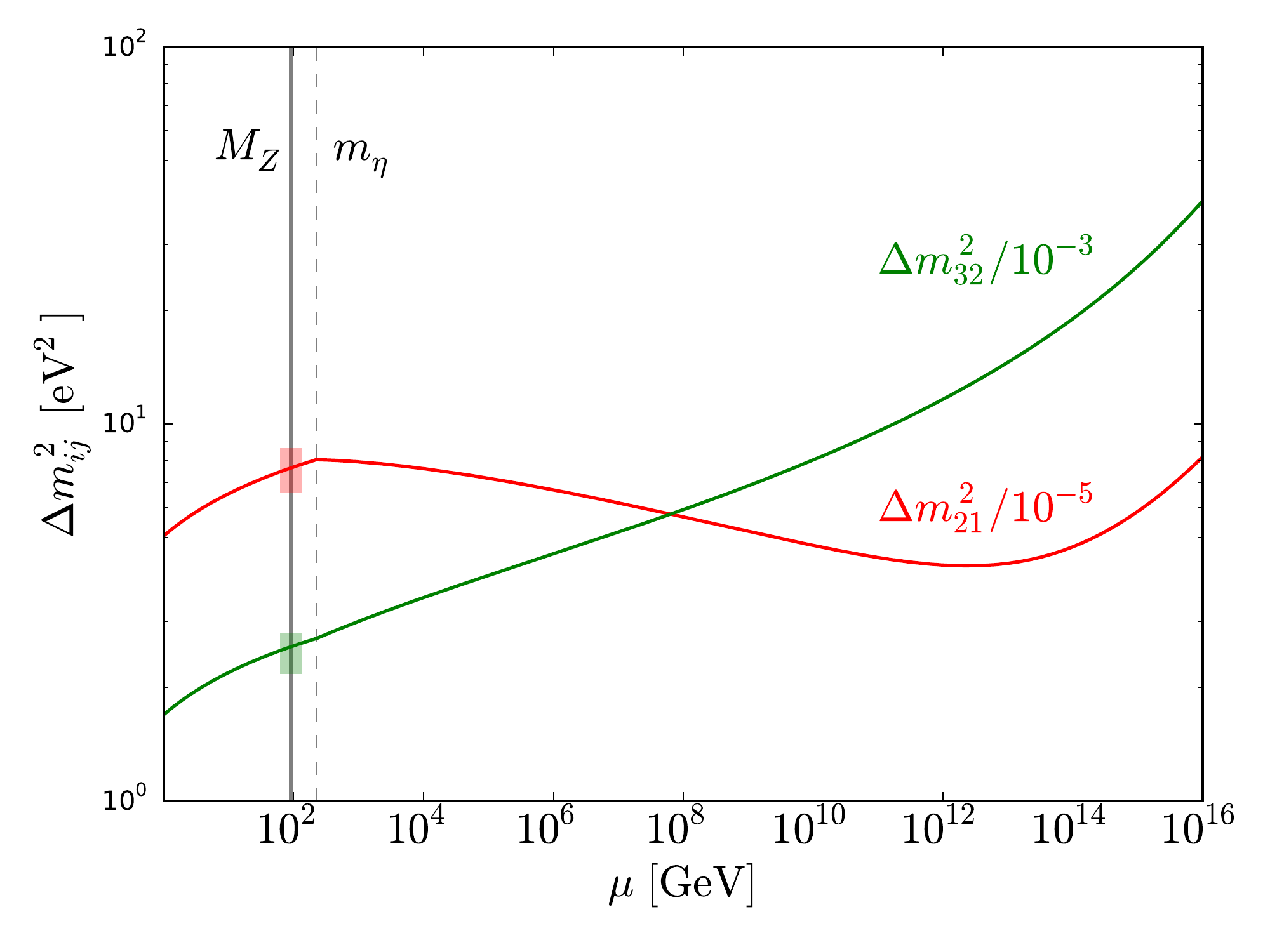}\label{fig:runningDiff_scalarDom3}}
  \subfigure[mass square differences (relative)]{\includegraphics[width=.5\textwidth]{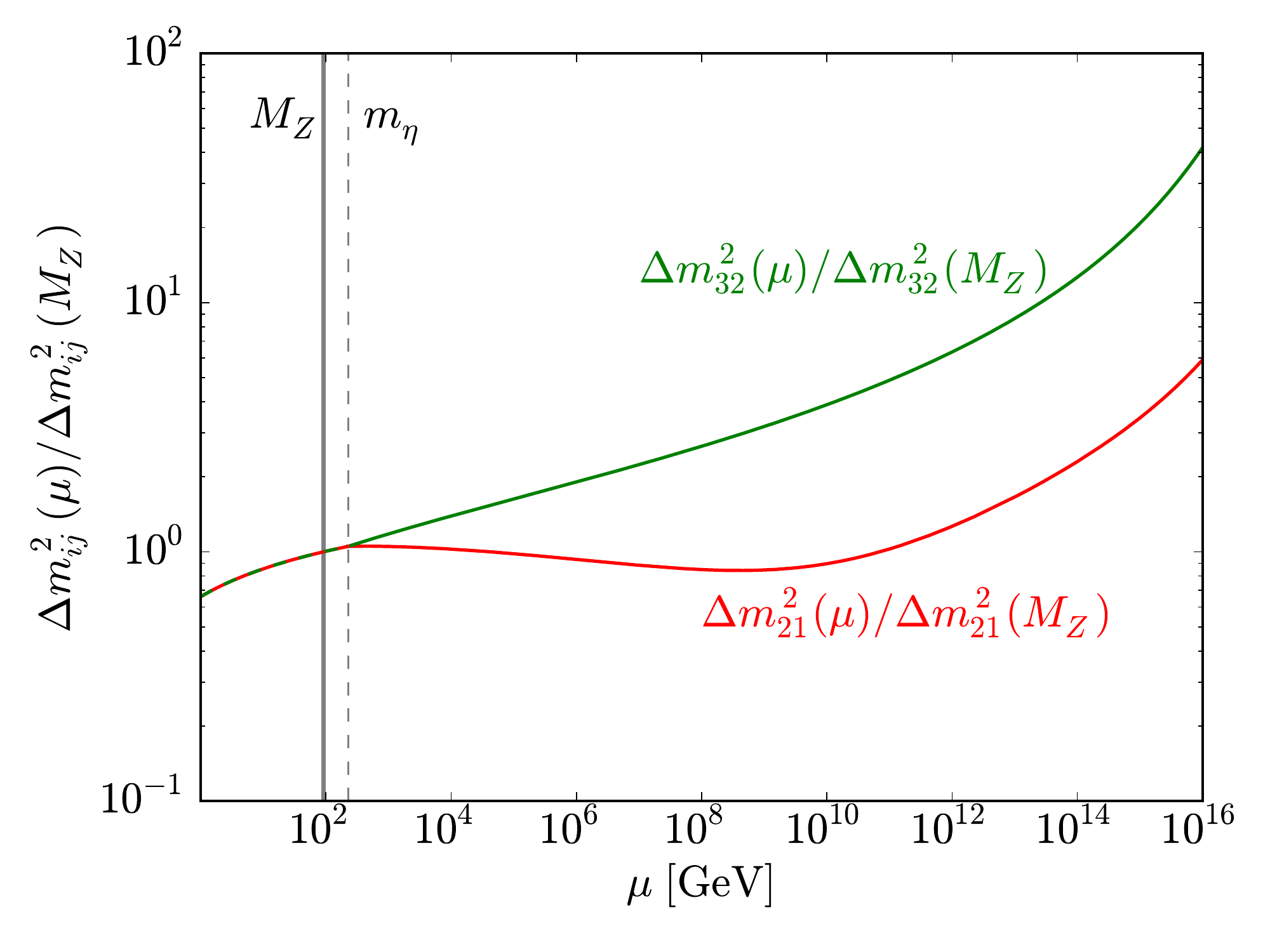}\label{fig:runningDiff_scalarDom4}}
  \caption{Running mixing angles and masses if the assumption $h^\dag h = \mathrm{diag}\left(h_1^2,h_2^2,h_3^2\right)$ is dropped. All mixing parameters agree with the experiments (lightly coloured boxes mark the $3\sigma$ ranges). Note the prominent running of the mass square differences, especially of $\Delta m_{32}^2$.}
\end{figure}

\begin{figure}[t]
  \subfigure[$h^\dag h$ diagonal]{\includegraphics[width=.5\textwidth]{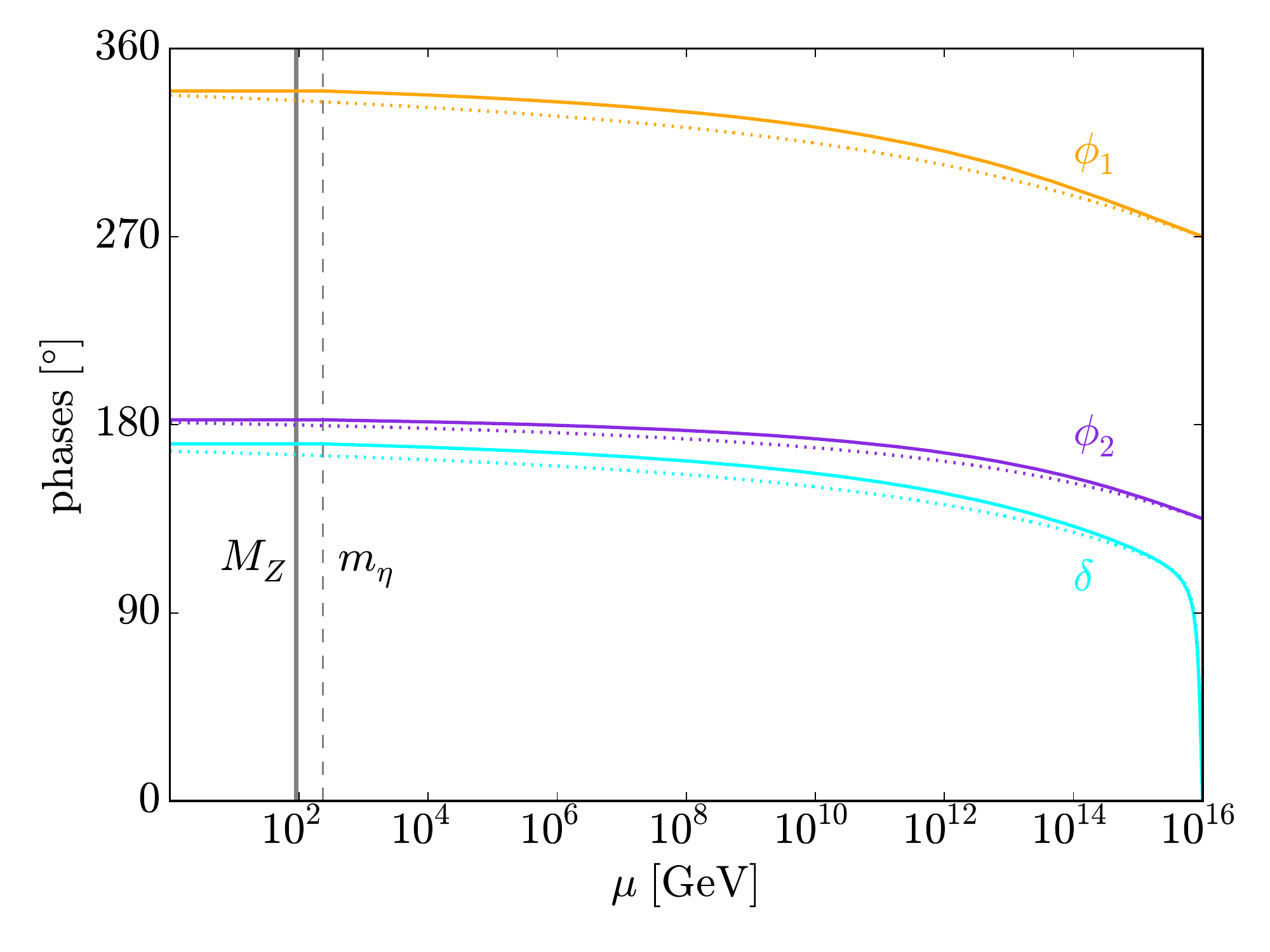}}
  \subfigure[Casas-Ibarra]{\includegraphics[width=.5\textwidth]{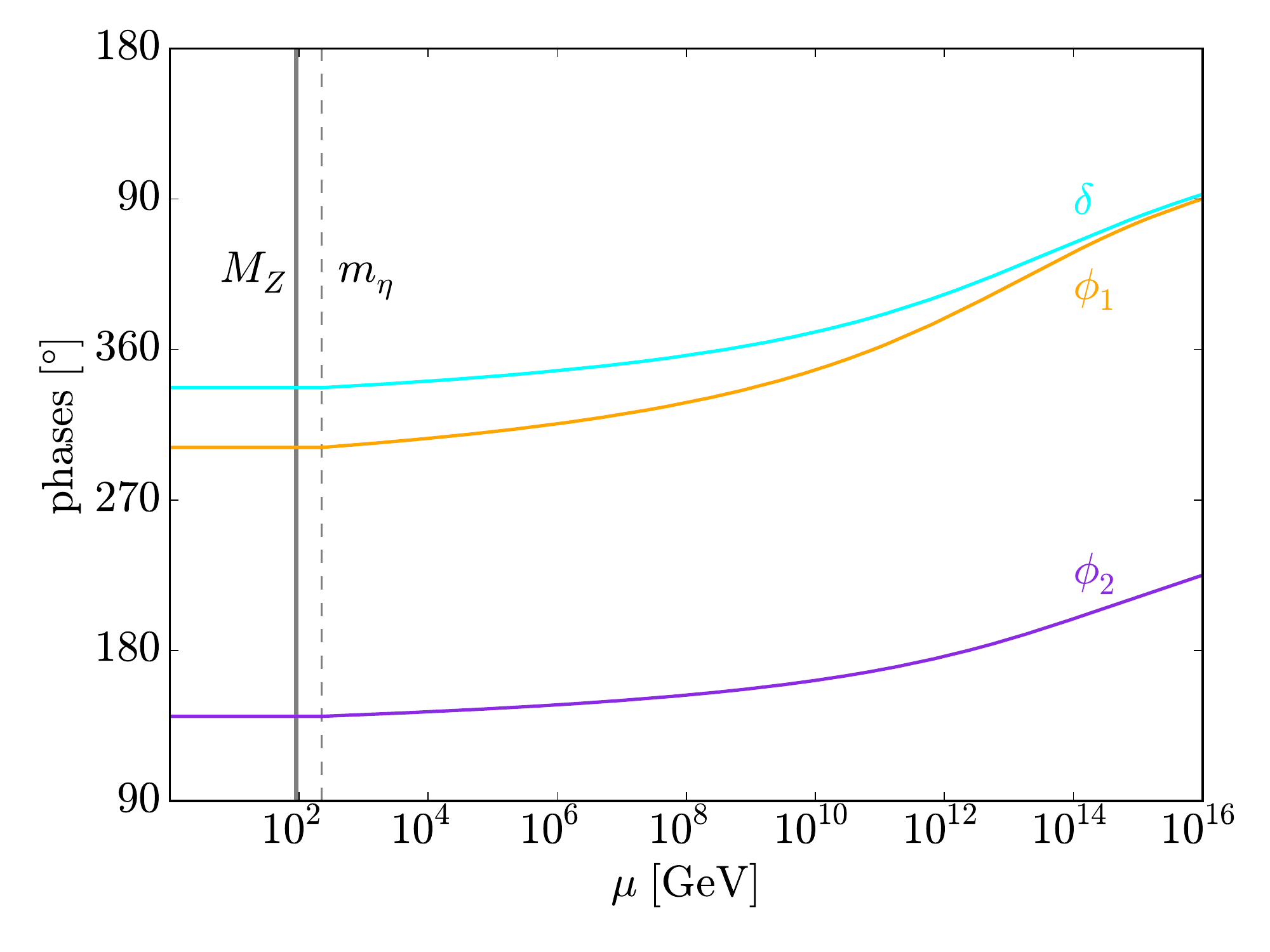}}
  \caption{\label{fig:phases}Running phases in the case of diagonal (left) and arbitrary (right) $h^\dag h$. Note that the phases can be restricted to the interval $[0,2\pi)$ in our parametrisation.}
\end{figure}

The analytical and numerical computations agree well as long as $\theta_{13}$ is small. However, as we go to smaller scales, $\theta_{13}$ must grow to reach its measured value, $\theta_{13} \sim 10^\circ$, and the results of the analytical treatment will inevitably deviate. Nevertheless, they provide a useful tool to estimate the values of masses and angles at the low scale given some mixing pattern at a high scale, or vice versa. 

From the example shown in Fig.~\ref{fig:angles_scalarDom1} we extract the following mixing angles at the low scale ($\mu=M_Z$):
\begin{equation}
  \theta_{12} =  35.78^\circ,\qquad \theta_{23}= 51.98^\circ,\qquad \theta_{13} = 7.85^\circ .
\end{equation}
The running of the mass square differences is shown in Fig.~\ref{fig:runningDiff_scalarDom1} and we can read off:
\begin{equation}
  \Delta m_{21}^2 = 7.5 \times 10^{-5} \textrm{ eV}^2,\qquad \Delta m_{32}^2 = 1.0 \times 10^{-3} \textrm{ eV}^2.
\end{equation}
While the angles are well within their experimental $3\sigma$ ranges, as is the solar mass square difference $\Delta m_{21}^2$~\cite{Gonzalez-Garcia:2014bfa}, the atmospheric $\Delta m_{32}^2$ is too small. In fact, for the case of dominant scalar mass and diagonal $h^\dag h$, we have not found a setting where all parameters fit the experimental results. This should, however, not be taken too seriously given the strong assumptions on the form of the Yukawa coupling matrix $h$ and the RH neutrino mass matrix $M$. Moreover, Fig.~\ref{fig:runningDiff_scalarDom2} teaches us that the running of the mass square differences is not only large in the full theory, but also in the SM as an effective theory below the scalar threshold. This is easily understood from Eqs.~\eqref{eq:MassesAnalytic}: the factor $C$ contains gauge couplings and the top Yukawa coupling, which are both large and also run significantly.

Finally, we have used the Casas-Ibarra parametrisation, Eq.~\eqref{eq:CasasIbarra}, to show that we can indeed reproduce the experimentally determined values of the mixing angles and mass square differences when relaxing our requirement that $h^\dag h$ be diagonal. Again, we choose $m_\eta = 350 \textrm{ GeV}$ and and RH neutrino masses $\left( M_1, M_2, M_3\right)= \left(50, 100, 125 \right) \textrm{ GeV}$ at $M_\textrm{GUT}$. All mixing angles, see Fig.~\ref{fig:angles_scalarDom2}, and mass square differences, see Fig.~\ref{fig:runningDiff_scalarDom3}, are in agreement with the experimental results. Note that the running of $\theta_{23}$ is reversed compared to the previous case, which can be explained by the appearance of off-diagonal elements in $h^\dag h$ with respect to the analytical RGE, Eq.~\eqref{eq:theta12Analytic}. We see that the running of $\Delta m_{32}^2$ is very strong, which is also nicely visible in Fig.~\ref{fig:runningDiff_scalarDom4}, where the mass square differences are plotted 
relative to their values at the $Z$ mass. The reason for this is that $m_3$ runs to very large values, as can be seen from Fig.~\ref{fig:masses_scalarDom2}.

We wish to conclude this subsection by discussing the running of the phases, which is shown in Fig.~\ref{fig:phases}. The left panel of this figure shows the running phases in the case where we impose $h^\dag h$ to be diagonal. In this case, in particular the running of $\delta$ is highly pronounced, and it is driven from its starting value $\delta=0$ to a value of almost $\pi$, such that the running interpolates completely between two regions of CP conservation, while passing through a region of maximal CP violation in between. This behaviour is correctly captured by the analytical estimates (in dotted lines). The right panel displays the results when using the Casas-Ibarra parametrisation, where the running is much weaker.

\subsection{No hierarchy among scalar and RH neutrino masses}

\begin{figure}[p]
  \subfigure[mixing angles]{\includegraphics[width=.5\textwidth]{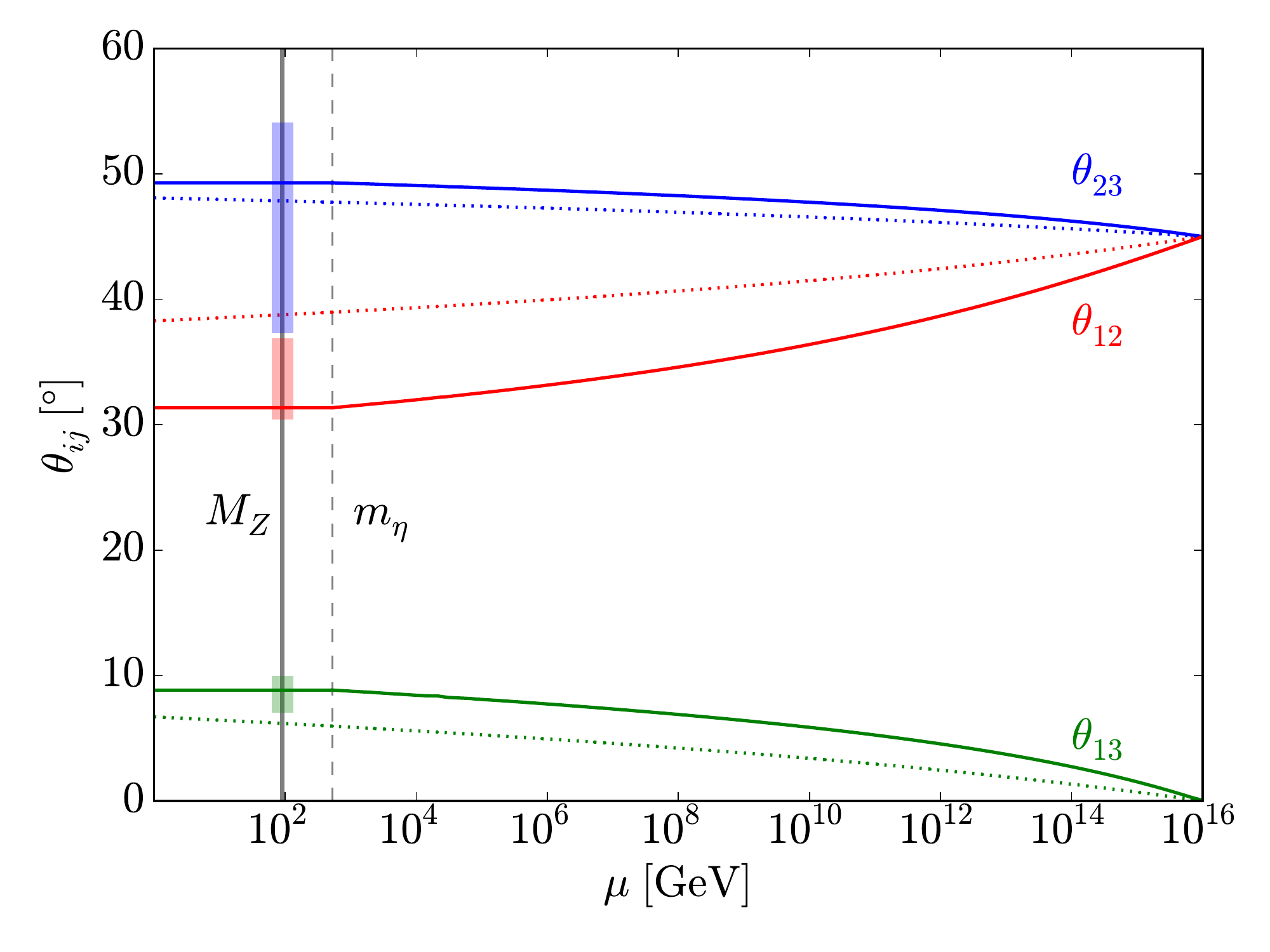}\label{fig:angles_noHierarchy1}}
  \subfigure[masses]{\includegraphics[width=.5\textwidth]{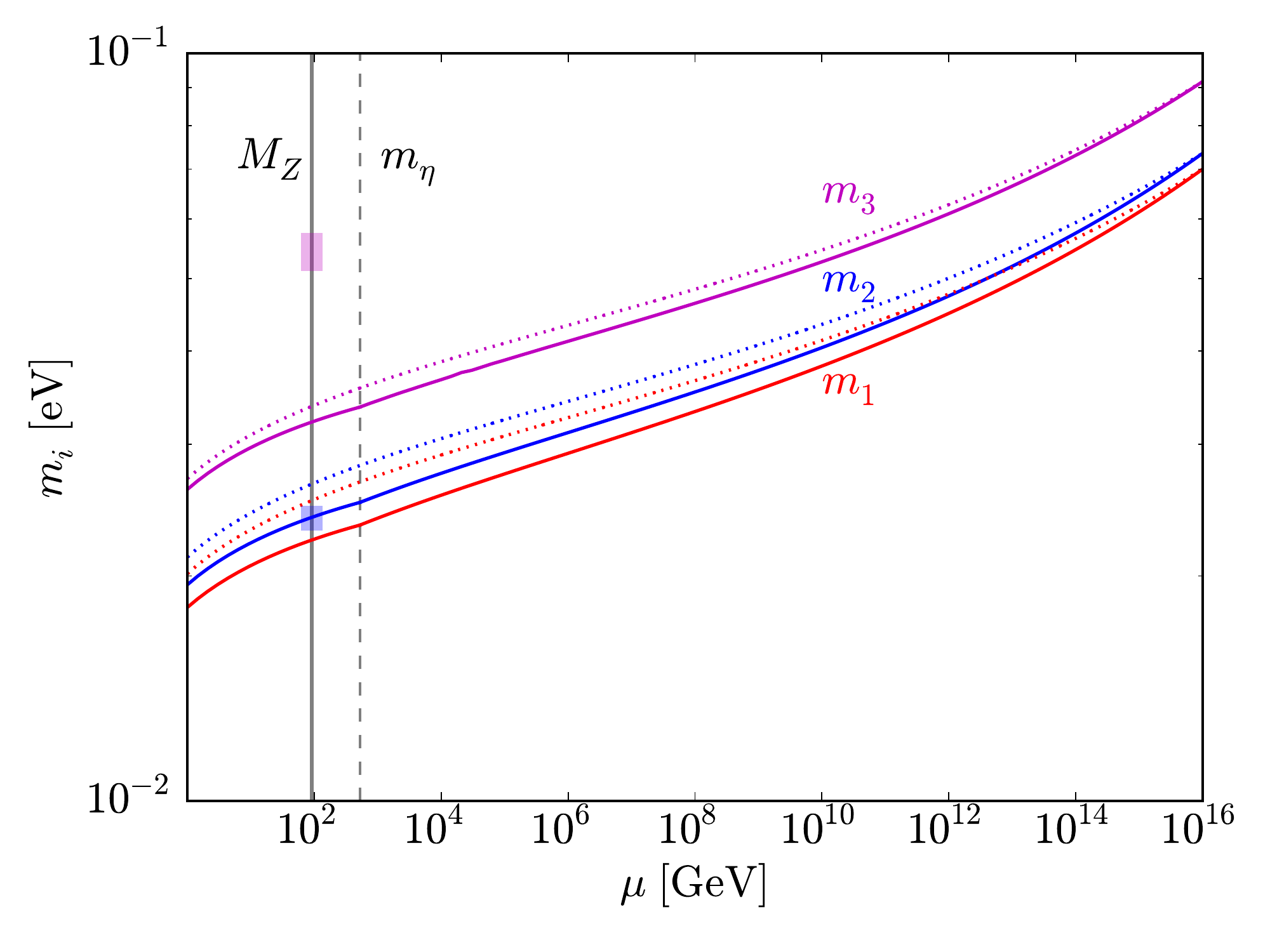}\label{fig:masses_noHierarchy1}}
  \subfigure[mass square differences (absolute)]{\includegraphics[width=.5\textwidth]{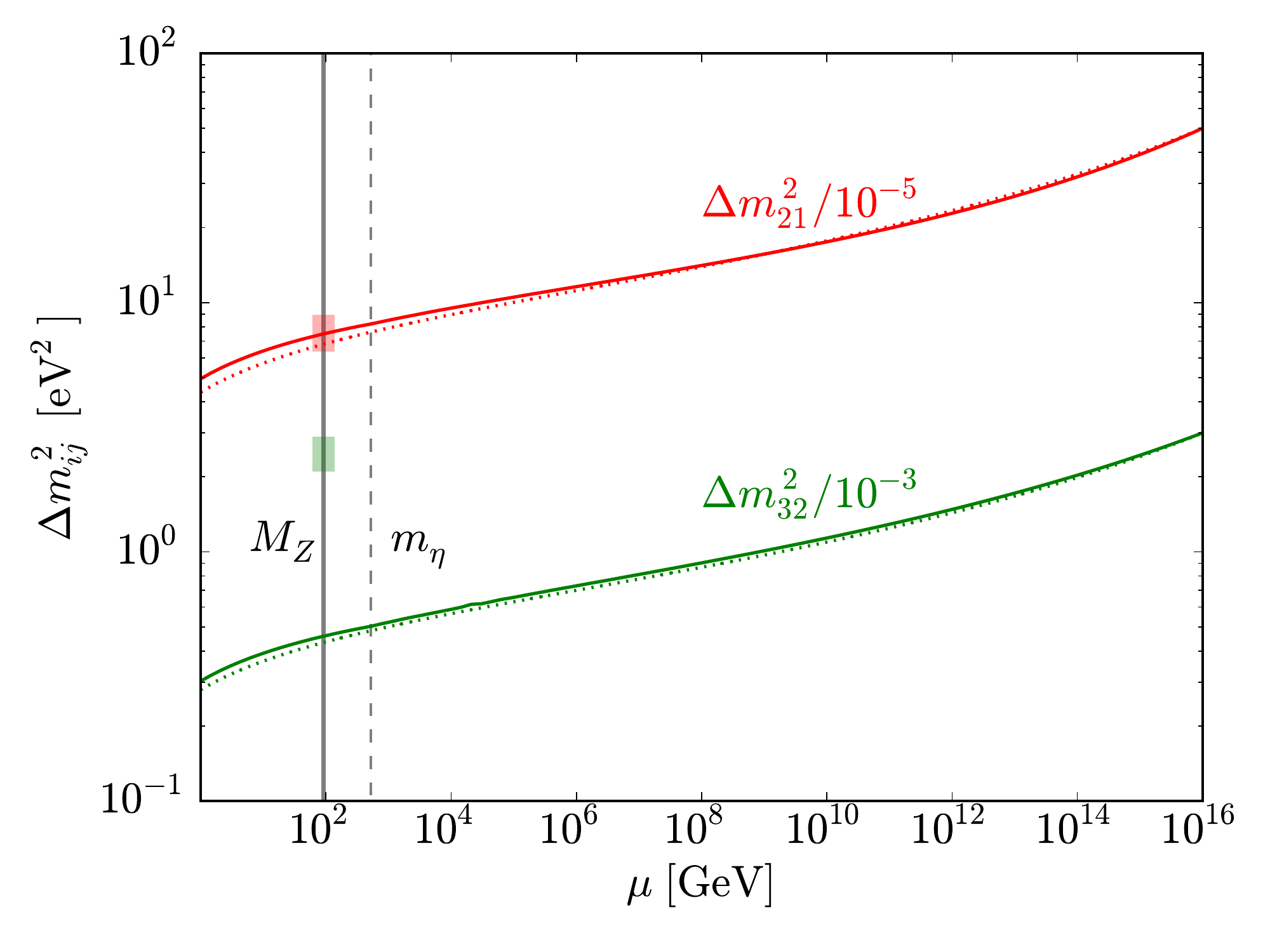}\label{fig:massDiff_noHierarchy1}}
  \subfigure[mass square differences (relative)]{\includegraphics[width=.5\textwidth]{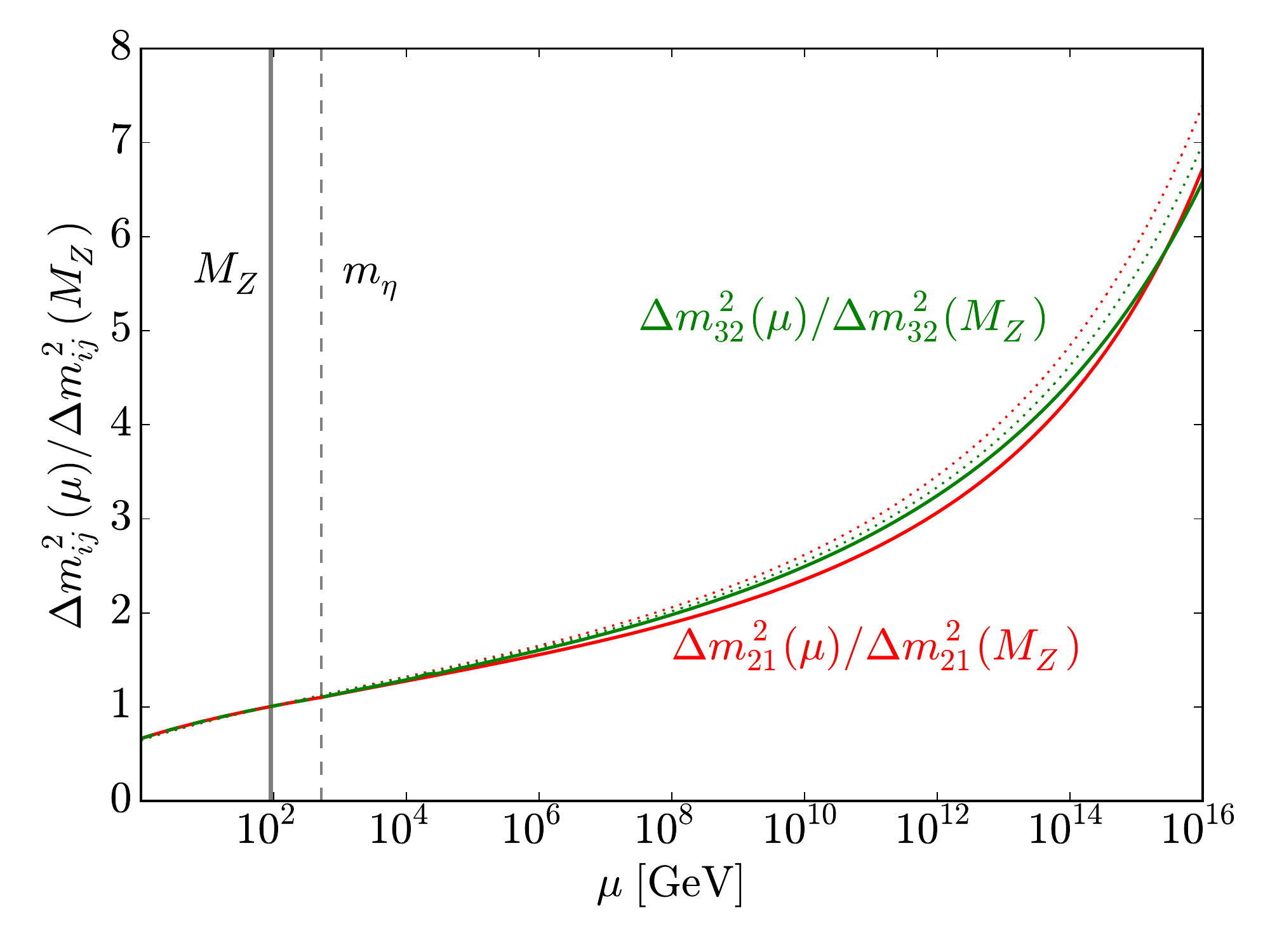}\label{fig:massDiff_noHierarchy2}}
  \caption{\label{fig:running_nohierarchy1}Running mixing angles and active neutrino masses for scalar masses comparable to those of the RH neutrinos [case~ii) in Sec.~\ref{sec:analytic}]. Solid lines are numerical results and the dotted lines originate from our analytical equations. The coloured areas indicate the experimental $3\sigma$ ranges for the mixing angles and mass square differences in normal mass ordering.}
\end{figure}

This limit, which corresponds to case~ii) in Sec.~\ref{sec:analytic}, is quite difficult to realise since, as we have learned from Sec.~\ref{sec:analytic}, we need $\mathcal{O}(1)$ Yukawa couplings to achieve large running effects and therefore at the same time we need a hierarchy among the RH masses to reproduce the desired mass square differences. This means that the RH neutrino masses cannot be exactly degenerate and the requirements $M_{1,2,3} \simeq m_\eta$ can hardly be met all at once. For our example shown in Fig.~\ref{fig:running_nohierarchy1}, we have chosen $m_\eta\left(M_\textrm{GUT}\right) = 700 \textrm{ GeV}$, which requires together with our $\mathcal{O}(1)$ Yukawa couplings and example phases of $\phi_1 = \frac{3\pi}{4}$, $\phi_2=\frac{5\pi}{4}$, and $\delta=0$, RH Majorana masses:
\begin{equation}
  \left( M_1, M_2, M_3 \right) = \left( 340, 407, 851 \right) \textrm{ GeV}.
\end{equation}
Clearly, we cannot expect the approximation to hold very accurately and it turns out that we find better agreement if we replace $M$ by $m_\eta$ in the formula for the mass matrix, as we had suggested in Sec.~\ref{sec:analytic}. This means that we use the results of the second row for case~ii) in Tab.~\ref{tab:analyticalParameters}. 

Glancing at Fig.~\ref{fig:running_nohierarchy1} we see that the tendency and the direction of the running are correctly captured by the analytical approximations, however, the running is largely underestimated. This can be explained by two effects. First of all, as mentioned above, approximating the mass matrix as we have done cannot be very accurate. And secondly, since $\theta_{13}$ is growing faster than before as we lower the renormalisation scale, the assumption of a small $\theta_{13}$ breaks down more rapidly.

This time at $\mu=M_Z$ we numerically find the mixing angles, see Fig.~\ref{fig:angles_noHierarchy1}:
\begin{equation}
  \theta_{12} =  31.36^\circ,\qquad \theta_{23}= 49.29^\circ,\qquad \theta_{13} = 8.85^\circ,
\end{equation}
and mass square differences, see Fig.~\ref{fig:massDiff_noHierarchy1}:
\begin{equation}
  \Delta m_{21}^2 = 7.5 \times 10^{-5} \textrm{ eV}^2,\qquad \Delta m_{32}^2 = 4.6 \times 10^{-4} \textrm{ eV}^2.
\end{equation}
As before, only the solar $\Delta m_{21}^2$ agrees with the value reported by experiments. The good agreement between numerics and analytical results for the mass square differences is rather surprising, since neither the angles nor the masses themselves suggest such good agreement. Most likely this is simply a lucky choice of input parameters.

\begin{figure}[p]
  \subfigure[mixing angles]{\includegraphics[width=.5\textwidth]{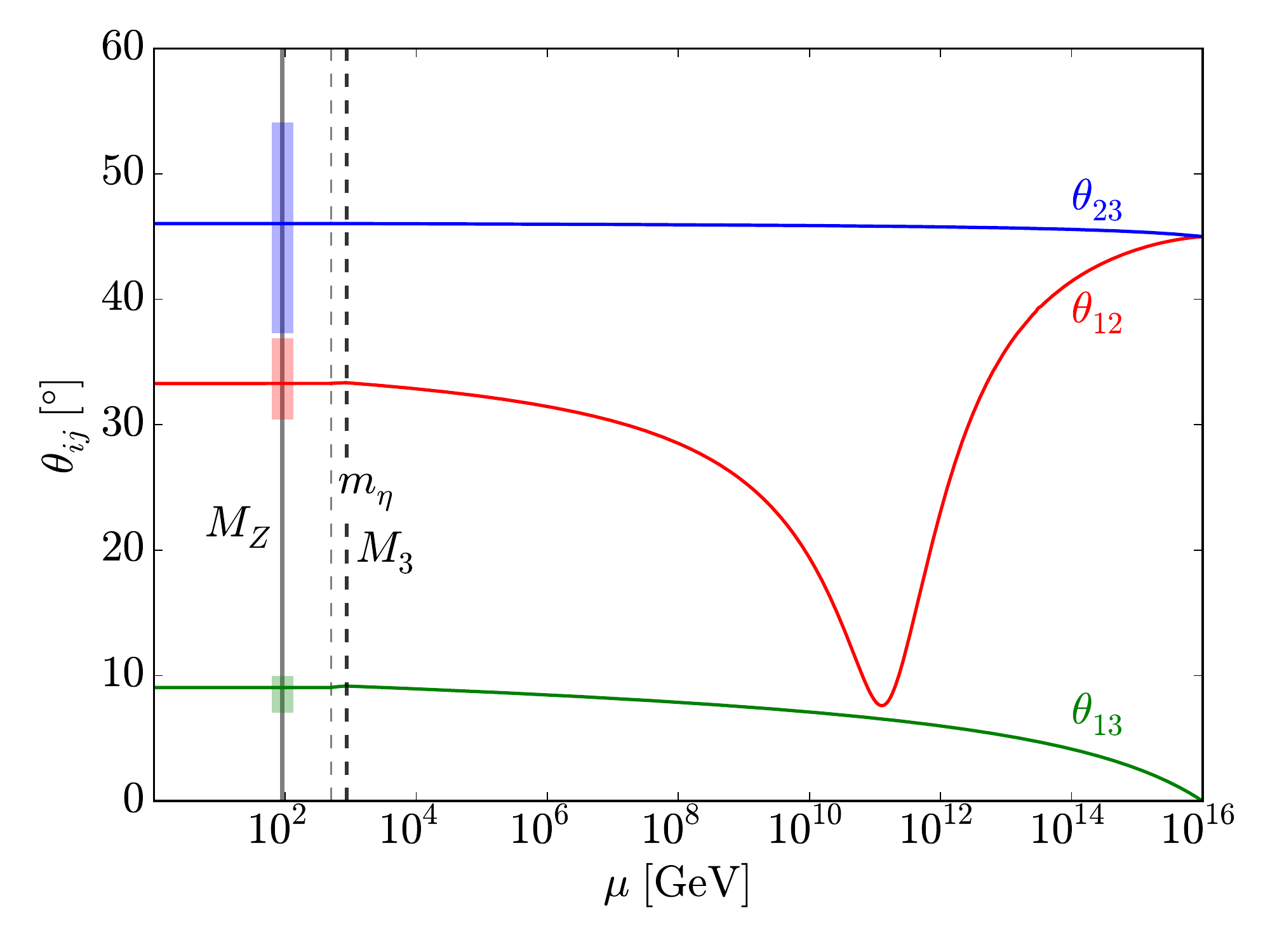}\label{fig:angles_noHierarchy2}}
  \subfigure[masses]{\includegraphics[width=.5\textwidth]{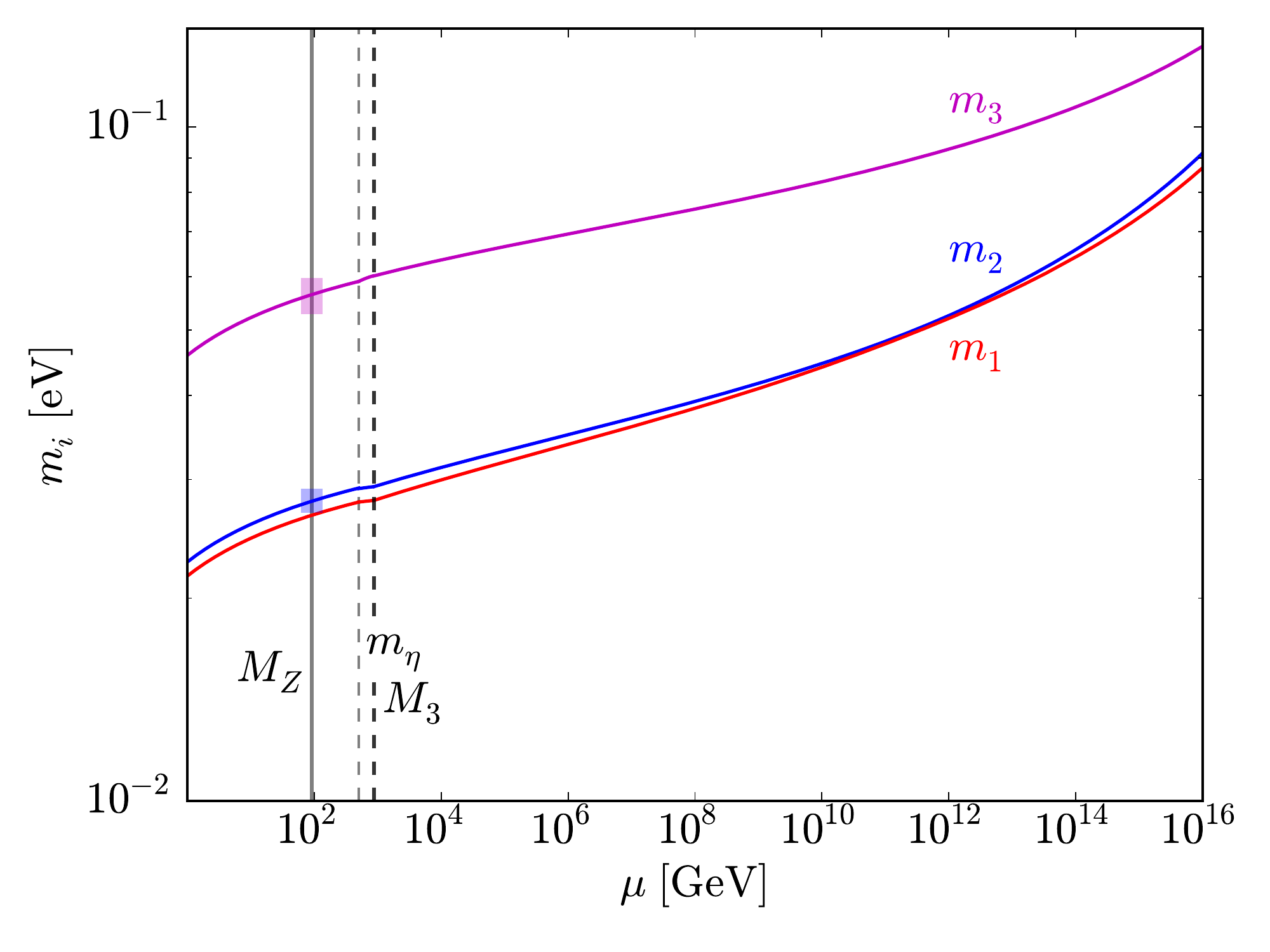}\label{fig:masses_noHierarchy2}}
  \subfigure[mass square differences (absolute)]{\includegraphics[width=.5\textwidth]{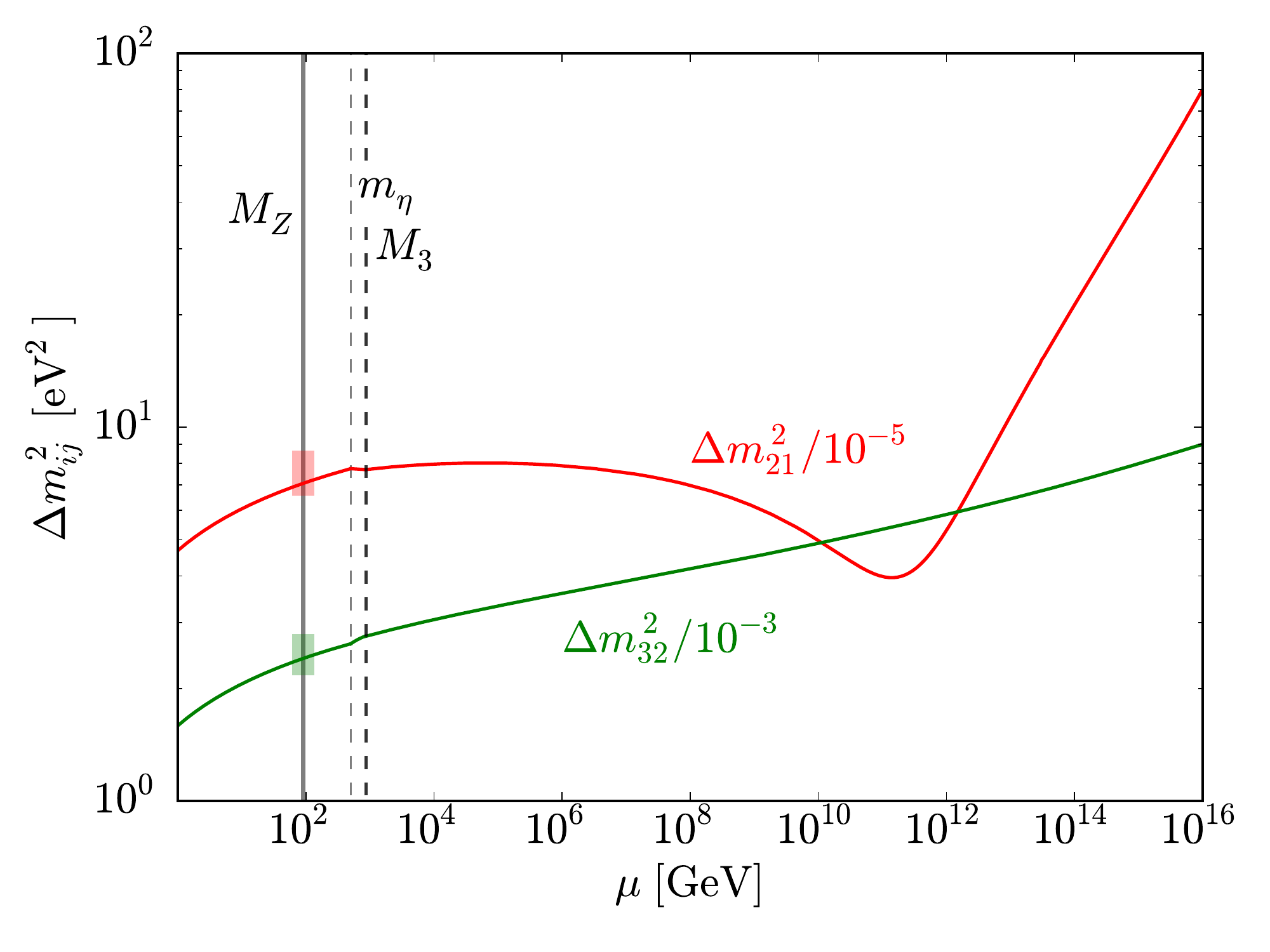}\label{fig:massDiff_noHierarchy3}}
  \subfigure[mass square differences (relative)]{\includegraphics[width=.5\textwidth]{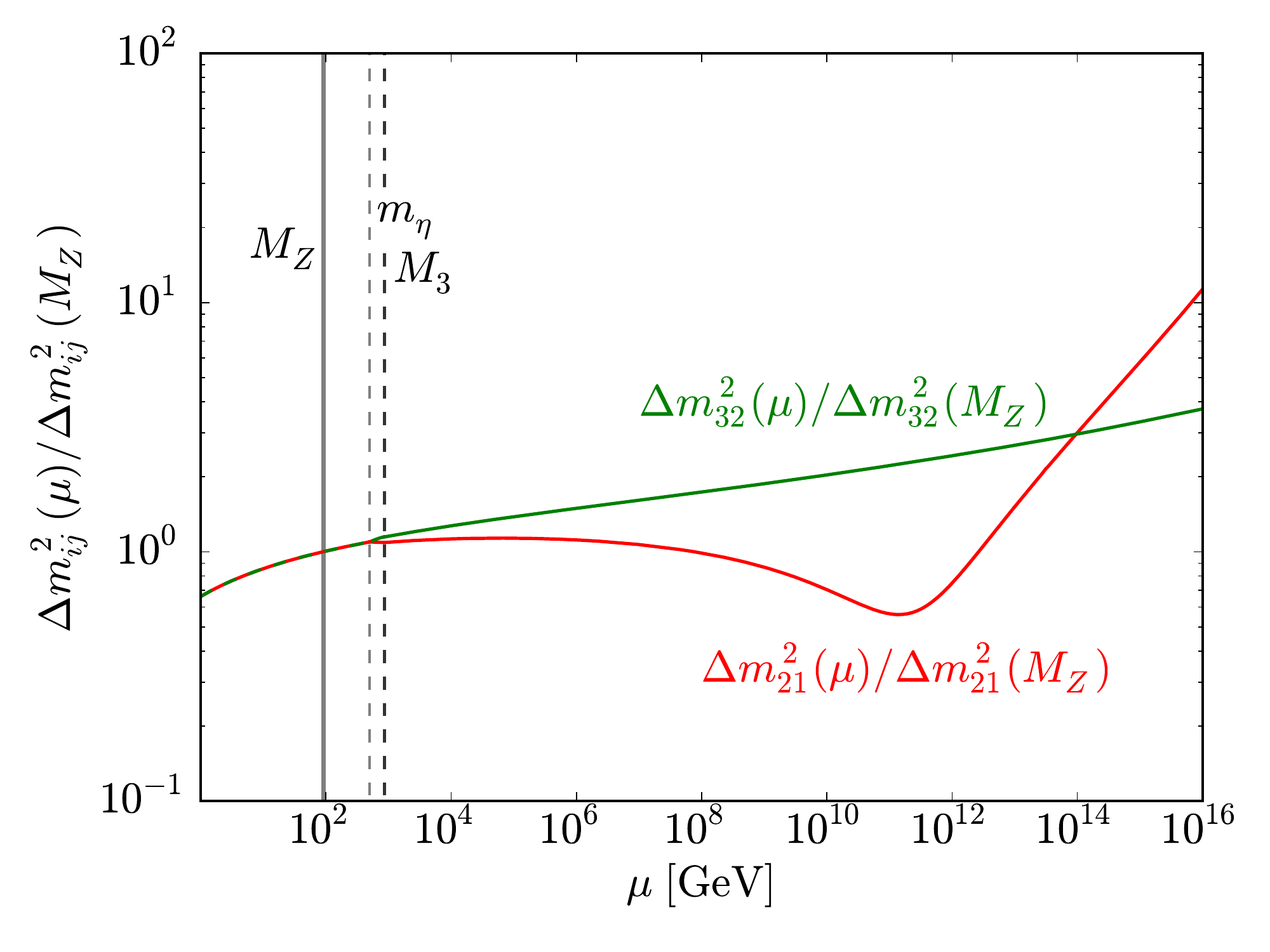}\label{fig:massDiff_noHierarchy4}}
  \caption{\label{fig:running_nohierarchy2}Running masses and mixing angles in case of no mass hierarchy among the new particles if the assumptions on $h^\dag h$ are relaxed.}
\end{figure}

Again, we have also found a less restrictive example for similar parameters, using the Casas-Ibarra parametrisation, see  Fig.~\ref{fig:running_nohierarchy2}, which reproduces all experimentally determined oscillation parameters. Here, we input $m_\eta=240\textrm{ GeV}$ and $( M_1, M_2, M_3 ) = (250, 320, 960) \textrm{ GeV}$ at $M_\textrm{GUT}$, as well as phases $\phi_1 \simeq \frac{\pi}{2}$, $\phi_2 \simeq \frac{5\pi}{8}$, and $\delta \simeq \frac{6\pi}{5}$. Two features are worth being highlighted at this point, one of which is that for the first time we have a RH neutrino threshold above the scalar threshold, which leads to observable but not overly strong threshold effects. Furthermore, two of the light neutrino masses almost meet at a scale $\mu \sim 10^{12} \textrm{ GeV}$ in Fig.~\ref{fig:masses_noHierarchy2}, which drives the running of $\theta_{12}$ to extremely small values, cf.\ Eq.~\eqref{eq:theta12Analytic}. Eventually, the difference between the masses grows again as does $\theta_{12}$. This is 
a nice example of how degeneracies (here generated by radiative effects, rather than by a choice of the input parameters) may strongly drive the running.

\subsection{Dominant RH neutrino masses}

\begin{figure}[p]
  \subfigure[mixing angles]{\includegraphics[width=.5\textwidth]{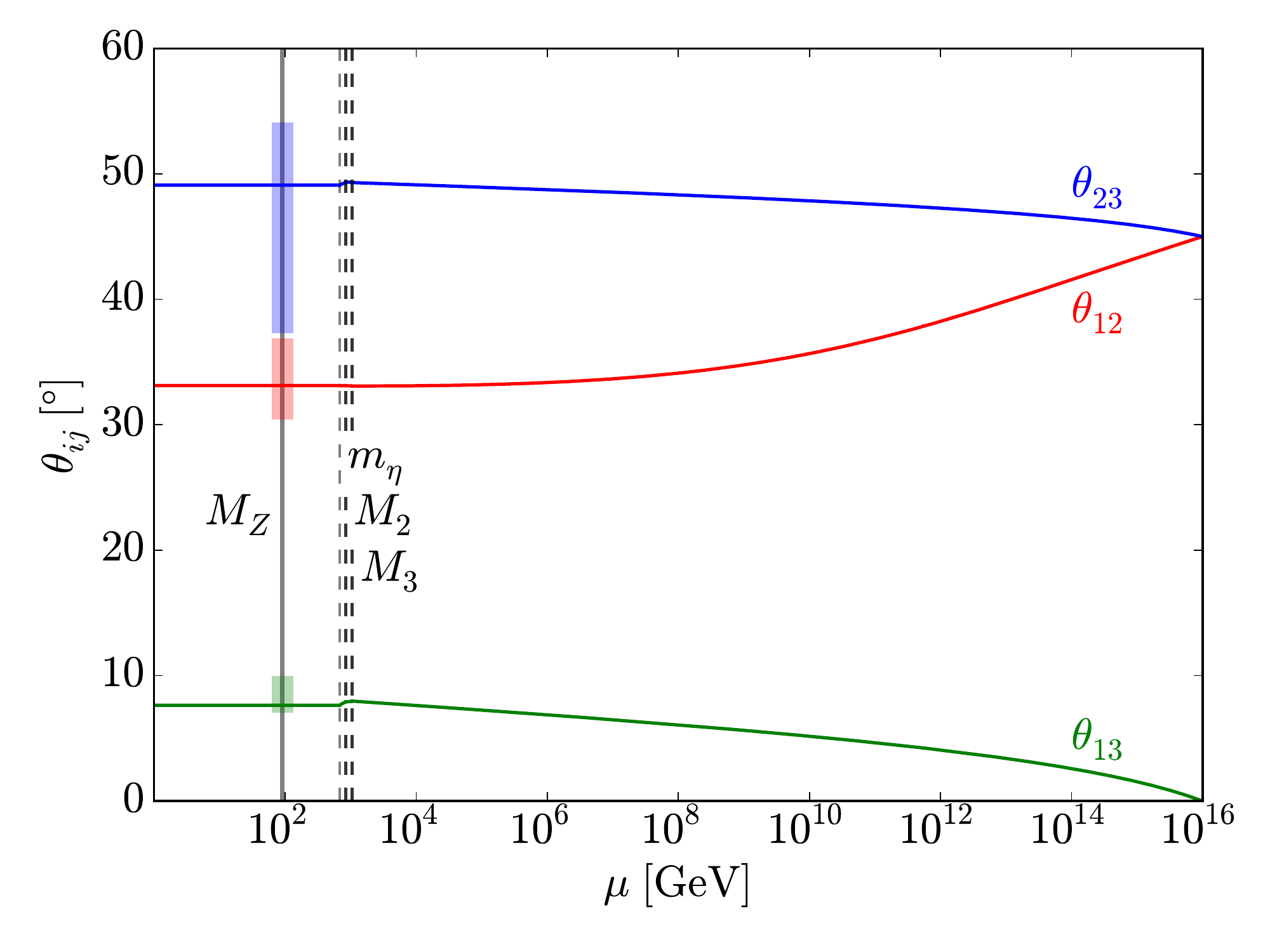}}
  \subfigure[masses]{\includegraphics[width=.5\textwidth]{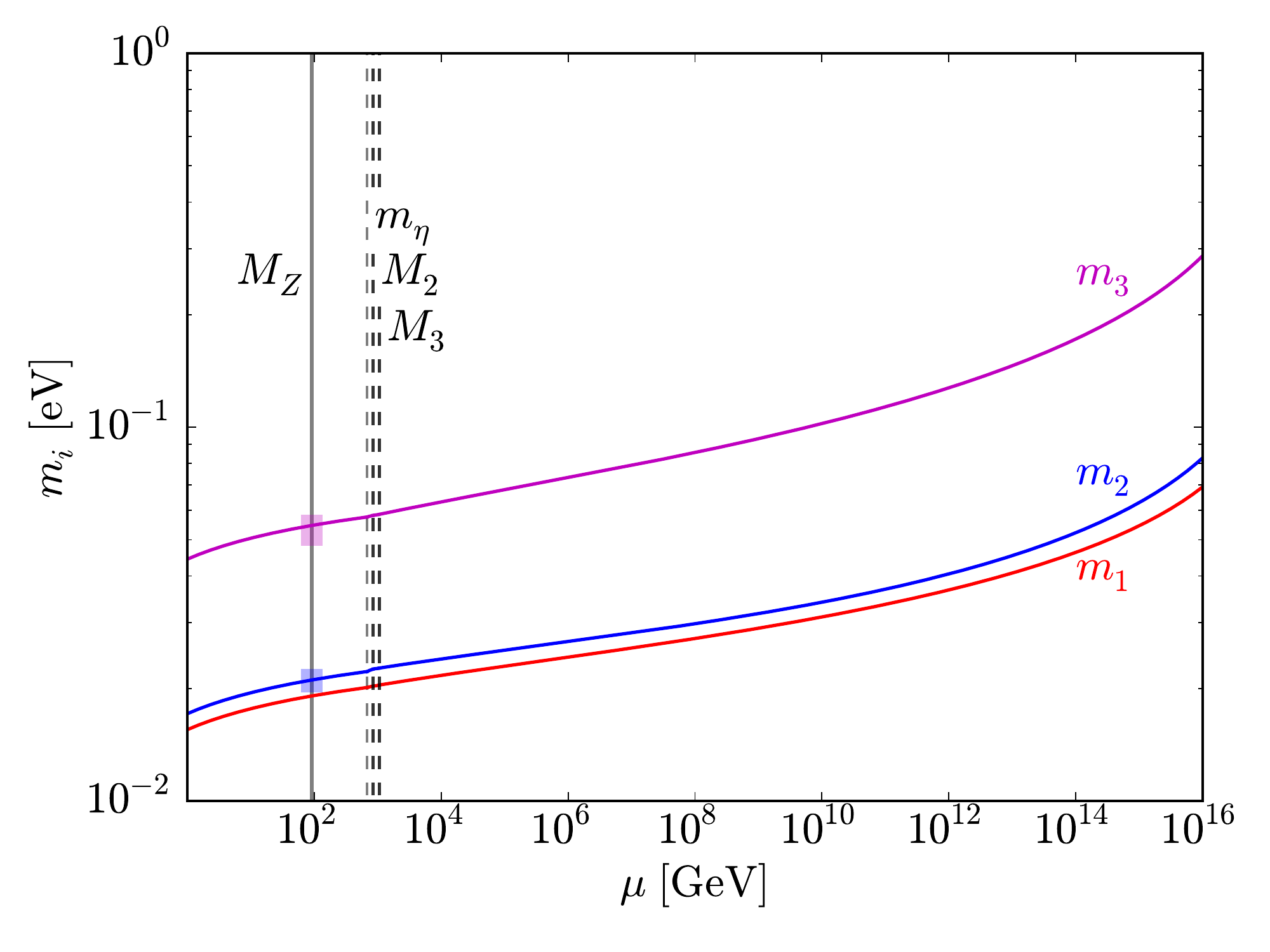} \label{fig:masses_RHDom}}
  \subfigure[mass square differences (absolute)]{\includegraphics[width=.5\textwidth]{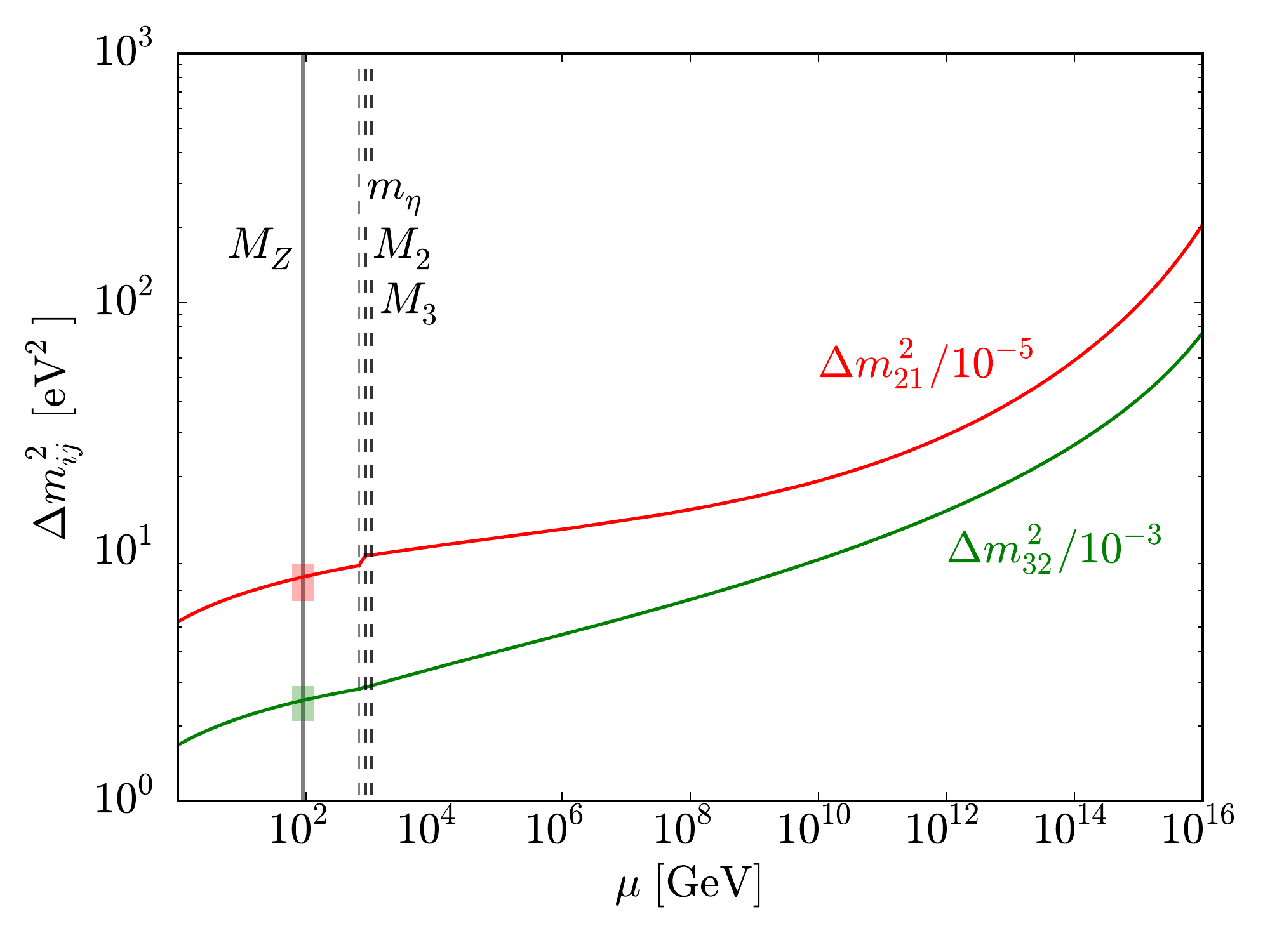} \label{fig:massDiff_RHDom1}}
  \subfigure[mass square differences (relative)]{\includegraphics[width=.5\textwidth]{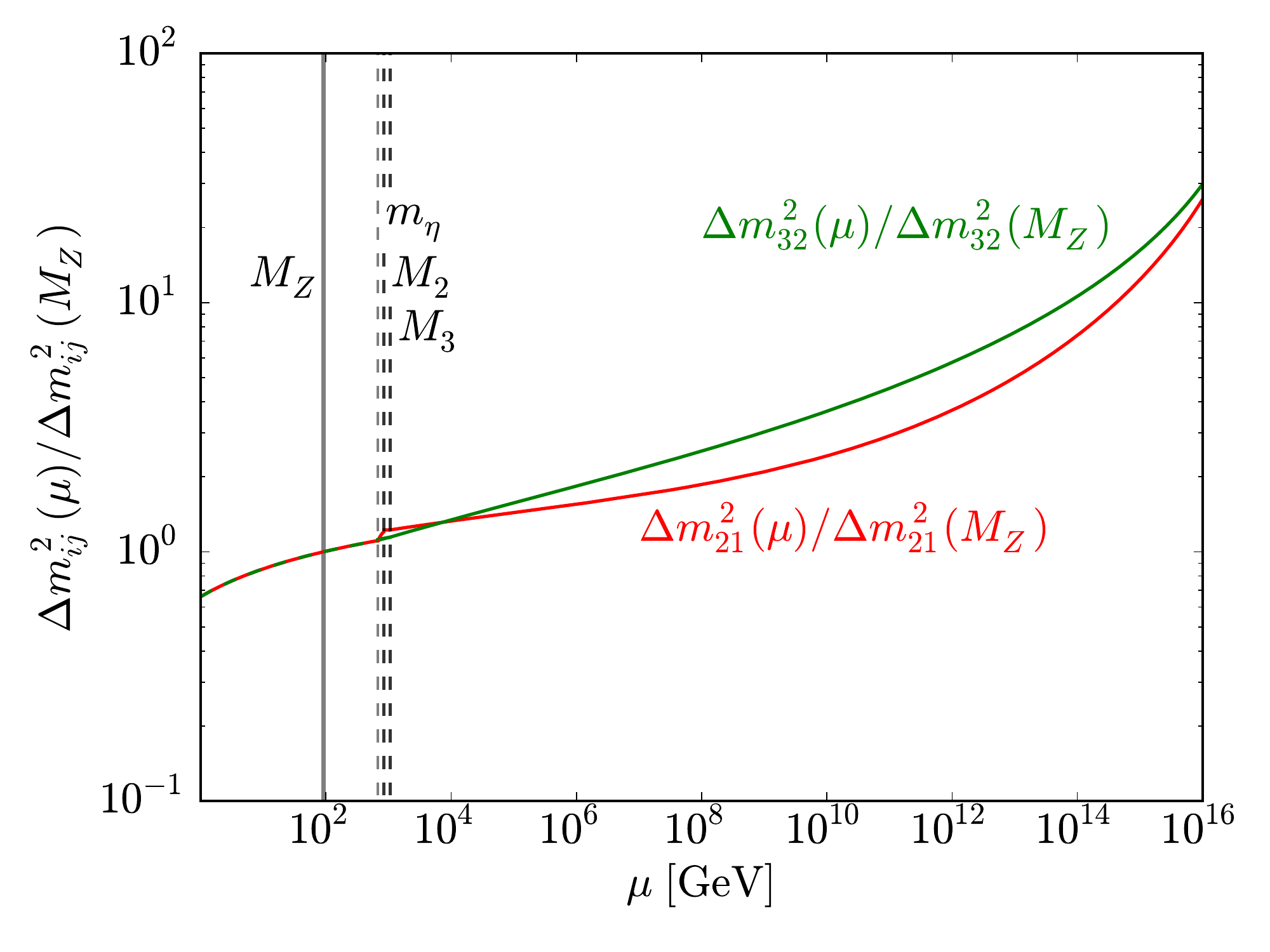} \label{fig:massDiff_RHDom2}}
  \caption{\label{fig:running_RHDom} Running of the mixing parameters for dominant RH masses. Note however that the scalar mass is ``attracted'' to the (TeV scale) RH masses via the last term in Eq.~\eqref{eq:m2RG}.}
\end{figure}

Finally, we discuss the case of large RH masses, however, we do not even attempt to compare this scenario to an analytical estimate since it is doomed to fail. This can be easily understood by recalling that we need to assume all quantities appearing inside the logarithm not to run. However, even though logarithmically suppressed, the running of $m_\eta^2$ cannot be ignored, since the last term in Eq.~\eqref{eq:m2RG} makes $m_\eta^2$ run over many orders of magnitude if $M_{1,2,3} \gg m_\eta$ and $h_{ij} \sim \mathcal{O}(1)$. Therefore, choosing a small $m_\eta$ at $M_\textrm{GUT}$ will result in an $m_\eta$ comparable to the $M_{1,2,3}$ at the electroweak scale -- and the approximation breaks down. A way out would be to choose $m_\eta^2<0$ at the high scale, but this may result in breaking the $\mathbb{Z}_2$ symmetry at high scales and hence the expression for the mass matrix being meaningless~\cite{Merle:2015gea}.

Fig.~\ref{fig:running_RHDom} shows an example for such a scenario. Here, we have chosen $m_\eta=240\textrm{ GeV}$, $M_1 \simeq 500 \textrm{ GeV}$, and $M_{2,3} \simeq 1 \textrm{ TeV}$. The phases are input as $\phi_1 \simeq \delta \simeq \frac{3\pi}{4}$ and $\phi_2 \simeq \frac{\pi}{20}$. The dashed grey lines indicate the thresholds as labelled in the plot. As expected, the scalar mass $m_\eta$ is very sensitive to the TeV scale RH masses, and at $\mu_* = m_\eta(\mu_*)$ it is of the same order (it even exceeds the smallest mass, which is why the threshold $M_1$ is not plotted). Note the strong running of the mass square differences in Figs.~\ref{fig:massDiff_RHDom1} and~\ref{fig:massDiff_RHDom2}. In spite of the aforementioned obstacles, the example shown reproduces the oscillation parameters correctly. It is also worth noting that in this case the threshold effects are quite significant, as discussed in Sec.~\ref{sec:EFT}.

\subsection{Inverted mass ordering}

\begin{figure}[p]
  \subfigure[mixing angles]{\includegraphics[width=.5\textwidth]{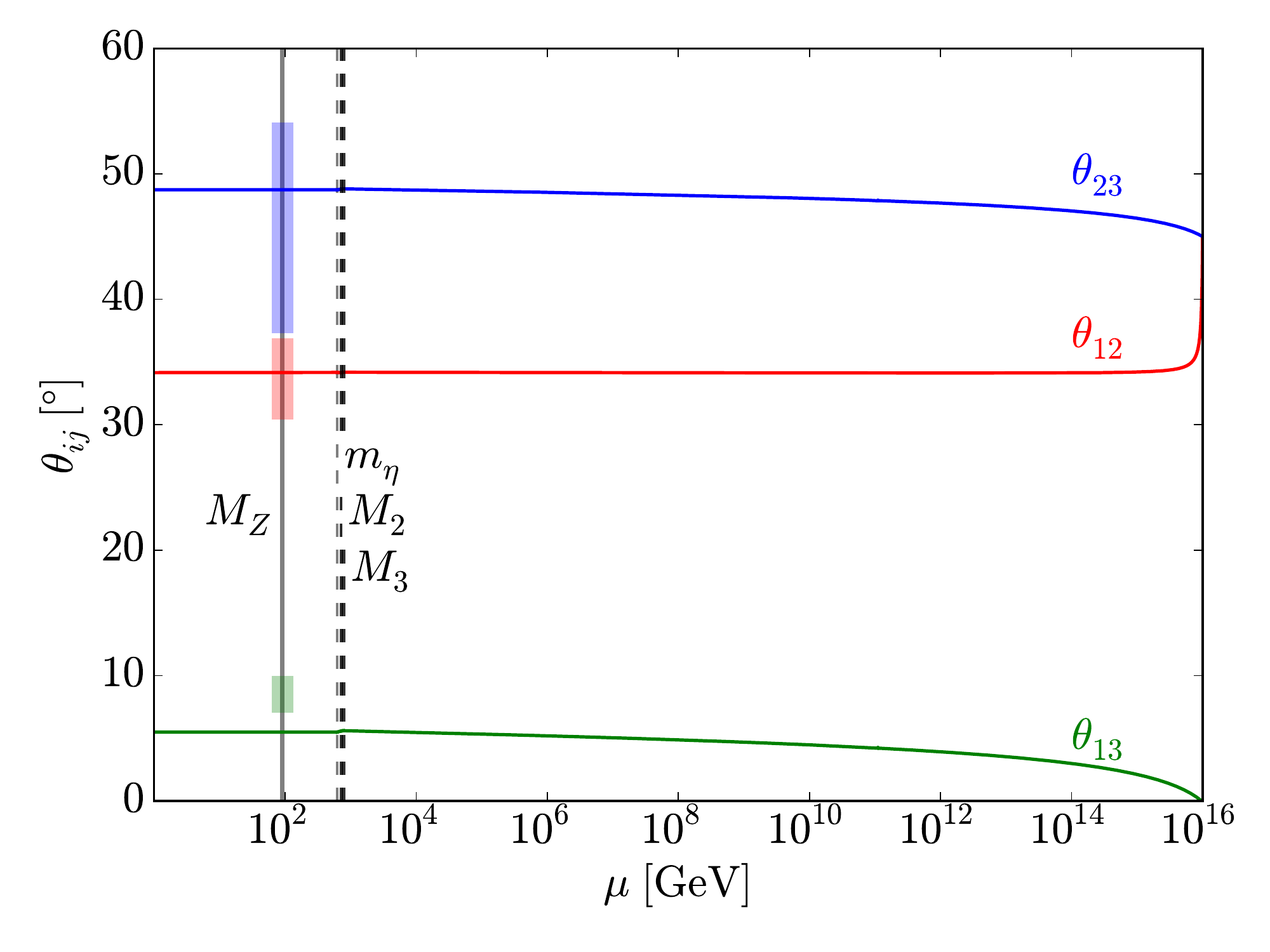} \label{fig:angles_IO}}
  \subfigure[masses]{\includegraphics[width=.5\textwidth]{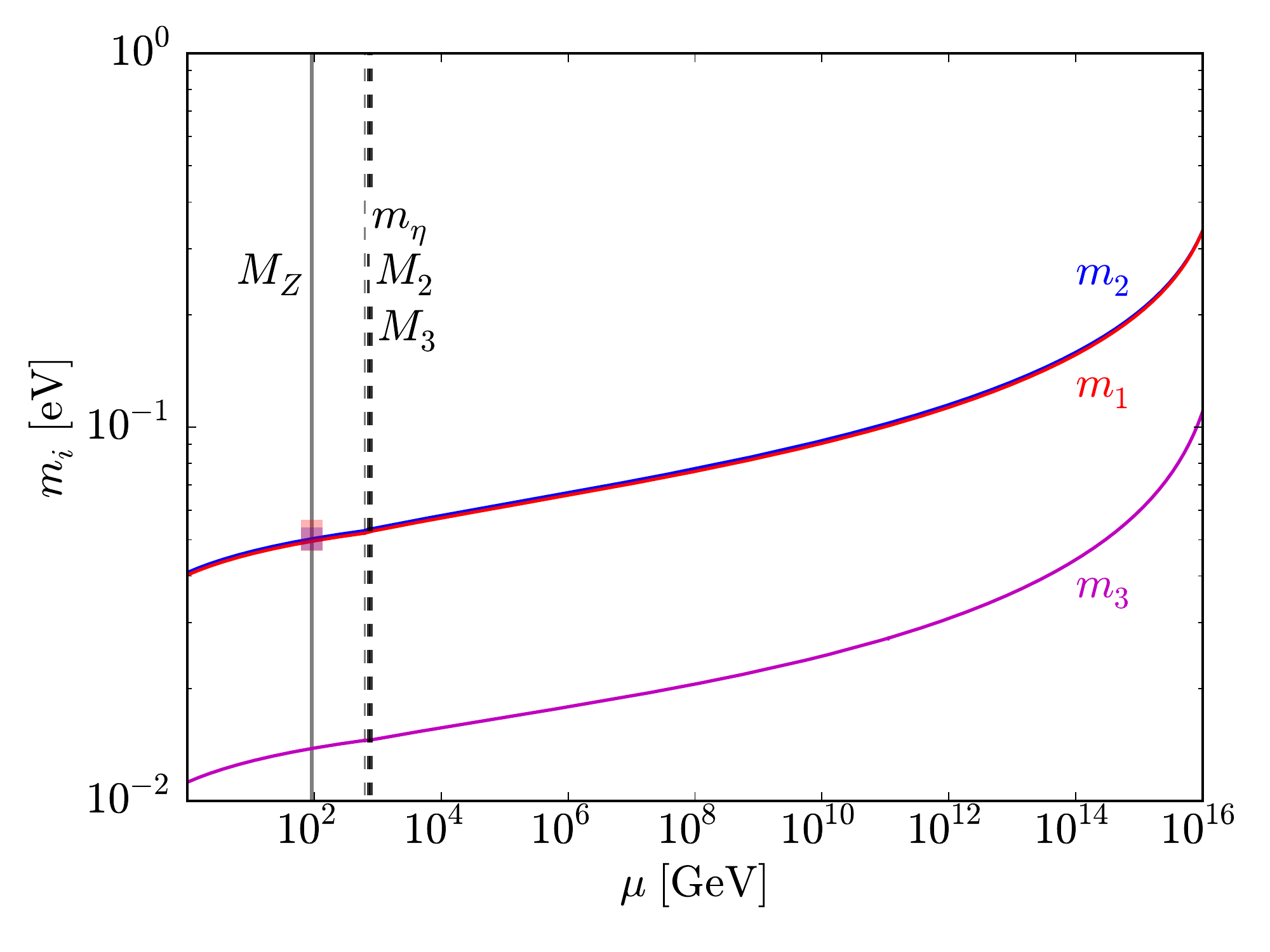} \label{fig:masses_IO}}
  \subfigure[mass square differences (absolute)]{\includegraphics[width=.5\textwidth]{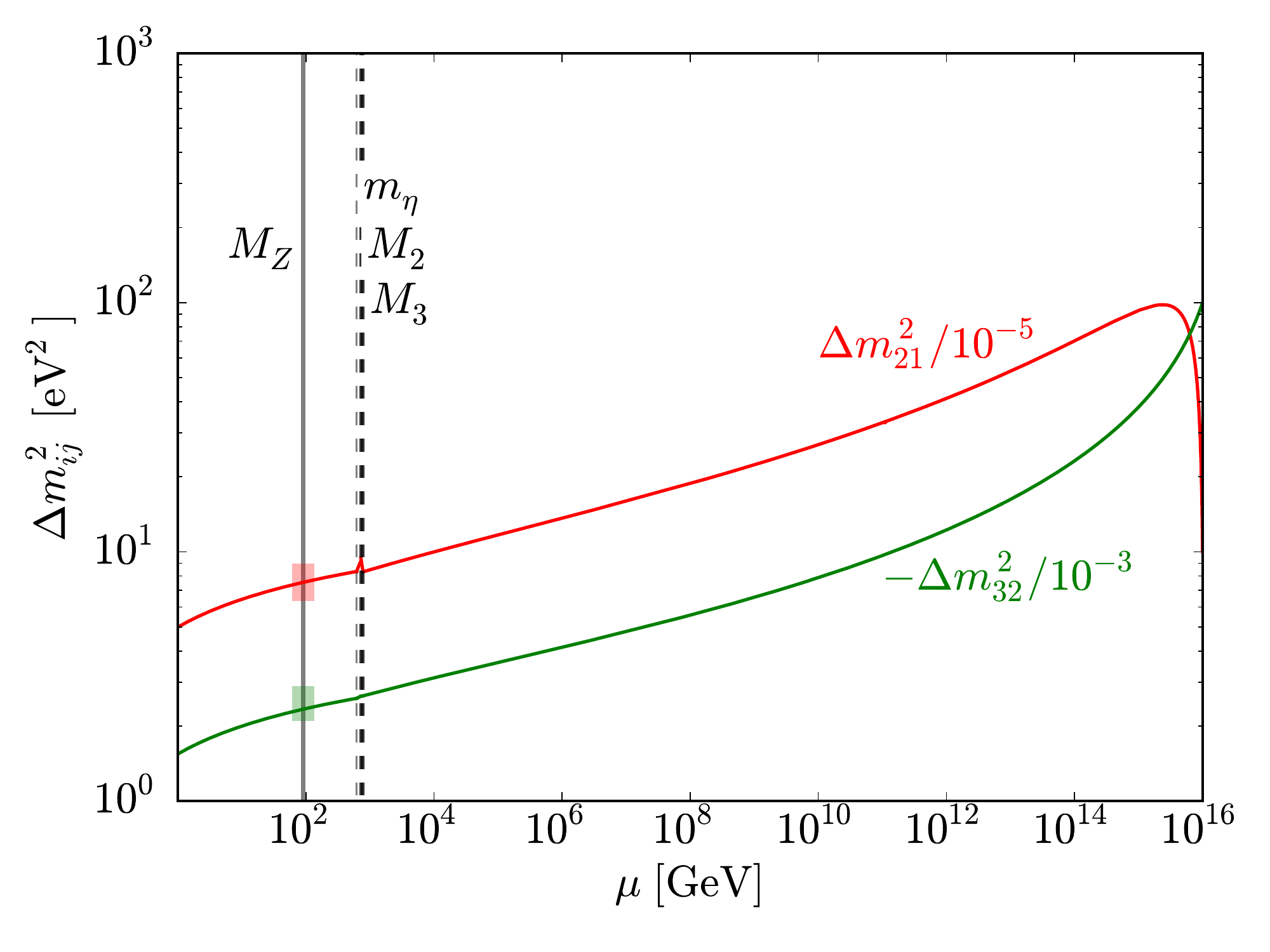} \label{fig:massDiff_IO1}}
  \subfigure[mass square differences (relative)]{\includegraphics[width=.5\textwidth]{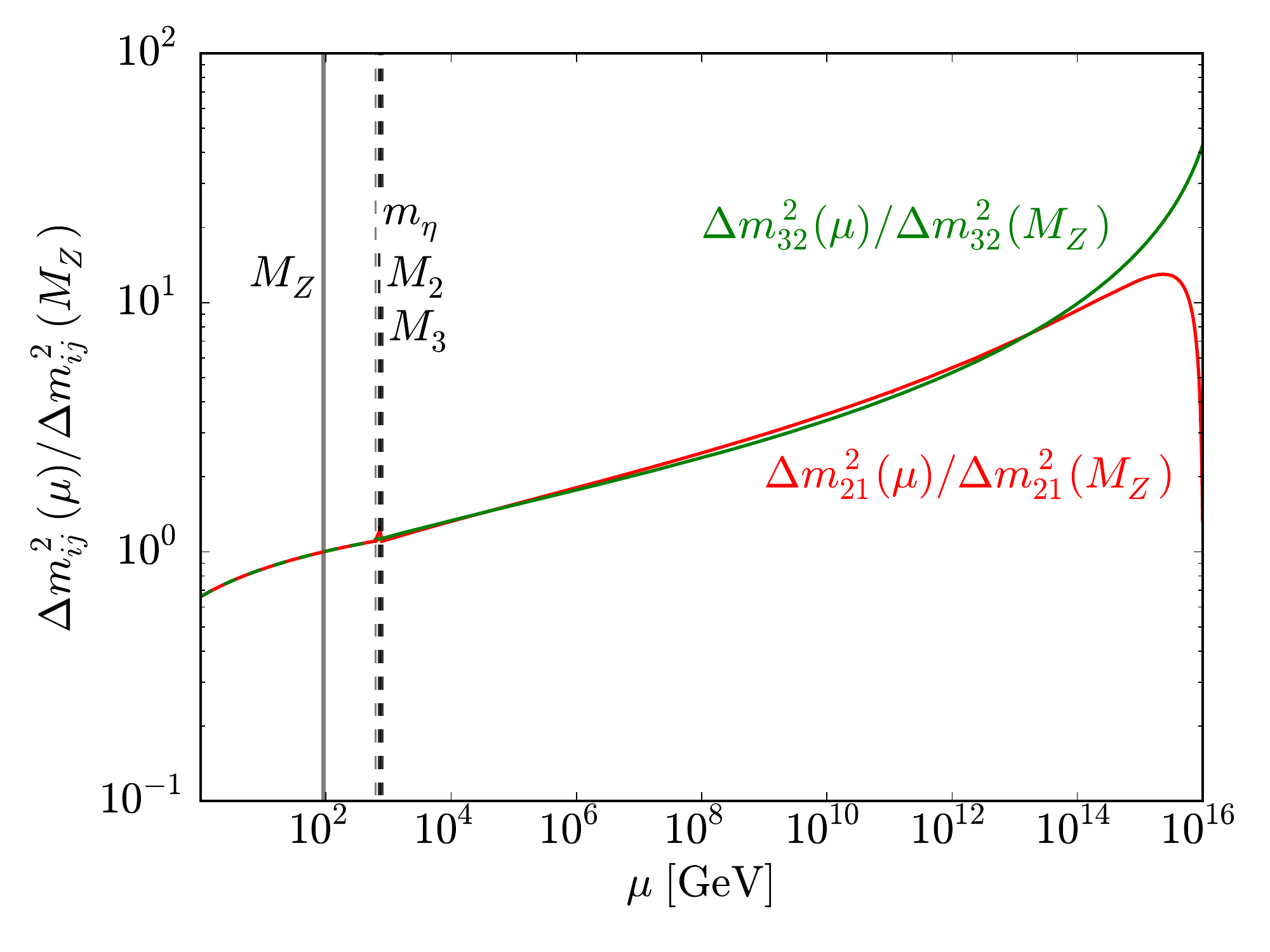} \label{fig:massDiff_IO2}}
  \caption{\label{fig:running_IO} Running of the mixing parameters for inverted mass ordering. Since $m_3$ is the lightest among the masses, the running of $\theta_{13}$ is not strong enough to reach the experimentally preferred region.}
\end{figure}

The previous analyses are of course equally applicable to the case of inverted mass ordering. It is, however, more difficult in our approach to find results in agreement with experiment because, since $m_3$ is the smallest mass, the running of $\theta_{13}$ is damped [cf.\ Eq.~\eqref{eq:theta12Analytic}]. Therefore, generating a non-zero $\theta_{13}$ exclusively via radiative corrections, as we have chosen to do so far, is very difficult. We show an example of this in Fig.~\ref{fig:running_IO}, where we have a similar setting as in the case of dominant RH neutrino masses. We input $m_\eta=120 \textrm{ GeV}$ and $\left( M_1,\, M_2,\, M_3 \right) =  \left( 450,\, 850,\, 900\right) \textrm{ GeV}$ at $M_\textrm{GUT}$,  and as before we find that the scalar mass is attracted to the scale of the RH neutrino masses at lower energy scales. All phases are zero at the input scale in order to achieve the largest possible value of $\theta_{13}$ (see the next subsection for a discussion of the effects the phases have on 
the running).

The qualitative features of Fig.~\ref{fig:running_IO} are easily understood. We start with a rather small $\Delta m_{21}^2 \sim 10^{-4} \textrm{ eV}^2$, which grows very fast and then slowly decreases to its value at $M_Z$ [Fig~\ref{fig:massDiff_IO1}]. This fast growth is driving the extreme running of $\theta_{12}$ in Fig.~\ref{fig:angles_IO} at scales just below $M_\textrm{GUT}$. Further lowering the scale $\mu$ has virtually no effect on $\theta_{12}$, because  $\Delta m_{21}^2$ is larger than before and the neutrino mass scale is monotonously decreasing, thus suppressing the running of all the mixing angles.

\subsection{An alternative approach}

\begin{figure}[t]
 \subfigure[$\phi_{1,2}=0$]{\includegraphics[width=.5\textwidth]{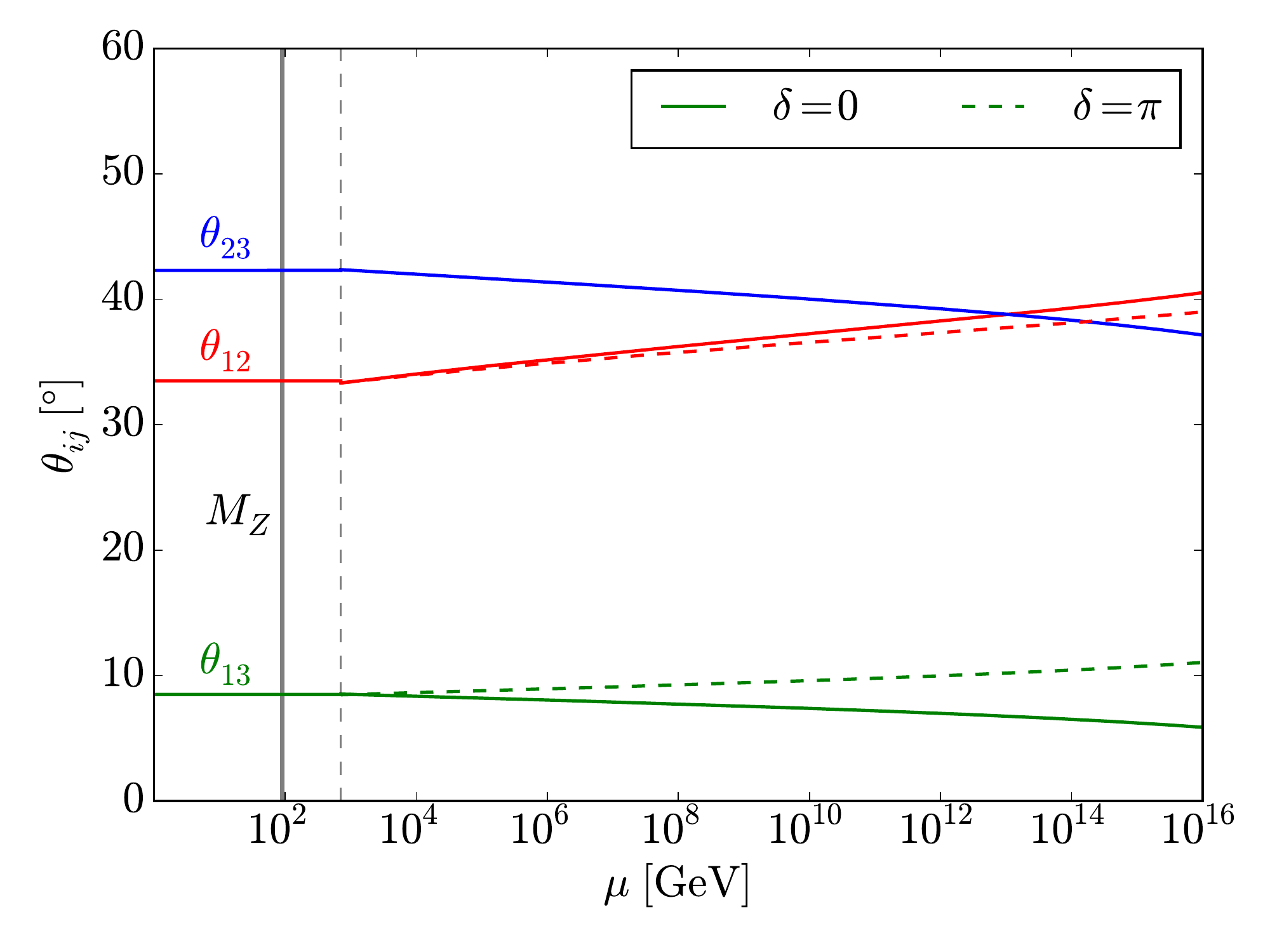}
\label{fig:BottomUpDIRAC}}
 \subfigure[$\phi_{1}=\delta=0$]{\includegraphics[width=.5\textwidth]{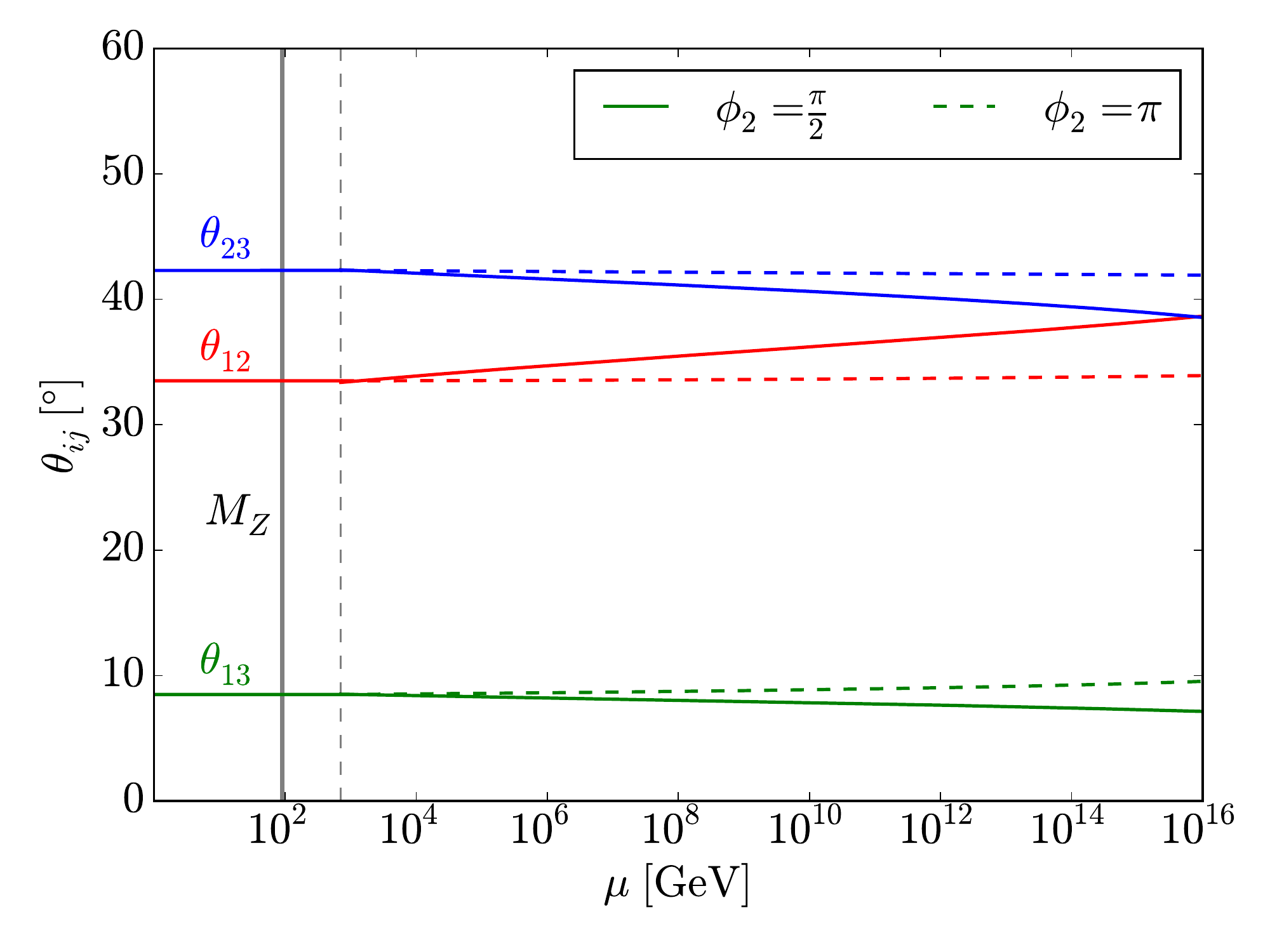}
\label{fig:BottomUpMAJORANA}}
 \caption{\label{fig:BottomUp} Running mixing angles in a bottom-up approach in normal ordering, where the values at the $Z$-boson mass have been fixed to their best-fit values~\cite{Gonzalez-Garcia:2014bfa}. The phases $\phi_1,\phi_2$, and $\delta$ are fixed at this scale, too, as indicated in the plots.}
\end{figure}

When studying the scotogenic model's parameter space, it may be more convenient to use a bottom-up approach as opposed to the so far employed top-down approach. The reason for this is simply that in principle \emph{all} model parameters should be considered as input at the high scale in the latter case.\footnote{In practice however it turns out that it is sufficient to treat the Higgs mass as a free input parameter at the high scale and run the remaining couplings from the low to the high scale and back. The changes due to this approach lie within the experimental uncertainties in almost all cases.} Obviously, this increases the size and dimensionality of the parameter space, which can be very costly in a phenomenological study. In contrast, a bottom-up approach allows one to fix all low-energy observables to their measured values. To ease the computation and remove some of the ambiguities of such an approach, we only consider the case where the three RH neutrinos and the inert scalars are almost of the same 
mass, slightly below $1\textrm{ TeV}$ (vertical dashed line in Fig.~\ref{fig:BottomUp}). A convenient choice for the scalar couplings is:
\begin{equation}
  \left.\left( \lambda_2,\ \lambda_3,\ \lambda_4,\ \lambda_5\right)\right|_{\mu=M_Z} = \left( 0.1,\
0.1,\ 0.2,\ 10^{-9} \right),
\end{equation}
which guarantees perturbativity, an intact $\mathbb{Z}_2$ symmetry, and the (meta-)stability of the scalar potential up to at least $10^{16} \textrm{ GeV}$. 

In Fig.~\ref{fig:BottomUp} the results of such an approach are shown. Since we expect no qualitatively new behaviour, we would like to use this to illustrate the role of the CP phases $\phi_1,\phi_2$, and $\delta$. From Eqs.~\eqref{eq:MixingAnglesAnalytic} we learn that a non-zero $\delta$ can only influence $\theta_{13}$ directly, which is confirmed by Fig.~\ref{fig:BottomUpDIRAC}: while $\theta_{23}$ is largely independent of the choice of $\delta$, $\theta_{12}$ is modified only indirectly via changes in $\theta_{13}$, which enter Eq.~\eqref{eq:theta12Analytic} through higher order corrections. 

Non-zero Majorana phases $\phi_{1,2}$ can cause a suppression of the running, as we can see from Eqs.~\eqref{eq:theta12Analytic} and~\eqref{eq:theta23Analytic}. For example, the running of $\theta_{12}$ is suppressed compared to the case of vanishing or equal Majorana phases if we have $|\phi_1-\phi_2|=\pi$, because in that case the decisive factor becomes:
\begin{equation}
  \frac{\left|m_1e^{i\phi_1}+m_2e^{i\phi_2}\right|^2}{\Delta m_{21}^2} \quad \rightarrow \quad
\frac{\left|m_1-m_2\right|^2}{\Delta m_{21}^2} < \frac{\left|m_1+m_2\right|^2}{\Delta m_{21}^2}.
\end{equation}
As an example, we direct the reader's attention to Fig.~\ref{fig:BottomUpMAJORANA}, which illustrates this for the case where only $\phi_2$ can take non-zero values at the low scale $\mu=M_Z$. As we vary $\phi_2$ from $0$ to $\pi$, we achieve a less pronounced running for all mixing angles and thereby (accidentally) achieve a unification of the mixing angles $\theta_{12}$ and $\theta_{23}$ at the GUT scale, for $\phi_2=\frac{\pi}{2}$. For $\phi_2=\pi$ we see that the running is essentially turned off due to the aforementioned suppression mechanism.

\section{Conclusions} \label{sec:Summary}

We have studied the scotogenic model, a particularly simple setting that generates active neutrino masses at 1-loop level and exhibits several Dark Matter candidates. In doing so, we have updated several previous results, namely the formula for the light neutrino mass matrix, the matching conditions at the particle thresholds, and the 1-loop RGEs of the model. We have derived analytical equations that allowed us to identify regions in parameter space which we suspected to show significant running behaviour. This was verified by comparing the analytical estimates to a numerical treatment. 

The point of this paper is not so much to perform a very detailed phenomenological study, but we rather aimed at an illustration of the effects the renormalisation group running can have on the scotogenic model, and how to understand them using approximate solutions of the evolution equations. Following this approach, we have intrinsically disregarded some points which could be very crucial in a realistic analysis. These include motivating the mass patterns used, explaining the leptonic mixing pattern by a concrete symmetry, or presenting an explicit justification for the scales used in the various examples. We nevertheless consider our approach to be valuable, because it will enable us --~and hopefully any reader~-- to make use of our general results in a concrete study.

The key results obtained in this work, apart from clearly illustrating how strong running effects can be in the scotogenic model, are the updated RGEs, their analytical approximate solutions (none of which have been known previously), and in particular the explicit discussion of all qualitatively different scenarios that can appear. Our work lays the foundation for a detailed study aiming at a determination of the full allowed parameter space in the scotogenic model. Given that such an endeavour can only be done in a purely numerical manner, having a picture of all running effects that could possibly appear as well as some limiting cases, which can be viewed as ``cornerstones'' for a numerical computation, will be highly useful.

The scotogenic model is the prime example for a setting with a radiative neutrino mass and it has been studied from all sides, from low-energy measurements over neutrino physics and Dark Matter constraints to collider phenomenology. However, there exists up to now no comprehensive work, which really tries to put together all constraints and work out which regions in the parameter space survive. A few attempts exists which e.g.\ confront collider data with Dark Matter signatures, but renormalisation group running is completely disregarded in such cases. This study will hopefully contribute to close all these gaps by our equations providing the basic tools for a fully comprehensive phenomenological analysis of the scotogenic model which can take into account the effects of having data available at several different energy scales.

\section*{Acknowledgements}

AM acknowledges partial support from the European Union FP7 ITN-INVISIBLES (Marie Curie Actions, PITN-GA-2011-289442). 

\appendix

\section{\label{app:Appendix_RGE}Renormalisation group equations}
\renewcommand{\theequation}{A-\arabic{equation}}
\setcounter{equation}{0}  

The 1-loop RGEs for the scotogenic model have first been computed in Ref.~\cite{Bouchand:2012}. We have re-derived those equations needed for the purpose of this paper, and have in passing taken the opportunity to update part of the earlier results.

For convenience, we define the differential operator $\mathcal{D} \equiv (4\pi)^2 \mu \frac{\textrm{d}}{\textrm{d}\mu}$. The 1-loop RGEs for the gauge couplings are those of a generic two Higgs doublet model (THDM)~\cite{Grzadkowski:1987wr}:
\begin{equation}
\mathcal{D} g_i = b_i g_i^3 \textnormal{ (no sum!)},
\label{eq:gauge-RGE}
\end{equation}
with $b = \left( 7,-3,-7 \right)$.

The quark sector of the scotogenic model is the same as that of the SM, such that the corresponding RGEs do not change.\footnote{Note the implicit changes in $g_{1,2}$, though, by virtue of Eq.~\eqref{eq:gauge-RGE}.} The RGEs for the leptonic Yukawa couplings are:
\begin{subequations}
\begin{align}
 \mathcal{D} Y_e &= Y_e \left\lbrace \frac{3}{2}Y_e^\dagger Y_e +\frac{1}{2} h^\dagger h + T - \frac{15}{4} g_1^2 - \frac{9}{4} g_2^2 \right\rbrace, \label{eq:leptonRG}\\
 \mathcal{D} h &= h \left\lbrace \frac{3}{2} h^\dagger h + \frac{1}{2} Y_e^\dagger Y_e + T_\nu - \frac{3}{4} g_1^2 - \frac{9}{4} g_2^2 \right\rbrace,
 \label{eq:neutrinoRG}
\end{align}
\end{subequations}
where $T_\nu \equiv \textrm{Tr}\left(h^\dagger h \right)$ and $T \equiv \textrm{Tr}\left(Y_e^\dagger Y_e + 3 Y_u^\dagger Y_u + 3 Y_d^\dagger Y_d\right)$. For the Majorana mass matrix, one finds~\cite{Antusch:2002rr,Bouchand:2012}:
\begin{equation}
 \mathcal{D} M = \left\lbrace \left(h\, h^\dagger\right) M + M \left(h\, h^\dagger \right)^* \right\rbrace.
 \label{eq:RHnuRG}
\end{equation}
For the quartic scalar couplings, we find the RGEs for a $\mathbb{Z}_2$ symmetric THDM~\cite{Hill:1985}:
\begingroup
\setlength{\jot}{8pt}
\begin{subequations}
\begin{align} 
 \begin{split} \label{eq:RGlambda1}
     \mathcal{D} \lambda_1 &= 12 \lambda_1^2 + 4 \lambda_3^2 + 4 \lambda_3 \lambda_4 + 2 \lambda_4^2 + 2 \lambda_5^2 
	+ \frac{3}{4} \left(g_1^4 + 2 g_1^2 g_2^2 + 3 g_2^4\right) \\
	&\qquad - 3 \lambda_1 \left(g_1^2 + 3g_2^2\right) + 4 \lambda_1 T - 4 T_4,
 \end{split}\\
 \begin{split} \label{eq:RGlambda2}
  \mathcal{D} \lambda_2 &= 12 \lambda_2^2 +4 \lambda_3^2 + 4 \lambda_3 \lambda_4 + 2 \lambda_4^2 + 2 \lambda_5^2 
	+ \frac{3}{4} \left(g_1^4 + 2 g_1^2 g_2^2 + 3 g_2^4\right) \\
	&\qquad - 3 \lambda_2 \left(g_1^2 + 3g_2^2\right) + 4 \lambda_2 T_\nu - 4 T_{4\nu},
 \end{split} \\
 \begin{split} \label{eq:RGlambda3}
  \mathcal{D} \lambda_3 &= 2 \left(\lambda_1 + \lambda_2 \right) \left( 3\lambda_3 + \lambda_4\right) + 4\lambda_3^2 + 2 \lambda_4^2 + 2 \lambda_5^2 
	+ \frac{3}{4} \left(g_1^4 - 2 g_1^2 g_2^2 + 3 g_2^4\right) \\
	&\qquad - 3 \lambda_3 \left(g_1^2 + 3g_2^2\right) + 2 \lambda_3 \left(T +T_\nu\right) - 4 T_{\nu e},
 \end{split} \\
 \begin{split} \label{eq:RGlambda4}
  \mathcal{D} \lambda_4 &= 2 \left(\lambda_1 + \lambda_2 \right) \lambda_4 + 8 \lambda_3 \lambda_4 + 4 \lambda_4^2 + 8 \lambda_5^2 + 3 g_1^2 g_2^2 \\
	&\qquad - 3 \lambda_4 \left(g_1^2 + 3g_2^2\right) + 2 \lambda_4 \left(T +T_\nu\right) + 4 T_{\nu e},
 \end{split} \\
  \mathcal{D}\lambda_5 &= \lambda_5 [ 2\left(\lambda_1 + \lambda_2\right) + 8\lambda_3 +12\lambda_4 
	- 3 \left(g_1^2 + 3g_2^2\right) + 2 \left(T + T_\nu\right)],  \label{eq:RGlambda5}
\end{align}
\end{subequations}
\endgroup
where we have used the abbreviations $T_{4}\equiv \mathrm{Tr} \left( Y_e^\dag Y_e Y_e^\dag Y_e + 3 Y_u^\dag Y_u Y_u^\dag Y_u + 3 Y_d^\dag Y_d Y_d^\dag Y_d \right)$, $T_{4\nu}\equiv \mathrm{Tr} \left( h^\dag h \, h^\dag h \right)$ and $T_{\nu e}\equiv \mathrm{Tr} \left( h^\dag h \,Y_e^\dag Y_e \right)$.

\noindent
The scalar mass parameters obey the following RGEs:
\begin{subequations}
\begin{align} \label{eq:m1RG}
  \mathcal{D}m_H^2 &= 6 \lambda_1 m_H^2 +2\left(2\lambda_3 + \lambda_4\right)m_\eta^2 + m_H^2\left[ 2T - \frac{3}{2} \left(g_1^2 + 3g_2^2\right)\right],\\
  \label{eq:m2RG}
    \mathcal{D}m_\eta^2 &= 6 \lambda_2 m_\eta^2 +2\left(2\lambda_3 + \lambda_4\right)m_H^2  
	+ m_\eta^2\left[ 2T_\nu - \frac{3}{2} \left(g_1^2 + 3g_2^2\right)\right] - 4 \sum_{i=1}^3 M_i^2\left(h \, h^\dagger\right)_{ii},
\end{align}
\end{subequations}
where the last term in Eq.~\eqref{eq:m2RG} is nothing but a trace and thereby invariant under the transformation that diagonalises $M$, such that we do not have to perform this diagonalisation explicitly. 

Finally, we obtain for the effective operators:
\begin{align}
  \begin{split}
    \mathcal{D} \eff{n}{1}{} &= \frac{1}{2}\left( \overset{(n)}{h}^\dag \overset{(n)}{h} - 3 Y_e^\dag Y_e \right)^* \eff{n}{1}{} + \frac{1}{2}\eff{n}{1}{} \left( \overset{(n)}{h}^\dag \overset{(n)}{h} - 3 Y_e^\dag Y_e \right) +\\
    & \qquad + \left( 2 T +2 \lambda_1 -3 g_2^2 \right) \eff{n}{1}{}+2\lambda_5 \eff{n}{2}{}, 
  \end{split}\\
  \begin{split}
    \mathcal{D} \eff{n}{2}{}&= \frac{1}{2}\left( \overset{(n)}{h}^\dag \overset{(n)}{h} + Y_e^\dag Y_e \right)^* \eff{n}{2}{} + \frac{1}{2}\eff{n}{2}{} \left( \overset{(n)}{h}^\dag \overset{(n)}{h} + Y_e^\dag Y_e \right) +\\
    &\qquad + \left( 2 T_\nu +2 \lambda_2 -3 g_2^2 \right) \eff{n}{2}{}+2\lambda_5 \eff{n}{1}{}.
    \end{split}
\end{align}

\section{Derivation of the Analytical Formulae}\label{app:AnalyticDerivation}

\setcounter{equation}{0}  
\renewcommand{\theequation}{B-\arabic{equation}}

In this appendix, we review the derivation of analytical equations for the running mixing angles and masses. As mentioned in Sec.~\ref{sec:analytic}, this is analogous to previous results found e.g.\ in~\cite{Antusch:2005gp, Mei:2005qp,Chao:2006ye,Schmidt:2007nq,Chakrabortty:2008zh,Bergstrom:2010qb,Casas:1999tg, Chankowski:1999xc, Antusch:2003kp}.

To arrive at an analytical expression, we define $t\equiv\log\left(\frac{\mu}{\mu_0}\right)$ (in the following a prime denotes the derivative w.r.t.~$t$). The light neutrino mass matrix RGE is most generally given by:
\begin{equation} \label{eq:MassMatrixRGE}
  (4\pi)^2 \mathcal{M}_\nu^\prime = P^T \mathcal{M}_\nu + \mathcal{M}_\nu P + C \mathcal{M}_\nu\,,
\end{equation}
with $C$ a factor that has no flavour structure, and $P$ a Hermitian matrix.

Since we generate Majorana neutrinos, the light neutrino mass matrix must be symmetric and can therefore be diagonalised by a unitary matrix $U$ according to:
\begin{equation} \label{eq:TakagiDecomp}
  U^T \mathcal{M}_\nu U = D_\nu \equiv \textrm{diag}(m_1,m_2,m_3),
\end{equation}
where we may right-multiply $U$ with a diagonal matrix of phases to render the $m_i$ real and positive.

Thus, we have:
\begin{equation}
  D_\nu^\prime = {U^\prime}^T \mathcal{M}_\nu U + U^T\mathcal{M}_\nu^\prime U + U^T \mathcal{M}_\nu U^\prime.
\end{equation}
We can always define an anti-Hermitian matrix $T$ such that $U^\prime=UT$, and  the remaining task is to determine the components of $T$ in terms of known quantities from the equation:
\begin{equation} \label{eq:ConstructT}
  D_\nu^\prime = D_\nu T - T^* D_\nu + \frac{1}{(4\pi)^2} \left[ C D_\nu + \widetilde{P}^TD_\nu + D_\nu \widetilde{P} \right].
\end{equation}
In this last expression, we have defined $\widetilde{P} \equiv U^\dag P U$.

Since we can choose the Majorana phases in $U$ such that $m_i \in \mathbb{R}^+$ for all $t$, we obtain the mass RGEs from the real parts of the diagonal entries in~\eqref{eq:ConstructT} (no sum over $i$):

\begin{equation} \label{eq:activeMassRGE}
  m_i^\prime = \frac{m_i}{(4\pi)^2} \left[ \Re(C) + 2 \Re(\widetilde{P}_{ii}) \right],
\end{equation}
and since the imaginary part of the diagonals must vanish:
\begin{equation}
  2 \Im (T_{ii}) = -\frac{1}{(4\pi)^2}\left[ \Im(C) + 2 \Im(\widetilde{P}_{ii})\right].
\end{equation}
The real part of $T_{ii}$ vanishes due to the anti-Hermiticity condition.

The off-diagonal elements of $T$ can be constructed from (again, no sum over $i$ or $j$):
\begin{equation}
  0 = D_{\nu\, ij}^\prime = m_i T_{ij} - T_{ij}^* m_j + \frac{1}{(4\pi)^2} \left[m_i \widetilde{P}_{ij} + m_j \widetilde{P}_{ji}\right].
\end{equation}
Taking the real an imaginary parts of this equation, we obtain for a Hermitian $P$:
\begin{subequations}
  \begin{align}
    \Re(T_{ij}) &= -\frac{1}{(4\pi)^2} \frac{m_i+m_j}{m_i-m_j}\Re(\widetilde{P}_{ij}),\\
    \Im(T_{ij}) &=-\frac{1}{(4\pi)^2} \frac{m_i-m_j}{m_i+m_j}\Im(\widetilde{P}_{ij}).
  \end{align}
\end{subequations}
The equations resulting from $U^\prime = U T$ can now be used to extract the running of the mixing angles and phases, while the running masses are obtained from Eq.~\eqref{eq:activeMassRGE}.

\section{Extraction of Mixing Angles \& Phases}\label{app:MixingAngles}

\setcounter{equation}{0}  
\renewcommand{\theequation}{C-\arabic{equation}}

The PMNS matrix can be parametrised in the following standard way~\cite{Agashe:2014kda}:
\begin{align}
  U&=
  \left(
  \begin{array}[c]{ccc}
    e^{i \delta_e} & 0 & 0 \\
    0 & e^{i \delta_\mu} & 0 \\
    0 & 0 & e^{i \delta_\tau} \\
  \end{array}
  \right)
V
  \left(
  \begin{array}[c]{ccc}
    e^{-i \phi_1/2} & 0 & 0 \\
    0 & e^{-i \phi_2/2} & 0 \\
    0 & 0 & 1 \\
  \end{array}
  \right)
\textnormal{, with}\\
  V&=
  \left(
  \begin{array}[c]{ccc}
    c_{12} c_{13} & s_{12}c_{13} & s_{13}e^{-i \delta} \\
    -c_{23}s_{12}-s_{23}s_{13}c_{12}e^{i \delta} & c_{23}c_{12}-s_{23}s_{13}s_{12}e^{i \delta} & s_{23}c_{13} \\
    s_{23}s_{12}-c_{23}s_{13}c_{12} e^{i \delta}& -s_{23}c_{12}-c_{23}s_{13}s_{12}e^{i \delta} & c_{23}c_{13} \\
  \end{array}
  \right),
\end{align}
where $c_{ij}\equiv\cos{\theta_{ij}}$ and $s_{ij}\equiv\sin{\theta_{ij}}$.

In our numerical treatment, we diagonalise the active neutrino mass matrix according to Eq.~\eqref{eq:TakagiDecomp} for which we have adapted the algorithm described in~\cite{Hahn:2006hr} to arrive at positive and real $m_{1,2,3}$. We can then extract the mixing angles and phases according to the following relations~\cite{Antusch:2005gp, MPTmanual}:
\begin{subequations}\allowdisplaybreaks
\begin{align}
  \theta_{12} =& 
  \begin{cases}
    \arctan\left|\frac{U_{12}}{U_{11}}\right| & \textrm{if } U_{11} \neq 0, \\ 
    \frac{\pi}{2} & \textrm{else},
  \end{cases}\\
  \theta_{23} =&
  \begin{cases}
    \arctan\left|\frac{U_{23}}{U_{33}}\right| & \textrm{if } U_{33} \neq 0, \\ 
    \frac{\pi}{2} & \textrm{else},
  \end{cases}\\
  \theta_{13} =& \arcsin\left|U_{13}\right|, \\
  \delta_\mu =& \arg\left(U_{23}\right), \\
  \delta_\tau =& \arg\left(U_{33}\right), \\
  \delta =&-\arg\left(\frac{\frac{U_{ii}^*U_{ij}U_{ji}U_{jj}^*}{c_{12}c_{13}^2c_{23}s_{13}}+c_{12}c_{23}s_{13}}{s_{12}s_{23}}\right), \textrm{ for }i\neq j, \\
  \delta_e =& \arg\left(e^{i\delta}U_{13}\right), \\
  \phi_1 =& 2\arg\left(e^{i\delta_e} U_{11}^*\right), \\
  \phi_2 =& 2\arg\left(e^{i\delta_e} U_{12}^*\right).
\end{align}
\end{subequations}
Note that the phases $\delta_e$, $\delta_\mu$, and $\delta_\tau$ are \emph{unphysical} in the sense that they can be absorbed into the three left-handed lepton doublets ${\ell_L}_{1,2,3}$ and do not appear in the analytical expressions for the running mixing angles [Eqs.~\eqref{eq:MixingAnglesAnalytic}] or masses [Eqs.~\eqref{eq:MassesAnalytic}].

\bibliographystyle{./apsrev}
\bibliography{./literature}

\end{document}